\documentclass[twocolumn,epjc3]{svjour3}  

\journalname{Eur. Phys. J. A}
\usepackage[utf8]{inputenc}
\usepackage{epsfig,amsfonts,amssymb,latexsym,amsmath,bm,color,graphics}

\usepackage{graphicx}
\usepackage{hyperref}
\hypersetup{colorlinks=true,linkcolor=blue,citecolor=blue,menucolor=black,urlcolor=blue,filecolor=blue}

\smartqed 
\RequirePackage{mathptmx} 
\RequirePackage[numbers,sort&compress]{natbib}

\usepackage[left=2cm,right=2cm,top=3cm,bottom=3cm]{geometry}
\graphicspath{{./Figures/}}

\begin{document}
\emergencystretch 3em

\title{Remarks on non-perturbative three--body dynamics and its application
to the $KK\bar K$ system}

\author{Xu Zhang\thanksref{e1,addr1,addr2,addr3}
        \and
        Christoph Hanhart\thanksref{e2,addr1} 
        \and
        Ulf-G. Mei{\ss}ner\thanksref{e3,addr4,addr1,addr5} 
        \and
        Ju-Jun Xie\thanksref{e4,addr2,addr3,addr6} 
}

\thankstext{e1}{e-mail: xu.zhang@fz-juelich.de}
\thankstext{e2}{e-mail: c.hanhart@fz-juelich.de}
\thankstext{e3}{e-mail: meissner@hiskp.uni-bonn.de}
\thankstext{e4}{e-mail: xiejujun@impcas.ac.cn}

\institute{Institute for Advanced Simulation, Institut f\"ur Kernphysik and\\ J\"ulich Center for Hadron Physics, Forschungszentrum J\"ulich, D-52425 J\"ulich, Germany\label{addr1}
          \and
         Institute of Modern Physics, Chinese Academy of Sciences, Lanzhou 730000, China\label{addr2}
                  \and
                  School of Nuclear Science and Technology, University of Chinese Academy of Sciences, Beijing 101408, China\label{addr3}
          \and
          Helmholtz-Institut f\"ur Strahlen- und Kernphysik and\\ Bethe Center for Theoretical Physics, Universit\"at Bonn, D-53115 Bonn, Germany\label{addr4}
                  \and
Tbilisi State University, 0186 Tbilisi, Georgia\label{addr5}
\and
School of Physics and Microelectronics, Zhengzhou University, Zhengzhou, Henan 450001, China
\label{addr6}                  
}

\date{}
\maketitle

\abstract{A formalism is discussed that allows for a straightforward treatment of the relativistic three-body problem while keeping the correct analytic structure. In particular it is demonstrated that sacrificing covariance for analyticity can be justified by the hierarchy of different contributions in the spirit of an effective field theory.
For definiteness the formalism is applied to the $KK\bar K$ system allowing for the
emergence of the $a_0(980)$ and the $f_0(980)$ as hadronic molecules. For simplicity all inelastic channels are switched off.}

\section{Introduction}
\label{intro}

Over the past decade, a large number of so-called exotic states have been observed experimentally, especially in the heavy quark sector~\cite{Zyla:2020zbs}, which can not be easily  explained either as $q\bar{q}$ or $qqq$ states. The interpretation of these states has motivated many theoretical studies~\cite{Guo:2019twa,Liu:2019zoy,Guo:2017jvc,Chen:2016spr,Brambilla:2019esw}. 
Understanding their structure will clearly extend our knowledge of the strong interaction dynamics. Many of these states, including the lightest scalar mesons, can be described as emerging from hadron-hadron dynamics and therefore qualify as hadronic molecules, structures analogous to the deuteron in nuclear physics. Although this interpretation
is not yet fully accepted in the literature, it is intriguing to ask, if also three- or even more-body bound states can emerge as well. This question is addressed, e.g., in
Ref.~\cite{Khemchandani:2008rk} and in a series of follow-up works. 
Here we put the focus on the implications of using relativistic kinematics in the scattering equations. For studies investigating this issue for different systems, however, employing non-relativistic kinematics
see, e.g.,  Refs.~\cite{Canham:2009zq,Baru:2011rs,Ma:2017ery,Wang:2018jlv,Wu:2019vsy}

If a two-body system gets embedded into a three-body system, the intrinsic variables need to be handled with care. In particular, the self-energy of a two-body subsystem needs to be evaluated at the subenergy available given the presence of the third particle. While in a non-relativistic system this is all well established and straightforward~\cite{gloecklebuch}, in relativistic systems usually certain
approximations are applied to deal with  the 
kinematic dependence of the subamplitudes. In this work we present the three-body scattering equations in a form close to what is known for two-body scattering that at the same time allow one to properly treat this kinematic dependence even when relativistic variables are employed. We also demonstrate that certain approximations to the choice of kinematic variables can lead to wrong conclusions regarding the emergence of three-body
bound states. The necessity to properly treat subsystem self-energies was stressed already, e.g.,
in Ref.~\cite{Filin:2010se} and the advantages of using relativistic kinematics in three particle systems are presented in Ref.~\cite{Epelbaum:2016ffd}. In particular, the emergence of Efimov states~\cite{Efimov:1971zz,Bedaque:1998kg} is in this way avoided.
 Here we extend the disucssion by studying the significance of the violation of covariance that comes with the equations.

For definiteness we focus on the $KK\bar K$ system in the absence of the $\pi \pi K$ and $\pi \eta K$ inelastic channels, while allowing for $f_0(980)$ and $a_0(980)$ intermediate states, which are included as bound states. While this clearly does not fully represent reality, it still allows us to address the issues mentioned above quantitatively and
to investigate the implications of certain choices of kinematics on the
appearance of three-body states.  As both the $f_0(980)$ and $a_0(980)$ are located close to the $K\bar{K}$ threshold, and couple strongly to this channel, they are widely regarded as bound states with dominant
$K\bar{K}$ component in their wave functions~\cite{Close:1992ay,Oller:1997ti,Nieves:1999bx,Baru:2003qq,Pelaez:2003dy,Kalashnikova:2004ta,Achasov:2005hm,Ambrosino:2006hb,Weinberg:1962hj,Matuschek:2020gqe,Hanhart:2007wa}.
Direct experimental support for this view comes from the observation of a very prominent isospin-violating signal in between the charged and neutral kaon thresholds~\cite{Ablikim:2018pik} that was predicted to emerge
due to $a_0(980)$-$f_0(980)$ mixing~\cite{Hanhart:2007bd,Wu:2008hx,Roca:2012cv}.
A natural candidate for a resulting $KK\bar K$ bound state is then the $K(1460)$ with
$I(J^P)=\frac{1}{2}(0^-)$~\cite{Zyla:2020zbs},
although it is not yet clearly established experimentally. 
First indications for this state were seen at SLAC in the $K\pi \pi$ channel in the reaction
$K^{\pm}p \to K^{\pm}\pi^+\pi^-p$~\cite{Brandenburg:1976pg}. 
The $J^P=0^-$ partial-wave analysis yields a mass around 1400~MeV and width around 250~MeV.  
Later, this state was reported at about 1460~MeV by the ACCMOR Collaboration in the diffractive
process $K^-p \to K^-\pi^+\pi^-p$~\cite{Daum:1981hb}. 
Recently, the LHCb collaboration showed further evidence for the $K(1460)$ in the $\bar{K}^*(892)^0\pi^-$
and $[\pi^+\pi^-]^{L=0}K^-$ channels 
with mass $M_{K(1460)}=1482.40 \pm 3.58 \pm 15.22$~MeV and width $\Gamma_{K(1460)}=335.60 \pm 6.20 \pm
8.65$~MeV~\cite{Aaij:2017kbo}. 
The $K(1460)$ is a good candidate for a three kaon bound state,
since its quantum numbers are consistent with all kaons in a relative $S$-wave and its mass is only a
few MeV below the three-kaon threshold.

The idea of the $K(1460)$ as a three-kaon bound state was investigated,
e.g., in Ref.~\cite{Albaladejo:2010tj}, where a certain triangle diagram was employed as the driving potential.
In Ref.~\cite{Torres:2011jt}, a study was carried out by solving the Faddeev equations for the
$KK\bar{K}$, $K\pi \pi$ and $K\pi \eta$ channels using the two-body inputs provided by unitarized
chiral perturbation theory (in the on-shell approximation). 
A three-body $KK\bar{K}$ quasibound state with $I(J^P)=\frac{1}{2}(0^-)$  was found with a
mass around 1420~MeV which was identified with the $K(1460)$. 
In a more recent work~\cite{Filikhin:2020ksv}, by solving the Faddeev equations in configuration space
within the Gaussian expansion method, 
the $K(1460)$ was identified with a three-body $KK\bar{K}$ bound state  with a mass of 1460~MeV.
We note that in Ref.~\cite{Longacre:1990uc}, within the isobar assumption, even the $Ka_0(980)$ interaction in 
$I(J^P)=\frac{3}{2}(0^-)$ channel generated a resonance above the $Ka_0(980)$ threshold with a mass
around 1500~MeV.  Note, however, that the $K(1460)$ was explained in a relativistic
quark model as the $2^1S_0$  excitation of the kaon~\cite{Godfrey:1985xj}.

In the present work as well as most of those mentioned above, 
the $KK\bar{K}$ three-body system in $I(J^P)=\frac{1}{2}(0^-)$ channel is studied using the isobar approach, 
where the two-body $K\bar{K}$ interaction is parametrised via the $f_0(980)$ and $a_0(980)$ poles.
Such a formalism satisfies two-body and three-body unitarity~\cite{Aaron:1969my,Mai:2017vot}.
The latter plays an import role as shown in Refs.~\cite{Janssen:1994uf,Sadasivan:2020syi} for
the $\rho \pi$ scattering. A very general formalism for  $3 \to 3$ scattering in the isobar approach was
presented in Ref.~\cite{Jackura:2018xnx}; decay amplitudes with three particles in the final-state 
may be calculated employing Khuri-Treiman equations~\cite{Khuri:1960zz} (those were used more recently in Refs.~\cite{Niecknig:2012sj,Niecknig:2015ija,Isken:2017dkw,Niecknig:2017ylb,Gasser:2018qtg,Albaladejo:2019huw,Albaladejo:2020smb}).
It turns out that the resulting equations are quite involved and demanding to solve. The goal we aim at here is in contrast
to this to present an easy to handle formalism that keeps relativistic kinematics, but sacrifices
covariance for the sake of simplicity. Our formalism is derived employing time ordered perturbation
theory (TOPT) rigorously. In particular, our amplitudes are constructed to keep track of the leading singularities of the amplitudes.
An alternative formulation, that leads to very similar expressions
is presented in Ref.~\cite{Mai:2017vot}. The formal covariance of this treatment is achieved
by modifications of some the contributions
to the potential. As stressed in Ref.~\cite{Dawid:2020uhn} this introduces unphysical singularities.
We demonstrate that avoiding those modifications removes the unphysical singularities
and at the same time only introduces a very mild violation of covariance that moreover can
be removed systematically order by order.

The paper is organized as follows. 
In Sec.~\ref{sec:effe}, the derivation of the interaction potential for the
concrete example employed here for illustration is given. 
In Sec.~\ref{sec:lipp}, we present the Lippmann-Schwinger-type equation which fulfills two-body and
three-body unitarity. 
The numerical results are discussed in Sec.~\ref{sec:result}. In this section we also study the impact
of different  approximations on the numerical results.

\section{Effective Potentials}
\label{sec:effe}

\subsection{The  Lagrangian and coupling constant}

For the $f_0(980)K\bar{K}$ and $a_0(980)K\bar{K}$ vertices, we use a scalar coupling~\cite{Lohse:1990ew}
(note that in a more sophisticated calculation, the Goldstone boson nature of the kaon should be
accounted for)
\begin{align}
\mathcal{L}=\frac{f_1}{\sqrt{2}}\, f_0\bar{K}K + \frac{f_2}{\sqrt{2}}\,
\bar K\left( \vec a_0\cdot \vec \tau\right)K \ ,
\end{align}
with
\begin{align}
K \equiv 
\begin{pmatrix}  K^+\\
K^0   
\end{pmatrix}, 
\qquad  \bar{K} \equiv (K^-,~ \bar{K}^0).
\end{align}
Here, $f_0$ and $\vec a_0$ denote the fields of the scalar-isoscalar $f_0(980)$ and
the scalar-isovector $a_0(980)$, respectively, where the three components of the latter
refer to the different charge states.
The coupling constants $f_S$ ($f_1$ and $f_2$) can be determined under the assumption 
that the $f_0(980)$ and $a_0(980)$ are pure bound states of $K\bar{K}$
system~\cite{Weinberg:1962hj,Kalashnikova:2004ta,Hanhart:2007wa}
\begin{align}
\frac{f_S^{2}}{4\pi}=32m_K\sqrt{m_K\varepsilon_S}~,
\label{eq:epsgrelat}
\end{align}
with $m_K$  the mass of $K$ meson, and $\varepsilon_S$ is the binding energy of
the scalar bound state $S$. 
It should be stressed in this context that Eq.~(\ref{eq:epsgrelat}) provides an upper bound for the coupling 
fixed by the normalisation of a bound state ---
if a state contains a non-molecular component, the coupling would be lower~\cite{Weinberg:1962hj,Guo:2017jvc}
(for an extension of the concept to virtual states see Ref.~\cite{Matuschek:2020gqe}).

Here we chose the binding energy equal for the two scalar states
and we use the following masses
\begin{align}
m_K=495 \   \text{MeV}, \quad m_{f_0(980)}=m_{a_0(980)}=980 \   \text{MeV}.
\end{align}
This leads to 
\begin{align}
 f_1= f_2=f_S=3.74\   \text{GeV},
\end{align}
which agrees with the experimental result of Ref.~\cite{Ambrosino:2006hb}. Clearly, for a realistic
calculation both the $\pi\pi K$ and the $\pi \eta K$ system need to be taken into account as well,
however, since our
focus lies on the formalism, what is introduced here is sufficient. For the same reason we also do not
try to better  constrain the input data for the $f_0(980)$ and the $a_0(980)$. All this will be
improved in a subsequent study.

\subsection{Coupled-channel matrix elements}

The formalism for the three-body scattering used here employs the scattering of two quasi-particles by 
means of a Lippmann-Schwinger
(LS) type equation. This implies that the scattering potential needs to contain the exchanges of the
constituents
of the quasi-particles.
Accordingly we may decompose the $S$-wave $Kf_0(980)$-$Ka_0(980)$ interaction 
to second order in the coupling $f_S$ as
\begin{align}
\bm{V}(E,p',p)=\bm{V}_t(E,p',p)  +  \bm{V}_s(E,p',p),
\end{align}
where $\bm{V}_t(E,p',p)$ and $\bm{V}_s(E,p',p)$ are the $t$- and $s$-channel one-kaon exchange, respectively.

\begin{figure*}[t]
\begin{center}
\includegraphics[scale=0.35]{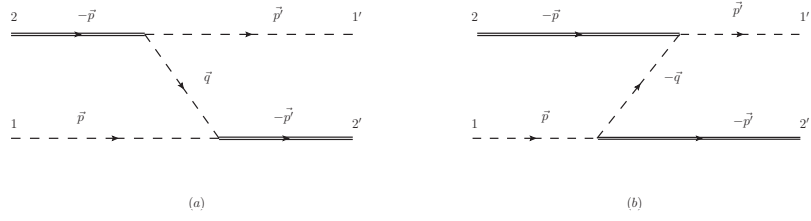}
\caption{Diagrams for the $t$-channel kaon exchange contribution. The double-solid and dashed line
  represent the $f_0(980)$ or $a_0(980)$ and the $K$ or $\bar{K}$ meson, respectively.}
\label{fig:tchannel}
\end{center}
\end{figure*}

The interaction potential $\bm{V}_t(E,p',p)$ reads in channel space
\begin{align}
\label{eq:vmatrix}
\bm{V}_t(E,p',p) =
\begin{pmatrix} 
V^{11}_{tS}(E,p',p) & \  V^{12}_{tS}(E,p',p)\\       \\  
V^{21}_{tS}(E,p',p) & \  V^{22}_{tS}(E,p',p)
\end{pmatrix} .
\end{align}
The same structure holds for $\bm{V}_s(E,p',p)$.
In each matrix element $V^{\lambda' \lambda}_{tS}(E,p',p)$ and ${V}_{sS}^{\lambda' \lambda}(E,p',p)$, the
index $\lambda (\lambda') = 1, 2$ denotes the particle channel ($Kf_0(980)=1$, $Ka_0(980)=2$) and the
$S$ denotes the $S$-wave projection of the potential~\cite{Lohse:1990ew} (see also Ref.~\cite{Gulmez:2016scm})
\begin{align}
V^{\lambda' \lambda}_{tS}(E, p',p)&= \frac{1}{2} \int_{-1}^1 V^{\lambda' \lambda}_t(E,\vec{p'},\vec{p}) \, dcos\theta \ .
\end{align}
In the expressions above $E$ denotes  the total energy of
the system and $\vec{p}$ and $\vec{p'}$ are the incoming and the outgoing
momenta.

Since we focus on a possible bound state with $I(J^P)=\frac{1}{2}(0^-)$, the flavor wave functions
of the $Kf_0(980)$ and $Ka_0(980)$ systems can be written as 
\begin{align}
\left|\frac{1}{2}, \frac{1}{2}\right\rangle&=-\left|K^+f_0(980)\right\rangle,  \nonumber \\
\left|\frac{1}{2}, \frac{1}{2}\right\rangle&=-\sqrt{\frac{2}{3}}\left|K^0a_0(980)^+\right\rangle-\sqrt{\frac{1}{3}}\left|K^+a_0(980)^0\right\rangle,  
\end{align}
using the convention that $|K^+\rangle=-|\frac{1}{2},\frac{1}{2}\rangle$.

\begin{figure*}[t]
\begin{center}
\includegraphics[scale=0.35]{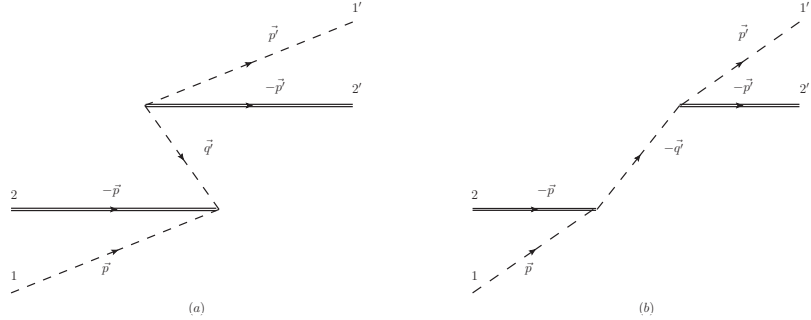}
\caption{Diagrams for the $s$-channel kaon exchange contribution. The double-solid and dashed
  line represent the $f_0(980)$ or $a_0(980)$ and the $K$ or $\bar{K}$ meson, respectively.}
\label{fig:schannel}
\end{center}
\end{figure*}

In TOPT, the $t$-channel one-kaon exchange
potential acquires two contributions depicted in Fig.~\ref{fig:tchannel}(a)~and~(b).
In the center-of-mass frame, for the scattering process $12\to 1'2'$ with $\vec{q}=-\vec{p}-\vec{p'}$,
the potential can be written as 
\begin{align}
V^{\lambda' \lambda}_t(E,\vec{p'},&\vec{p})=f_S^2{\cal I} N \frac{1}{2\omega_K(\vec{q})}  \nonumber \\
&\times \left[\frac{1}{E-\omega_{1'}(p')-\omega_K(\vec{q})-\omega_1(p)} \right. \nonumber\\
&\qquad + \left.\frac{1}{E-\omega_{2'}(p')-\omega_K(\vec{q})-\omega_2(p)} \right],
\label{eq:Vt}
\end{align}
with 
\begin{align}
\omega_i(p)&=\sqrt{m_i^2+\vec{p}^2}, \quad  i=1^{(')}, 2^{(')},\nonumber\\
\omega_K(\vec{q})&=\sqrt{m_K^2 +  p^2+ {p'}^2+2 p p' cos\theta },
\end{align}
where $\theta$ is the angle between $\vec{p}$ and $\vec{p'}$. 
The isospin factors ${\cal I}$ are listed in Table~\ref{tab:isf}. 

Since we work in TOPT, all the potentials contain the normalization factor
\begin{align}
\label{eq:norfac}
N=\frac{1}{\sqrt{16\omega_1(p)\omega_2(p)\omega_{1'}(p')\omega_{2'}(p')}}.
\end{align}


Analogously, the $s$-channel one-kaon exchange potential acquires the two contributions
depicted in Fig.~\ref{fig:schannel}(a)~and~(b). The potential can be written as 
\begin{align}
V^{\lambda' \lambda}_s&(E,\vec{p'},\vec{p})= {\cal I} N\frac{1}{2 m_K}\nonumber \\
&\times\left[\frac{f_S^2}{E-\omega_1(p)-\omega_{1'}(p')-m_K-\omega_2(p)-\omega_{2'}(p')} \right. \nonumber\\
&\qquad\qquad \qquad\qquad \qquad\qquad 
+ \left. \frac{f_{\lambda'}^{(0)}f_{\lambda}^{(0)}}{E-m_K^{(0)}}\right] \ ,
\label{eq:Vsch}
\end{align}
where again the isospin factors are listed in Table~\ref{tab:isf}. In the expression above it is
already used that the scattering equation is solved in the overall center-of-mass frame.
Moreover, the notation distinguishes explicitly between the physical parameters $f_S$ and $m_K$,
 {and the bare parameters $f_\lambda^{(0)}$ and $m_K^{(0)}$ that get renormalized by the scattering equation.}
How the latter parameters are determined is described in the next section. 
We can see that  $V^{\lambda' \lambda}_s(E,\vec{p'},\vec{p})$ is independent of the scattering
angle $\theta$ and thus does not need
to be partial-wave projected.

\begin{table*}
\caption{Isospin factors for one-kaon exchange potentials and box diagram contributions}
\begin{center}
\begin{tabular}{cccc}
\hline
~~~~~~~~~                                            &  $Kf_0(980) \to Kf_0(980)$        & $Kf_0(980) \to Ka_0(980)$          &  $Ka_0(980) \to Ka_0(980)$\\
\hline 
$t$-channel                                                           &             $\frac{1}{2}$                    &      $ \frac{\sqrt{3}}{2} $               &      $-\frac{1}{2}$                           \\
\hline
$s$-channel                                                          &             $\frac{1}{2}$                    &      $ \frac{\sqrt{3}}{2} $                &      $\frac{3}{2}$                              \\
\hline 
single-channel:  stretched  boxes in Fig.~\ref{fig:stret},~\ref{fig:stret-negs},~\ref{fig:stret-negk} and \ref{fig:stret-negks}.  
                                                                              &             $\frac{1}{4}$                    &     ---                                            &          ---                                         \\
\hline
single-channel:   crossed  boxes in Fig.~\ref{fig:crossed},~\ref{fig:crossed-negs},~\ref{fig:crossed-negk} and \ref{fig:crossed-negks}.
                                                                              &             $\frac{1}{4}$                    &      ---                                             &        ---                                          \\\hline
coupled-channel:   stretched  boxes in Fig.~\ref{fig:stret},~\ref{fig:stret-negs},~\ref{fig:stret-negk} and \ref{fig:stret-negks}.  
                                                                                  &             1                                   &      0                                             &              1                                         \\
\hline
coupled-channel:  crossed  boxes in Fig.~\ref{fig:crossed},~\ref{fig:crossed-negs},~\ref{fig:crossed-negk} and \ref{fig:crossed-negks}.                     
                                                                                  &             1                                   &      0                                             &              1                                         \\
\hline
\end{tabular}
\end{center}
\label{tab:isf}
\end{table*}

\section{Lippmann-Schwinger-type equation}
\label{sec:lipp}

The partial wave decomposed LS equation can be written as 
\begin{align}
\label{eq:lst}
\bm{T}(E,p',p)=&\bm{V}(E, p', p)\nonumber\\
&+\int_0^{\Lambda} \frac{4\pi k^2dk}{(2\pi)^3}\bm{V}(E, p', k)\bm{G}(E,k)\bm{T}(E,k,p),
\end{align}
with the definitions
\begin{align}
\bm{G}(E,k) = 
\begin{pmatrix} 
                           G_r^1(E,k)&  \   0             \\          \\  
                              0         &  \  G_r^2(E,k) 
\end{pmatrix}, 
\end{align}
 {where the renormalized isobar-$K$ propagators are 
\begin{align}
\label{eq:pro}
 G_{r}^{\lambda}(E,k)= \Big[&Z\Big(E-\omega^{(\lambda)}(k)-\omega_K(k)\Big) \nonumber\\
&\qquad\qquad-\alpha f_S^2\Sigma_R^{(\lambda)}(E,k)\Big]^{-1} ,
 \end{align}
as we show in~\ref{app:f0renorm}.}

The full potential contains all contributions free of $(f_0(980)/a_0(980)) K$ cuts.
The part of it to second order in the coupling was defined in the previous
section. In the formalism employed here the matrix $\bm{G}$ describes the propagation of an
$a_0(980)/f_0(980) K$ intermediate state. Accordingly, $\omega^{(1)}(k)$ ($\omega^{(2)}(k)$) is the 
energy of an on mass-shell $f_0(980)$ ($a_0(980)$) with momentum $k$. 
The self-energy $f_S^2\Sigma_R^{(\lambda)}(E,k)$ captures the effect of the two-meson
loops on the resonance propagators.  {The renormalization factor $Z$ is defined as $Z=f_S^2/{f_S^0}^2$,
which is introduced in Eq.~(\ref{eq:pro}) to satisfy the condition that
the residue of the renormalized propagator $G_{r}(E,k)$ is one at $E=\omega^{(\lambda)}(k_{on})
+\omega_K(k_{on})$.} (We will come back to this issue in following.)
A key study of this work is to investigate the impact of
the self-energy, and in particular certain approximations thereof, on the potentially emerging
three-body bound states. For that study we introduced the parameter $\alpha$ that will eventually be
varied  $0\le \alpha \le1$, with $\alpha=1$ representing the fully unitary treatment, while $\alpha=0$
leads above the three-kaon threshold to a violation of unitarity and accordingly below this
threshold  to a violation of analyticity. The reason for this is that above the three-$K$ threshold 
the self-energy generates an imaginary part that is necessary for the equations to be unitary.
Below threshold this term needs to be continued analytically as otherwise the amplitude 
suffers from unphysical non-analyticities.

\begin{figure}[t]
\begin{center}
\includegraphics [scale=0.35] {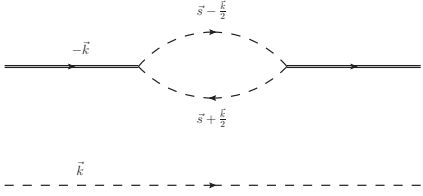}
\caption{Time-ordering for the self-energy correction of $f_0(980)$ and $a_0(980)$ mesons.
The double-solid line represents $f_0(980)$ or $a_0(980)$ meson, and the dashed line represents $K(\bar{K})$.
}
\label{fig:bubble}
\end{center}
\end{figure}

In Fig.~\ref{fig:bubble}, all relevant momenta are shown explicitly for the self-energy
correction of the $f_0(980)$ or the $a_0(980)$ meson in TOPT.

The expression corresponding to Fig.~\ref{fig:bubble} may be written in the following form
\begin{align}
{f_S^0}^2\Sigma^{(\lambda)}(E,k)= & \frac{{f_S^0}^2}{2\omega^{(\lambda)}(k)} \int \frac{d^3s}{(2\pi)^3}\frac{1}{4\omega_K(s_+)\omega_K(s_-)}
\nonumber \\
&\times\frac{1}{E-\omega_K(k)-\omega_K(s_+)-\omega_K(s_-) +i \varepsilon} \ ,
\label{signonren}
\end{align}
where $s_\pm = s\pm k/2$.
Since in this exploratory study the inelasticities of the $a_0(980)$ and $f_0(980)$ are neglected, both states appear as stable bound states. To ensure that the amplitudes 
generate the  $a_0(980)/f_0(980) K$ branch cuts 
correctly, instead of Eq.~(\ref{signonren}) in the three-body equations we need to employ a
renormalized self-energy as shown in~\cite{Hanhart:2010wh} and~\ref{app:f0renorm} 
\begin{align}
\label{eq:regu}
f_S^2\Sigma_R^{(\lambda)}(E,k)= f_S^2\Sigma^{(\lambda)}(E,k)-\mbox{Re}\left(f_S^2\Sigma^{(\lambda)}(E,k_{on})\right),
\end{align}
where 
\begin{align}
k_{on}=\frac{\sqrt{[E^2-(m^{(\lambda)}+m_K)^2][E^2-(m^{(\lambda)}-m_K)^2]}}{2E}.
\end{align}
Note that this procedure needs to be generalized when  inelastic channels [for our example
  $\pi\pi$ for the $f_0(980)$ and $\pi\eta$ for the $a_0(980)$] are switched on, since
the mentioned cut does not disappear, but moves into the complex plane of the unphysical
sheet~\cite{Doring:2009yv}, with the location of the corresponding branch point being related to
the complex pole position of the resonance involved.

The expression of the self-energy as defined in Eq.~(\ref{signonren}) contains both $k$ and $E$
in a non-trivial way. In case of non-relativistic kinematics, one finds
\begin{equation}
\omega_K(s_+)+\omega_K(s_-) = 2\omega_K(s) + k^2/(4m_K) \ , 
\label{eq:energies}
\end{equation}
where the $k^2$ term captures the kinetic energy of the two-kaon system in the overall 
center-of-mass frame. With this, the $k$ and $E$ dependence of the self-energy can be absorbed into
an effective energy
$$
E^{\rm eff}(E,k) = E - \omega_K(k) - k^2/(4m_K) \ 
$$
and the self-energy depends on this single variable only.
However, for relativistic kinematics this appears to be not possible and both the  $k$
and the $E$ dependence need to be kept explicitly.
While  $k\ll m_K$ and $s\ll m_K$  might be a good approximation for the system studied here in the
absence of inelastic
channels, it is certainly invalid as soon as the $\pi\pi$ and $\pi\eta$ channels are included. 
Because of this and the reason presented in the introduction,
we  proceed using relativistic kinematics. Note that, to speed up the numerical solution
of the integral equation, the
self-energy can be calculated for the energies of interest outside the routine that fills the potential
directly on the grid employed for the discretisation of the integral in the LS equation.

\begin{figure*}[t]
\begin{center}
\includegraphics [scale=0.48] {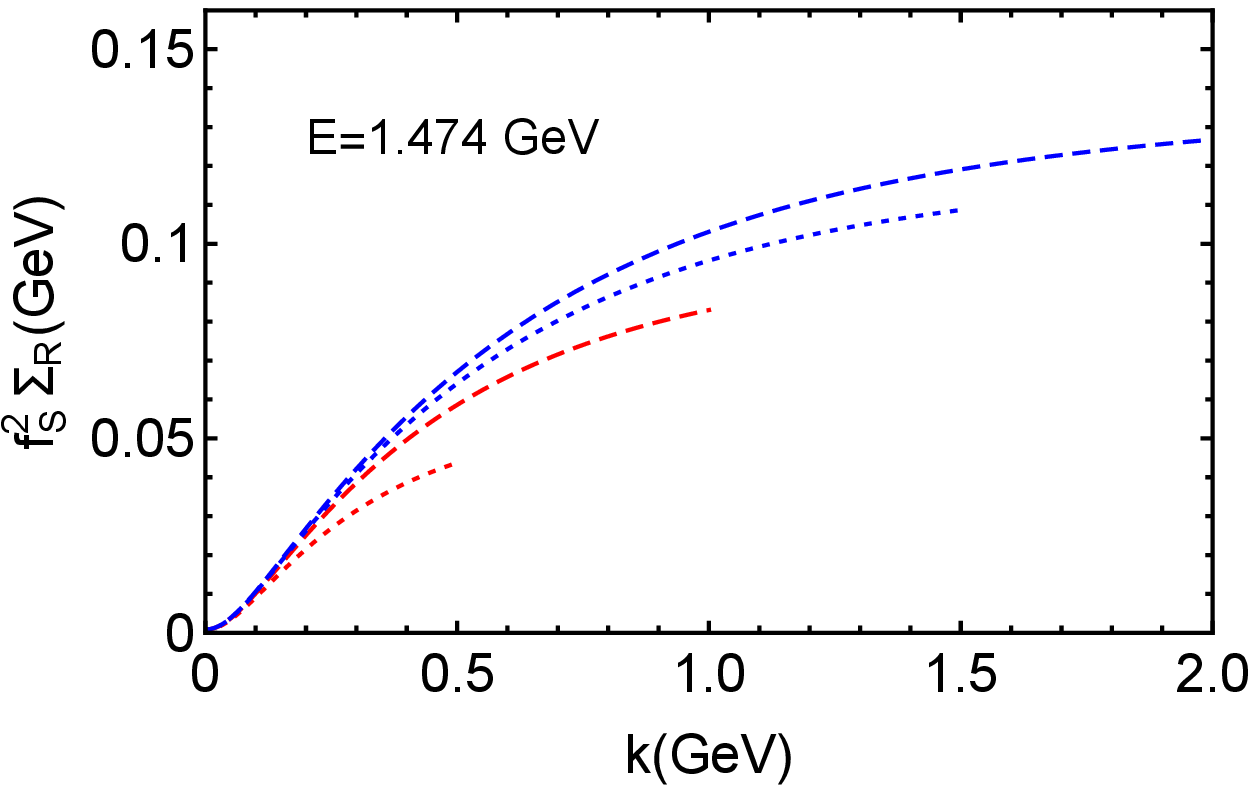}
\includegraphics [scale=0.48] {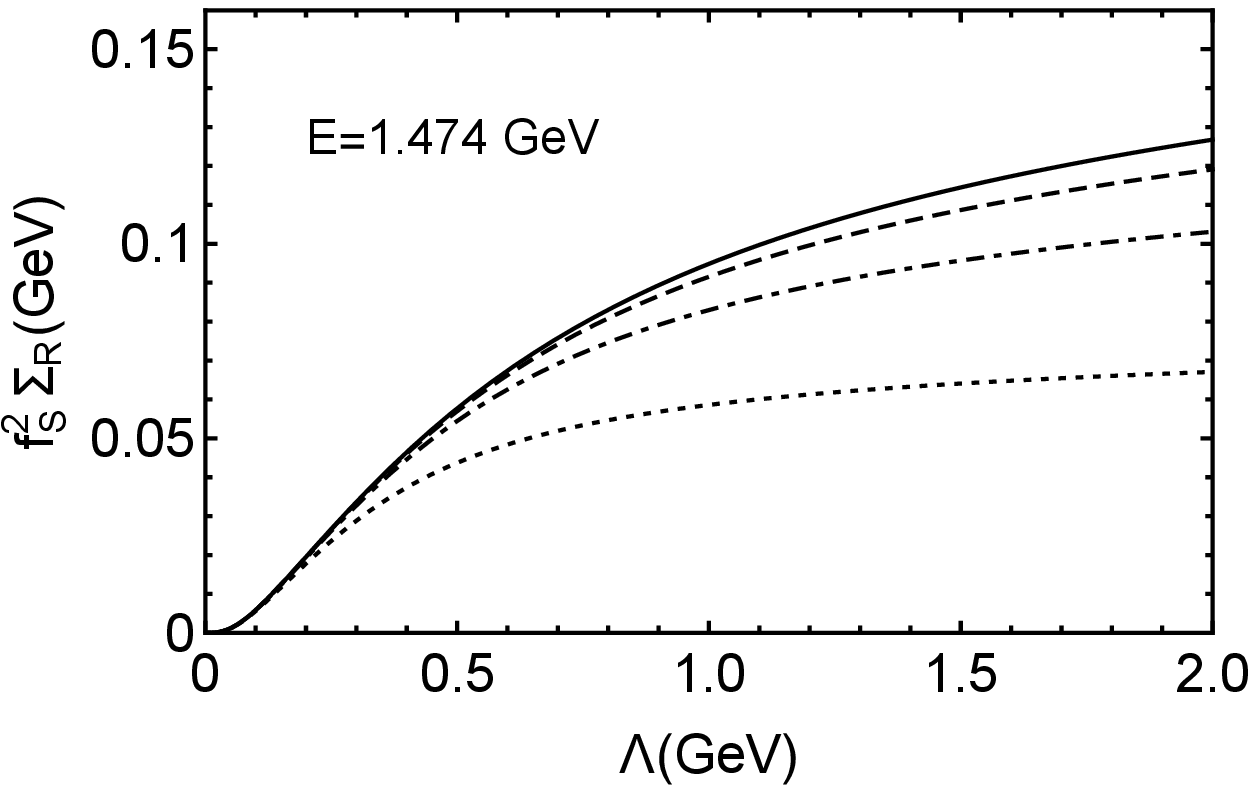}
\caption{Illustration of the momentum (left panel) and cutoff dependence of the renormalized
  self-energy $f_S^2\Sigma_R(E,k)$.
Left panel:
The red dotted, red dashed, blue dotted and blue dashed lines correspond to $\Lambda= 0.5$, 1.0, 1.5 and 2.0~GeV,
respectively.
Right panel: The black dotted, black dotted-dashed, black dashed and black lines correspond to $k= 0.5$,
1.0, 1.5 and 2.0~GeV, respectively.
}
\label{fig:renor}
\end{center}
\end{figure*}

To render the LS equation in Eq.~\eqref{eq:lst} 
well defined, it is regularized by a finite momentum cutoff $\Lambda$.
For consistency the same cutoff is employed for the self-energy defined in Eq.~\eqref{eq:regu},
although the regularized expression is formally convergent.
In our calculation,  we vary the value of
 $\Lambda$ in the range from 0.5~GeV to 2.0~GeV. To illustrate the resulting 
cutoff dependence of $f_S^2\Sigma_R(E,k)$, the numerical results corresponding to $E=1.474$~GeV are
shown in Fig.~\ref{fig:renor}.  
 
\begin{figure}[t]
\begin{center}
\includegraphics [scale=0.48] {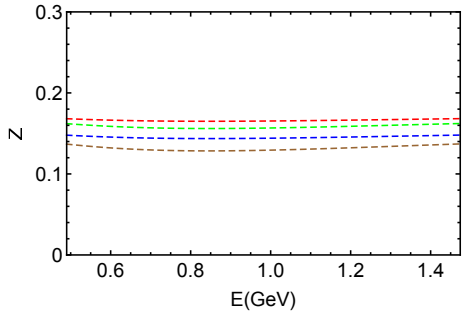}
\caption{The $E$ dependence of the renormalization constant $Z$.
The red (green) dashed and blue (brown) dashed lines correspond to $\Lambda= 1.0$ and 2.0~GeV,
respectively. For the green dashed and brown dashed lines, the subleading contribution to the 
self energy, Fig.~\ref{fig:retardbubble}, was also included.
}
\label{fig:ConstZ}
\end{center}
\end{figure}

From the condition that the residue of the renormalized propagator $G_{r}(E,k)$ in Eq.~(\ref{eq:pro})
is one at 
 \begin{equation}
 E=E_{on}=\omega^{(\lambda)}(k_{on})+\omega_K(k_{on}) \ ,
 \label{Eondef}
 \end{equation}
we get 
\begin{align}
Z=1+\frac{d}{dE}f_S^2\Sigma^{(\lambda)}(E,k_{on}) \big |_{E=E_{on}} \ .
\end{align}
The derivation of this expression is presented in~\ref{app:f0renorm}.
The physical value of $f_S$ is fixed to be 3.74~GeV. The values of $Z$ corresponding calculated in this way
are quoted in Tab.~\ref{tab:Zconst}. In addition we
show their energy dependence for $\Lambda=1$~GeV
and 2~GeV
in Fig.~\ref{fig:ConstZ}. Thus we find a 
negligible energy dependence of the $Z$ factor as it
should be in general, however, this feature could have been distorted
here by the non-covariance of the formalism.

\begin{table}
\caption{The renormalization factor $Z$. In all
cases $f_S=3.74$~GeV was employed. The $\dagger$-symbol is added
to the $\alpha$-value when the subleading contribution to the 
self energy, Fig.~\ref{fig:retardbubble}, was also included.}
\begin{center}
\begin{tabular}{cccc}
\hline
$E$ [GeV] & $\alpha$ & $\Lambda$ [GeV] & $Z$                  \\
\hline
0.495       & 1              & 0.5                         & 0.223            \\
\hline
0.495       & 1              & 1                         & 0.168            \\
\hline
0.495       & 1              & 1.5                         & 0.154            \\
\hline
0.495       & 1              & 2                         & 0.148            \\
\hline
0.495       & $1^\dagger$              & 0.5                         & 0.221            \\
\hline
0.495       & $1^\dagger$              & 1                         & 0.162            \\
\hline
0.495       & $1^\dagger$              & 1.5                         & 0.145            \\
\hline
0.495       & $1^\dagger$              & 2                        & 0.137           \\
\hline
1.474       & 1              & 0.5                         & 0.224            \\
\hline
1.474       & 1              & 1                         & 0.168            \\
\hline
1.474       & 1              & 1.5                         & 0.154            \\
\hline
1.474       & 1              & 2                         & 0.148            \\
\hline
1.474       & $1^\dagger$              & 0.5                         & 0.222            \\
\hline
1.474       & $1^\dagger$              & 1                         & 0.162            \\
\hline
1.474       & $1^\dagger$              & 1.5                         & 0.145            \\
\hline
1.474       & $1^\dagger$              & 2                        & 0.137           \\
\hline
\end{tabular}
\end{center}
\label{tab:Zconst}
\end{table}

Most of the integrals entering the LS equation are formally convergent. Only those
that contain the $s$--channel diagrams lead to a divergence and
correspondingly may lead to a sizeable regulator dependence. However, at least the divergence
in the one particle reducible diagrams introduced
via the kaon pole diagram
can be absorbed into mass and wave function regularization.
 In this procedure the bare parameters $f_1^{(0)}$, $f_2^{(0)}$ and $m_K^{(0)}$, introduced in Eq.~(\ref{eq:Vsch}),
are determined from
\begin{align}
{f_1^{(0)}}^2=\frac{1}{2}f_S^2\Big/\Big[&(\Gamma_{11}+ R\Gamma_{12} ) (\Gamma_{11}^{T}+ R\Gamma_{21}^{T} )+\frac{1}{2}f_S^2 \nonumber \\
&\times({\Sigma_{11}^{(3)}}'+2R{\Sigma_{12}^{(3)}}' +R^2{\Sigma_{22}^{(3)}}')\Big],
\label{eq:Kwffrenorm}
\end{align}
\begin{align}
\label{eq:Kwfarenorm}
f_2^{(0)}=f_1^{(0)}R,
\end{align}
and
\begin{align}
m_K^{(0)}= m_K -\Big(&{f_1^{(0)}}^2{\Sigma_{11}^{(3)}}+2f_1^{(0)}f_2^{(0)}{\Sigma_{12}^{(3)}}+{f_2^{(0)}}^2{\Sigma_{22}^{(3)}}\Big), 
\end{align}
with
\begin{align}
R=\frac{\sqrt{3}\Gamma_{11}-\Gamma_{21}}{-\sqrt{3}\Gamma_{12}+\Gamma_{22}},
\end{align}
and
\begin{align}
{\Sigma_{\lambda'\lambda}^{(3)}}'=\frac{d}{dE}{\Sigma_{\lambda'\lambda}^{(3)}}(E) \big |_{E=m_K} .
\end{align}
Here we used the short hand notation $\Gamma_{\lambda'\lambda}=\Gamma_{\lambda'\lambda}(m_K,0)$ and $\Sigma_{\lambda'\lambda}^{(3)}=\Sigma_{\lambda'\lambda}^{(3)}(m_K)$.
Explicit expressions for the self energy of the kaon pole  and the dressed vertex function
 as well as the derivation
of Eq.~(\ref{eq:Kwffrenorm}) and Eq.~(\ref{eq:Kwfarenorm}) are presented in~\ref{app:Krenorm}.



The solution of the LS equation is found by straightforward numerical matrix inversion.
For this we use the method given in Ref.~\cite{Haftel:1970zz}. In our calculation, a 40-point
Gaussian quadrature 
yields stable results. Note that since we only study energies below the $a_0/f_0 K$ threshold, no three-body
singularities need to be dealt with numerically.

The requirement that Eq.~\eqref{eq:lst} has a pole at some energy $E$ is equivalent to the condition 
 \begin{align} 
\text{ det}
\left [\uppercase\expandafter{\romannumeral1} - \bm{V(E)}\bm{ G(E)} \right ]=0 \ .
\end{align}
For a given pole the binding energy is $E_B=m^{(1)}+m_K-E$, since we measure
the energy relative to the $f_0(980) K$ threshold (which for the parameters employed here
equals to the $a_0(980) K$ threshold).



\section{Numerical Results for the potential quadratic in $f_S^2$}
\label{sec:result}


For our study all parameters are fixed as discussed above,
however, we still quote the
bare, calculated parameters for the single and the
coupled channel calculation in Tab.~\ref{tab:para}
to illustrate that the renormalization effects can
in fact be quite sizeable. Moreover, note that
for the coupled channel case, although the dressed couplings of $a_0(980)$ and $f_0(980)$
to kaon-antikaon are equal, the corresponding bare couplings are different due to the different
isospin factors in the different channels.

\begin{figure}[t]
\begin{center}
\includegraphics [scale=0.48] {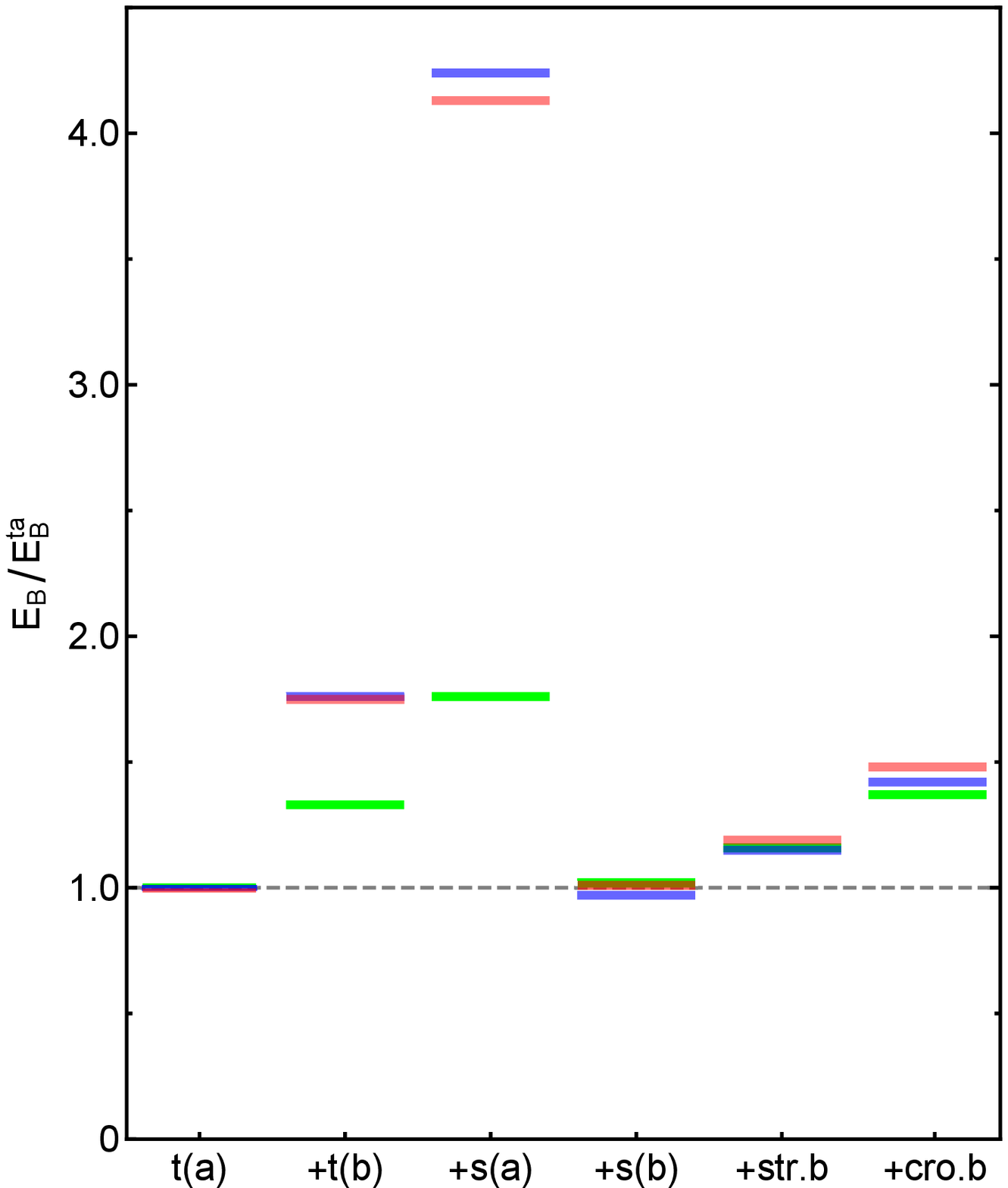}
\caption{The numerical results for the binding energies
for the  coupled-channel $Kf_0(980)$-$Ka_0(980)$ formalism for the
individual contributions of the potential added in one-by-one with $f_S=3.74$~GeV and $\Lambda=1$~GeV.
The green bars correspond to $\alpha=0$, the blue bars correspond to $\alpha=1$ and  the red bars correspond 
to $\alpha=1^\dagger$ where the subleading contribution to the 
self energy, Fig.~\ref{fig:retardbubble}, was also included. The labels for the first 4 contributions refer to those of Figs.~\ref{fig:tchannel} and \ref{fig:schannel}. The last two contributions are labeled collectively via the types of the diagrams.}
\label{fig:SkeCut1000}
\end{center}
\end{figure}

We start the discussion by omitting the effect of the self-energy $\Sigma_R^{(\lambda)}(E,k)$ in
Eq.~(\ref{eq:pro}) by setting 
$\alpha=0$ and $Z=1$. In this case  the three-body scattering
generates a bound state pole very close to the threshold on the physical
sheet as soon as we study the coupled-channel $Kf_0(980)$-$Ka_0(980)$ formalism is used --- the corresponding
binding energies that arise when more and more terms
are added to the potential are shown
by the columns $f(a)-s(b)$ in the lines
marked by $\alpha=0$ in Tab.~\ref{tab:result-three-body}
and as the first four green bars in Fig.~\ref{fig:SkeCut1000}, where the relative importance of the different contributions calculated for $\Lambda=1$ GeV is illustrated.
For the single-channel $Kf_0(980)$ formalism, the three-body scattering does not generate a bound
state, reflecting that the coupled-channel effect has a strong influence on the scattering process. Here, we employed $f_S=3.74$~GeV, fixed via Eq.~(\ref{eq:epsgrelat}) by the masses
of $f_0(980)$ and $a_0(980)$.

\begin{table}
\caption{The calculated bare parameters 
for the single-channel (s)
and coupled-channel (cc) calculation. In all
cases $f_S=3.74$~GeV was employed. The $\dagger$-symbol is added
to the $\alpha$-value when the subleading contribution to the 
self-energy, Fig.~\ref{fig:retardbubble}, was also included.}
\begin{center}
\begin{tabular}{cccccc}
\hline
type & $\alpha$ & $\Lambda$ [GeV] & $f_1^{(0)}$  [GeV] & $f_2^{(0)}$  [GeV] & $m_K^{(0)}$ [GeV]                  \\
\hline
s & $0$ & 0.5 & 3.72     & $-$                                          &         0.500     \\
\hline
s & $0$ & 1 & 3.70     & $-$                                          &         0.513     \\
\hline
s & $0$ & 1.5 & 3.68     & $-$                                          &         0.526     \\
\hline
s & 0 & 2 &
 3.68     & $-$                                          &         0.537      \\
 \hline
s & 1 & 0.5 &
  3.67           & $-$                                    &         0.513       \\
\hline
s & 1 & 1 &
  3.50           & $-$                                    &         0.571       \\
 \hline
s & 1 & 1.5 &
  3.38          & $-$                                    &         0.628       \\
\hline
s & 1 & 2 &
  3.31          & $-$                                     &          0.676      \\
\hline
s & $1^\dagger$ & 0.5 &  3.66              & $-$                                     &         0.513     \\
\hline
s & $1^\dagger$ & 1 &  3.48              & $-$                                     &         0.570      \\
\hline
s & $1^\dagger$ & 1.5 &  3.34              & $-$                                     &         0.626      \\
\hline
s & $1^\dagger$ & 2 &  3.25            & $-$                                       &          0.671      \\
\hline
cc & $0$ & 0.5 &  3.67                         &              3.75          &         0.515      \\
\hline
cc & $0$ & 1 &  3.57                         &              3.76          &         0.569      \\
\hline
cc & $0$ & 1.5 &  3.51                        &              3.76          &         0.623      \\
\hline
cc & $0$ & 2 & 3.48                         &              3.75          &         0.669      \\
\hline
cc & $1$ & 0.5 & 3.44                        &              3.73        &         0.567     \\
\hline
cc & $1$ & 1 & 2.75                        &              3.59        &         0.798     \\
\hline
cc & $1$ & 1.5 & 2.28                        &              3.40        &         1.005     \\
\hline
cc & $1$ & 2 & 1.99                        &              3.24         &         1.155      \\
\hline
cc & $1^\dagger$ & 0.5& 3.43                        &              3.72         &         0.567 \\
\hline
cc & $1^\dagger$ & 1 & 2.69                        &              3.52         &         0.787  \\
\hline
cc & $1^\dagger$ & 1.5 & 2.18                       &              3.26         &         0.963 \\
\hline
cc  & $1^\dagger$ & 2 & 1.87                         &             3.04        &         1.077     \\
\hline
\end{tabular}
\end{center}
\label{tab:para}
\end{table}

\begin{table*}
\begin{center}
\caption{The numerical results for the binding energies
for the  coupled-channel $Kf_0(980)$-$Ka_0(980)$ formalism for the
individual contributions of the potential added in one-by-one with $f_S=3.74$~GeV.
The labels for the first 4 contributions refer to those
of Figs.~\ref{fig:tchannel} and \ref{fig:schannel}. The last two
columns are labeled collectively via the types of the diagrams. The numbers quoted in the columns 3 and up are the binding energy in MeV
as well as in brackets those binding energies in units of the leading contribution, $t(a)$, for the given calculation. The $\dagger$-symbol is added
to the $\alpha$-value when the subleading contribution to the 
self energy, Fig.~\ref{fig:retardbubble}, was also included. The numbers in bold face denote
the full result of the leading calculation and the calculation with higher order interactions and
those that correct for Lorenz symmetry violation are included.}
\begin{tabular}{|cc|c|c|c|c|c|c|c|}
\hline
$\alpha$ & $\Lambda$ [GeV]         & $t(a)$&$+t(b)$&$+s(a)$&$+s(b)$&$+streched$&$+crossed$                    \\
 &       &  & & & &$boxes$&$boxes$                    \\
\hline 
\hline 
0 & 0.5 & $0.51$ $(1)$ &$0.66$ $(1.29)$ &$0.83$ $(1.63)$&$0.52$ $(1.02)$&$0.58$ $(1.14)$& $0.68$ $(1.33)$           \\
\hline
1 & 0.5                          &$1.63$ $(1)$ &$2.51$ $(1.54)$ &$3.93$ $(2.41)$ &$\mathbf{ 1.63}$ $(1)$&$1.86$ $(1.14)$ &$2.21$   $(1.36)$        \\
\hline
1$^\dagger$ & 0.5                           &$1.58$ $(1)$ &$2.44$ $(1.54)$ &$3.81$ $(2.41)$&$1.59$ $(1.01)$&$1.81$ $(1.15)$&$\mathbf{2.16}$     $(1.37)$     \\
\hline 
\hline 
0 & 1 & $0.51$ $(1)$ &$0.68$ $(1.33)$ &$0.90$ $(1.76)$&$0.52$ $(1.02)$&$0.59$ $(1.16)$& $0.70$ $(1.37)$           \\
\hline
1 & 1                          &$1.72$ $(1)$ &$3.03$ $(1.76)$ &$7.29$ $(4.24)$ &$\mathbf{1.67}$ $(0.97)$&$1.97$ $(1.15)$ &$2.45$   $(1.42)$        \\
\hline
1$^\dagger$ & 1                           &$1.59$ $(1)$ &$2.79$ $(1.75)$ &$6.57$ $(4.13)$&$1.60$ $(1.01)$&$1.90$ $(1.19)$&$\mathbf{2.36}$     $(1.48)$     \\
\hline 
\hline 
0 & 1.5    & $0.51$ $(1)$ &$0.68$ $(1.33)$ &$0.91$ $(1.78)$& $0.51$ $(1)$&$0.59$ $(1.16)$& $0.70$     $(1.37)$     \\
\hline
 1 & 1.5                      &$1.67$ $(1)$ &$3.02$ $(1.81)$ &$9.05$ $(5.42)$ &$\mathbf{1.68}$ $(1.01)$ &$1.99$ $(1.19)$&$2.49$    $(1.49)$      \\
\hline
1$^\dagger$ & 1.5                        &$1.49$ $(1)$ &$2.66$ $(1.79)$ &$7.65$ $(5.13)$ &$1.62$ $(1.09)$ &$1.94$ $(1.30)$ &$\mathbf{2.43}$ $(1.63)$          \\
\hline 
\hline 
0 & 2    & $0.51$ $(1)$ &$0.68$ $(1.33)$ &$0.92$ $(1.80)$& $0.51$ $(1)$&$0.59$ $(1.16)$& $0.70$     $(1.37)$     \\
\hline
 1 & 2                      &$1.63$ $(1)$ &$2.95$ $(1.81)$ &$9.85$ $(6.04)$ &$\mathbf{1.70}$ $(1.04)$ &$2.02$ $(1.24)$&$2.54$    $(1.56)$      \\
\hline
1$^\dagger$ & 2                        &$1.41$ $(1)$ &$2.51$ $(1.78)$ &$7.92$ $(5.62)$ &$1.66$ $(1.18)$ &$1.99$ $(1.41)$ &$\mathbf{2.51}$ $(1.78)$          \\
\hline
\end{tabular}
\end{center}
\label{tab:result-three-body}
\end{table*}


The dependence of the resulting binding energy on the 
 four different cut-offs for the coupled channel case is also shown in Table~\ref{tab:result-three-body}.  
 Clearly for the contributions discussed so
 far the cut-off dependence is rather weak. Although not
 directly reflected in the numbers reported in the table, it turns out that the most dominant contribution
 to the emergence of the bound state
 comes from the first diagram of the $t$-channel kaon exchange
 labled as $t(a)$: 
 When using only individual contributions in solving the LS equation,
 only this part of the potential generates binding.
 This is expected, since this contribution contains the leading
 three-body singularity. 
The second $t$-channel contribution, although not binding by
itself, is still important quantitatively as it increases the binding energy by about 
30\%. Also the two $s$-channel contributions are large
individually, however, there are quite effective cancellations among
$t(b)$, $s(a)$ and $s(b)$ such that the binding energy deduced
from the full potential to order $f_S^2$ for all cut-offs
is within 2\% of the one calculated from $t(a)$ only
($cf.$ the sixth column
of Tab.~\ref{tab:result-three-body}).


Since both $f_0(980)$ and $a_0(980)$ couple strongly to the $K\bar{K}$ system, the implications
of three-body unitarity above the three kaon-threshold and its impact on  analyticity 
should play an important role in the three-body dynamics. We thus 
repeat the calculation with $\alpha=1$ and the values
of $Z$ given in Fig.~\ref{fig:ConstZ}.
As expected the binding energies calculated for the four different cut-offs become lager by
more than a factor of 3, see the lines marked with $\alpha=1$ in Tab.~\ref{tab:result-three-body}.
Moreover, also the relative contributions
from the individual other diagrams get enhanced
(e.g. when diagram $t(b)$ is added the binding energy
get enhanced by almost a factor of 2; the contributions
from the $s$-channel diagrams are even larger).
Also the cut-off dependence is now larger. However,
the binding energies deduced from the sum of the full 
potential to order $f_S^2$ remains very close to that
calculated from $t(a)$ only as shown by the numbers in bold face in the sixth column of Tab.~\ref{tab:result-three-body} as well
as the fourth bar in Fig.~\ref{fig:SkeCut1000}.

Thus, our numerical results 
suggest that a $I(J^P)=\frac{1}{2}(0^-)$ bound state can be generated from 
from coupled-channel $Kf_0(980)$-$Ka_0(980)$ interactions, at least
as long as inelastic channels are omitted. The binding energies
deduced from the interactions discussed so far are below 2~MeV.
We also show that for reliable quantitative results the full potential to order $f_S^2$ needs
to be included, since the individual pieces of the potential undergo significant cancellations.




\section{Study of the violation of covariance}

The potentials introduced in Sec.~\ref{sec:effe} have only physical singularities, but are not
invariant under Lorentz transformations. In this section we investigate how much the results
change when also the contributions of the streched boxes
are included in the potential that restore covariance at the one-loop level. We demonstrate
below that their effect on the results is rather small --- in any case
of the same order a other contributions to the potential higher order in the couplings $f_S$. 

To show how the violation of covariance emerges in  time--ordered perturbation theory,
we start from the expression for the t-channel potential given in Eq.~(\ref{eq:Vt}). The two terms
can be combined to
\begin{align}
V^{\lambda' \lambda}_t(E,\vec{p'},\vec{p})= \frac{f_S^2{\cal I} N}{\omega_K(\vec{q})}  \left(
 \frac{\omega_K(\vec{q})-E^{\rm off}}{\Delta E^2-(\omega_K(\vec{q})-E^{\rm off})^2}\right) ,
 \label{eq:TOPTt}
\end{align}
where 
$$
E^{\rm off}=E-(\omega_{1'}(p')+\omega_{2'}(p')+\omega_{1}(p)+\omega_{2}(p))/2
$$
is the average off-shellness of the initial and final state for any give pair of momenta $p$ and $p'$
and
$$
\Delta E = (\omega_{1'}(p')-\omega_{2'}(p')+\omega_{1}(p)-\omega_{2}(p))/2
$$
denotes the energy transfer for initial and final particles on their mass shell. 
For both particles entering and leaving the potential being 
on the energy shell, $E^{\rm off}$
vanishes 
and the potential reduces to the well known, covariant Feynman amplitude
\begin{equation}
V^{\lambda' \lambda}_t(E,\vec{p'},\vec{p})= f_S^2{\cal I} N \left(\frac{1}{t-m_K^2}\right) \ ,
\label{eq:Ft}
\end{equation}
where $t$ is the four-momentum transfer squared, as it should be. However, when being put
into the LS equation, both $p$ and $p'$ are integration variables and thus the potential is
typically evaluated off-shell. Then clearly there is a difference between the two potentials.
To keep covariance also then, the formalism of Ref.~\cite{Mai:2017vot} calls for putting
Eq.~(\ref{eq:Ft}) into an LS type
equation. This keeps formal covariance, however, it introduces unphysical
singularities~\cite{Dawid:2020uhn}.
We propose to use the full potential of Eq.~(\ref{eq:TOPTt}) or equivalently Eq.~(\ref{eq:Vt}) instead.
This avoids unphysical
singularities, but violates covariance, since the potential depends on the particle energies which are not
invariant under Lorentz transformations.

\begin{figure*}[ht]
\begin{center}
\includegraphics[scale=0.25]{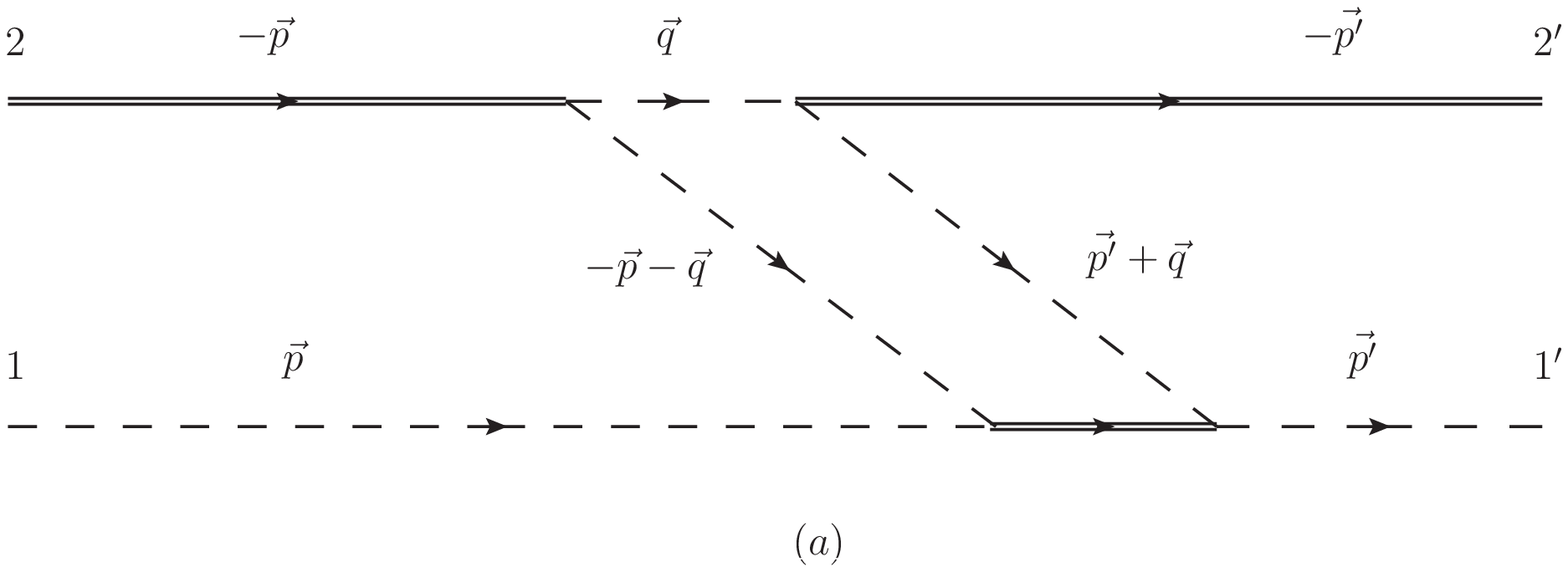}
\includegraphics[scale=0.25]{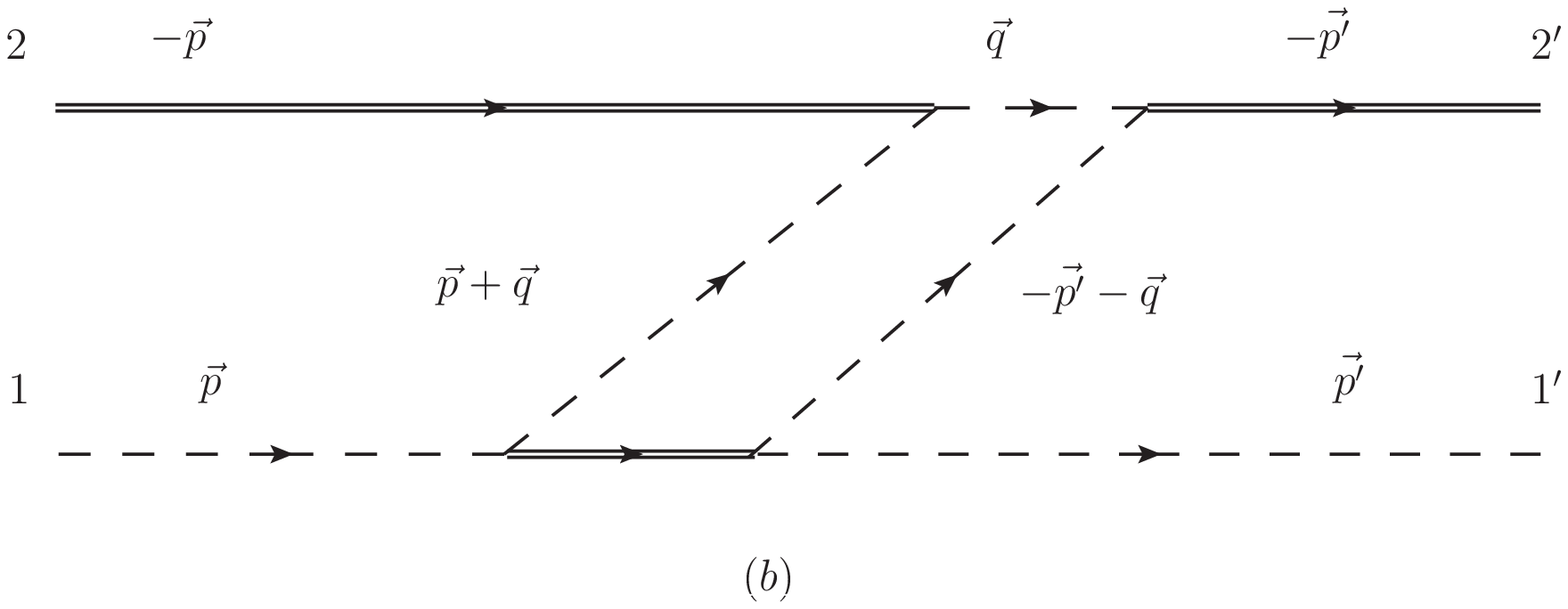}
\caption{The stretched boxes contribution.}
\label{fig:stret}
\end{center}
\end{figure*}

\begin{figure*}[ht]
\begin{center}
\includegraphics[scale=0.25]{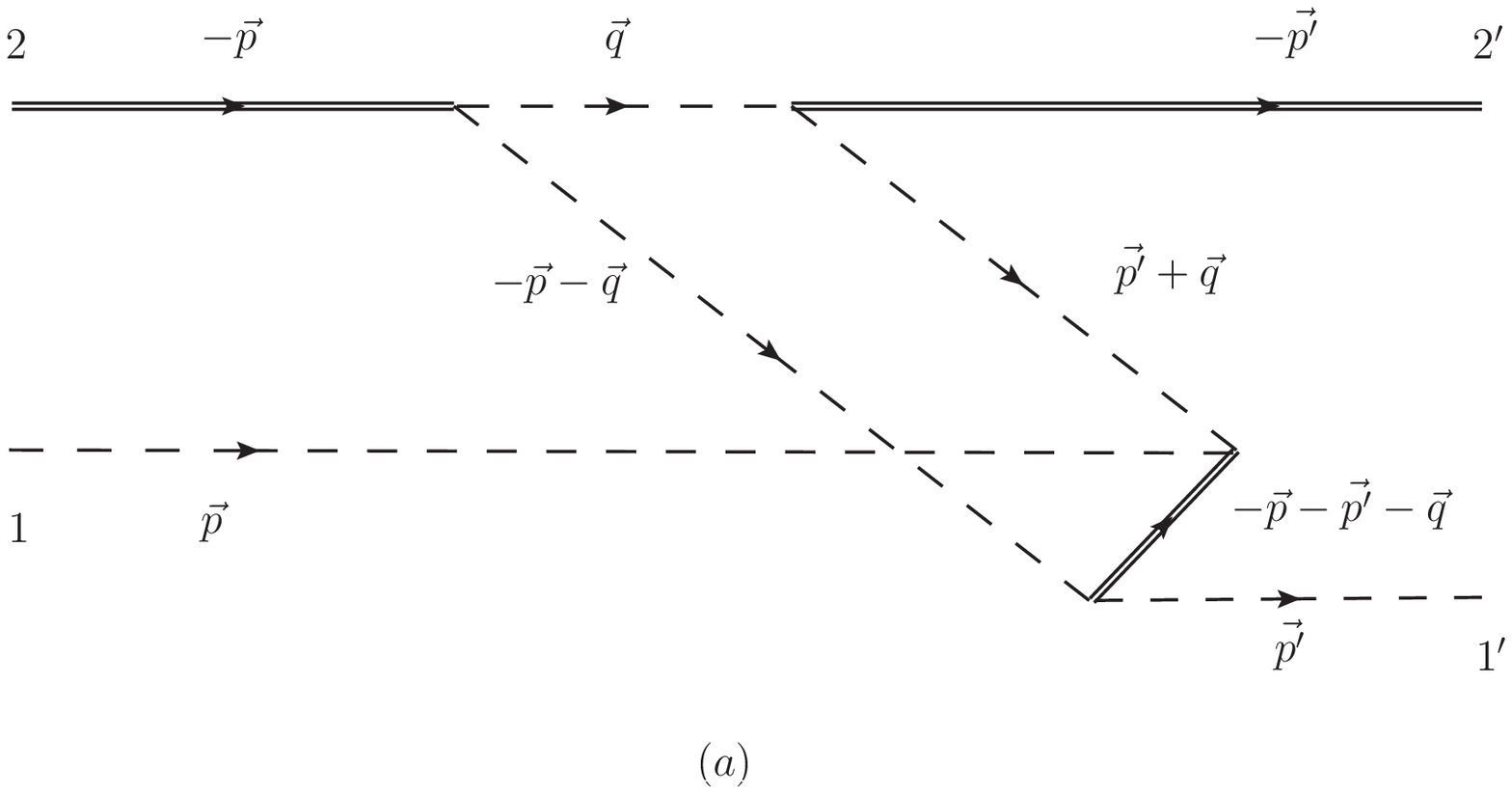}
\includegraphics[scale=0.25]{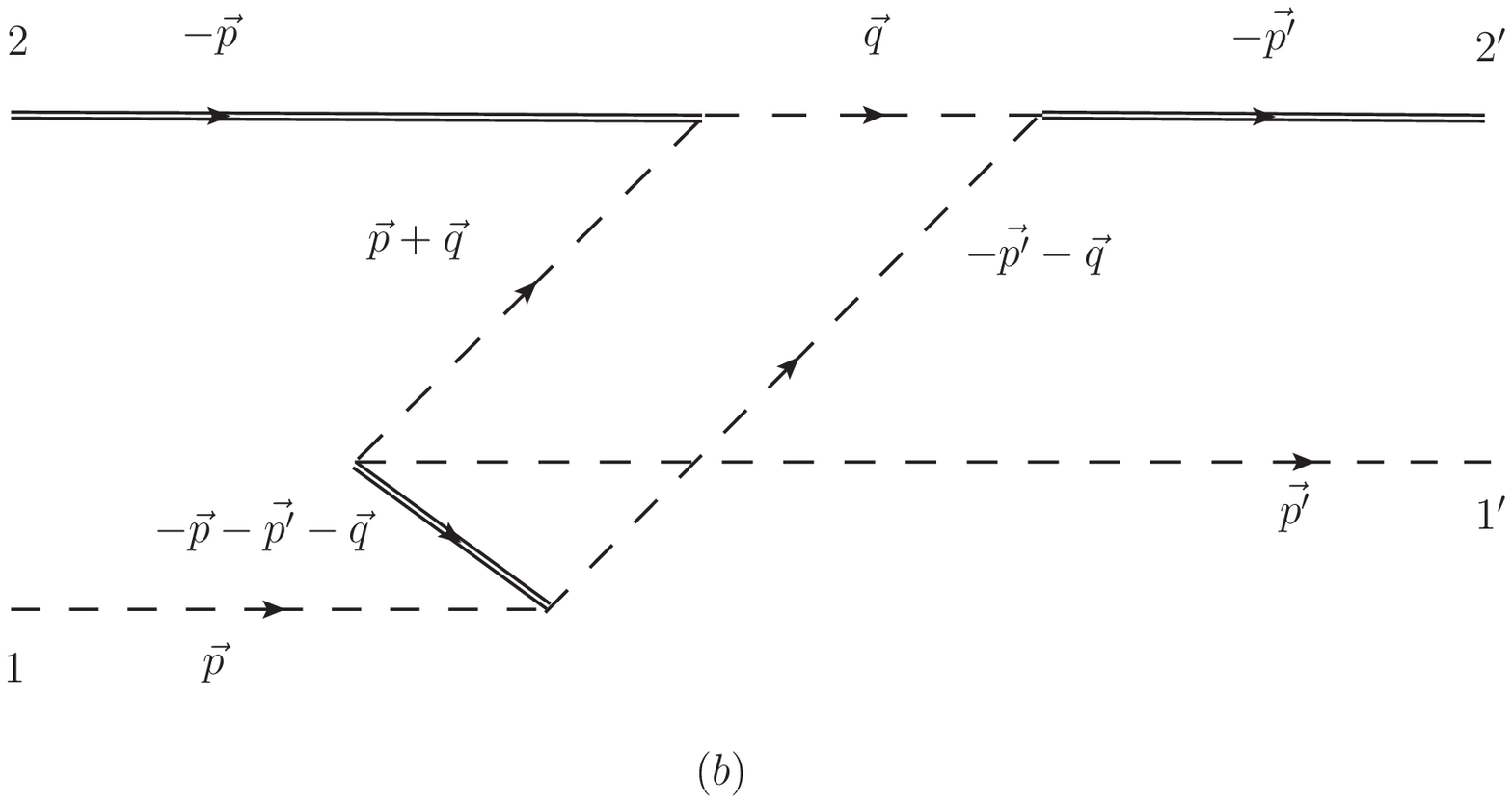}
\includegraphics[scale=0.25]{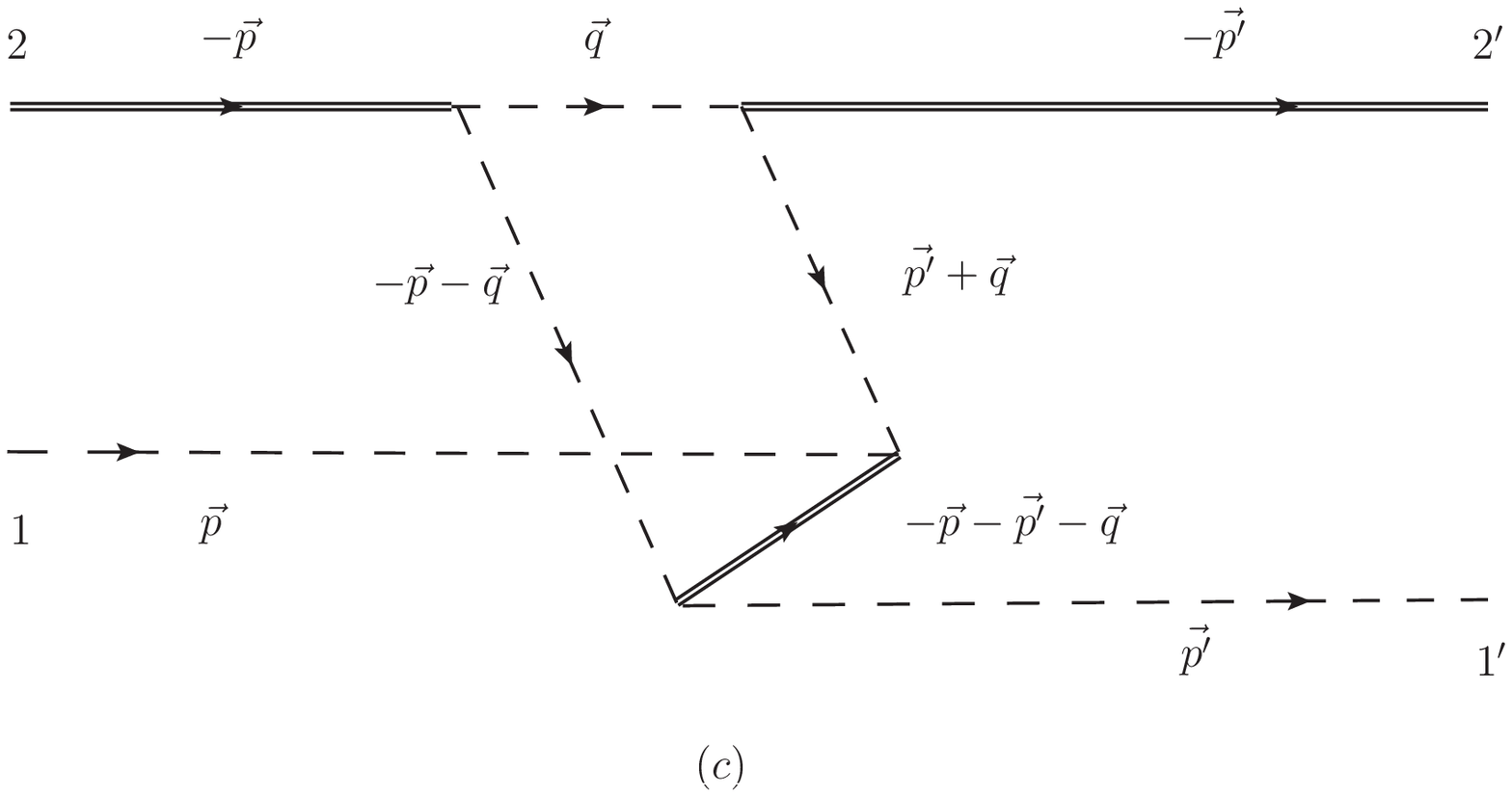}
\includegraphics[scale=0.25]{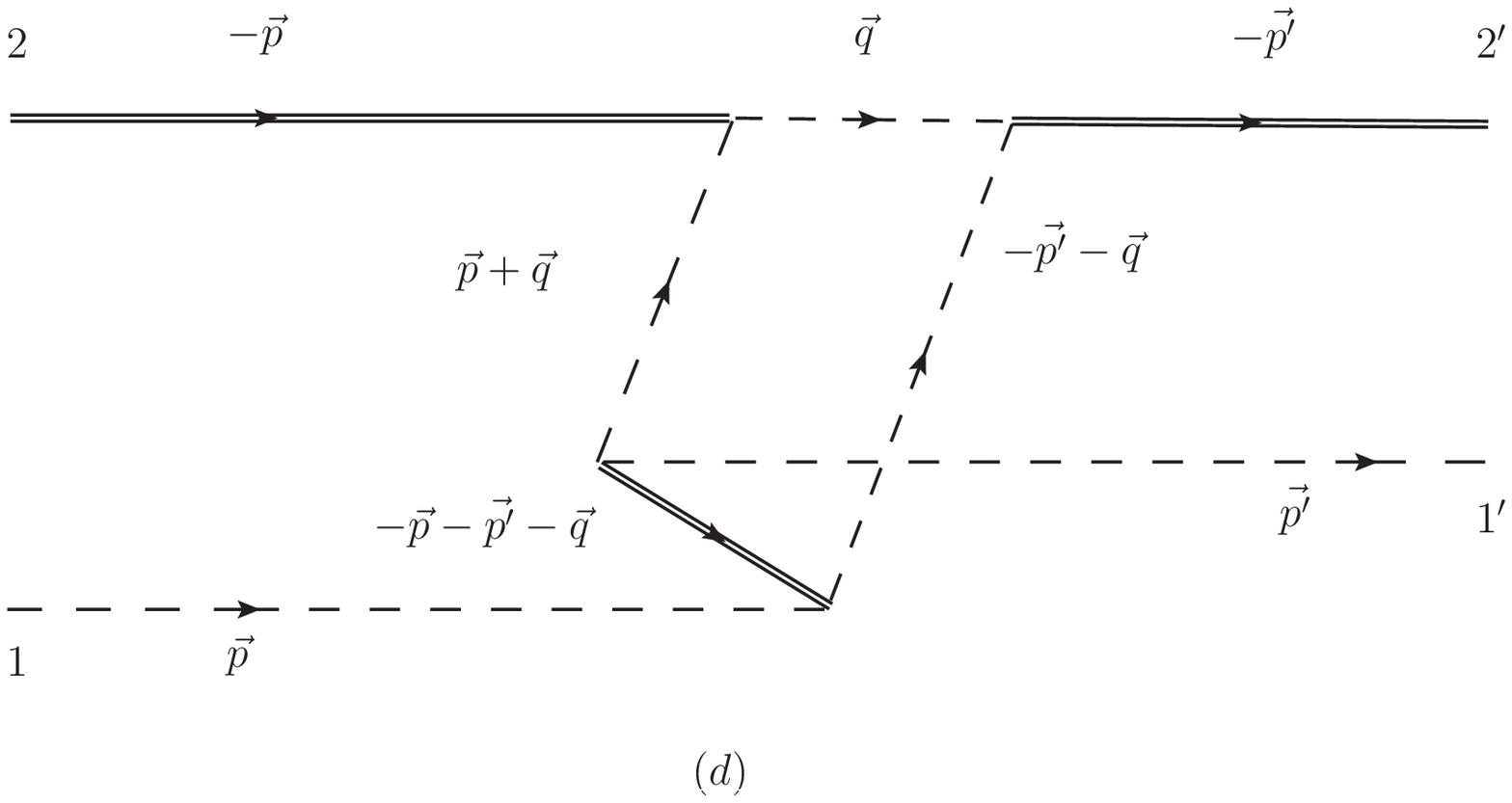}
\includegraphics[scale=0.25]{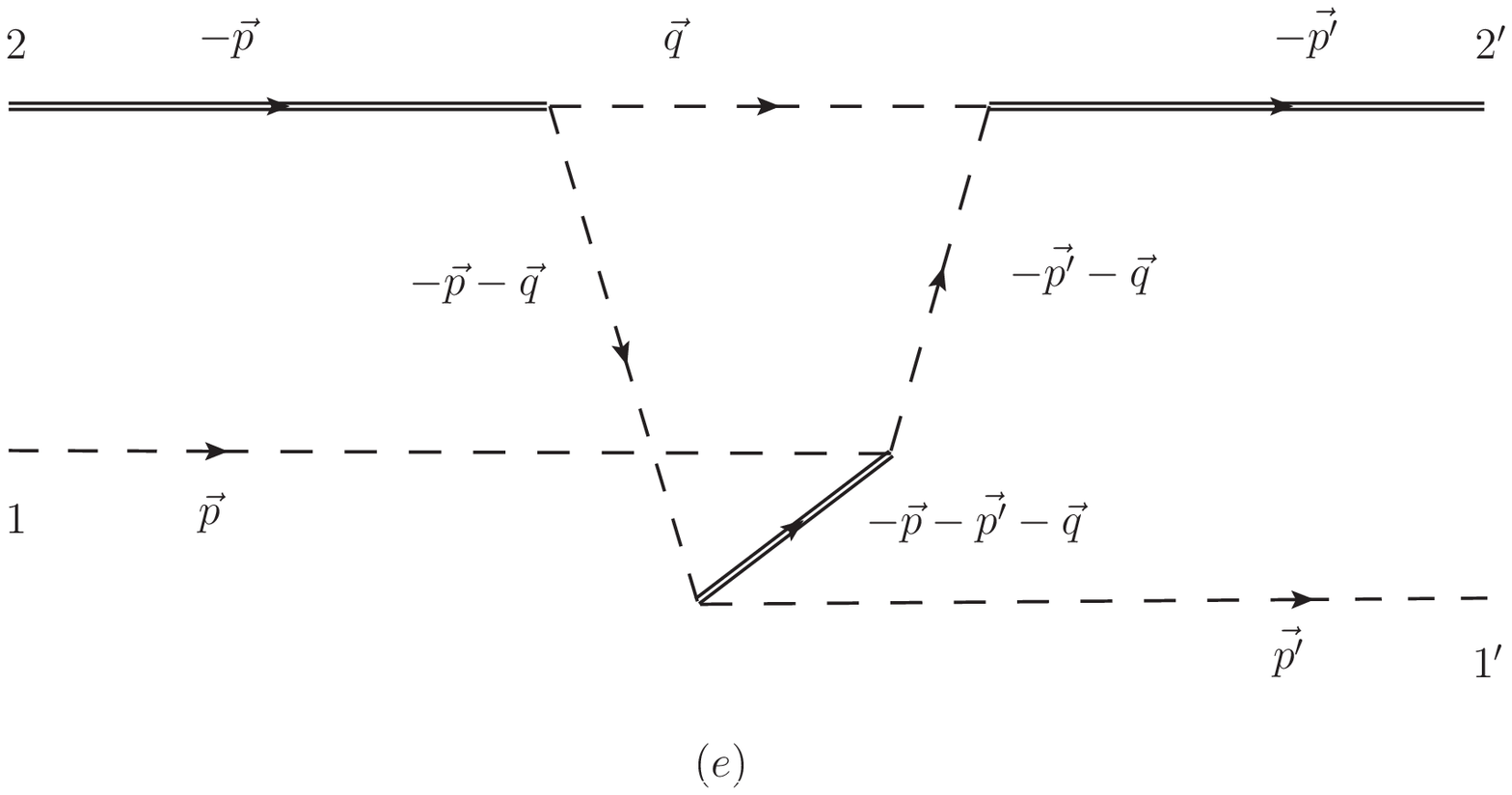}
\includegraphics[scale=0.25]{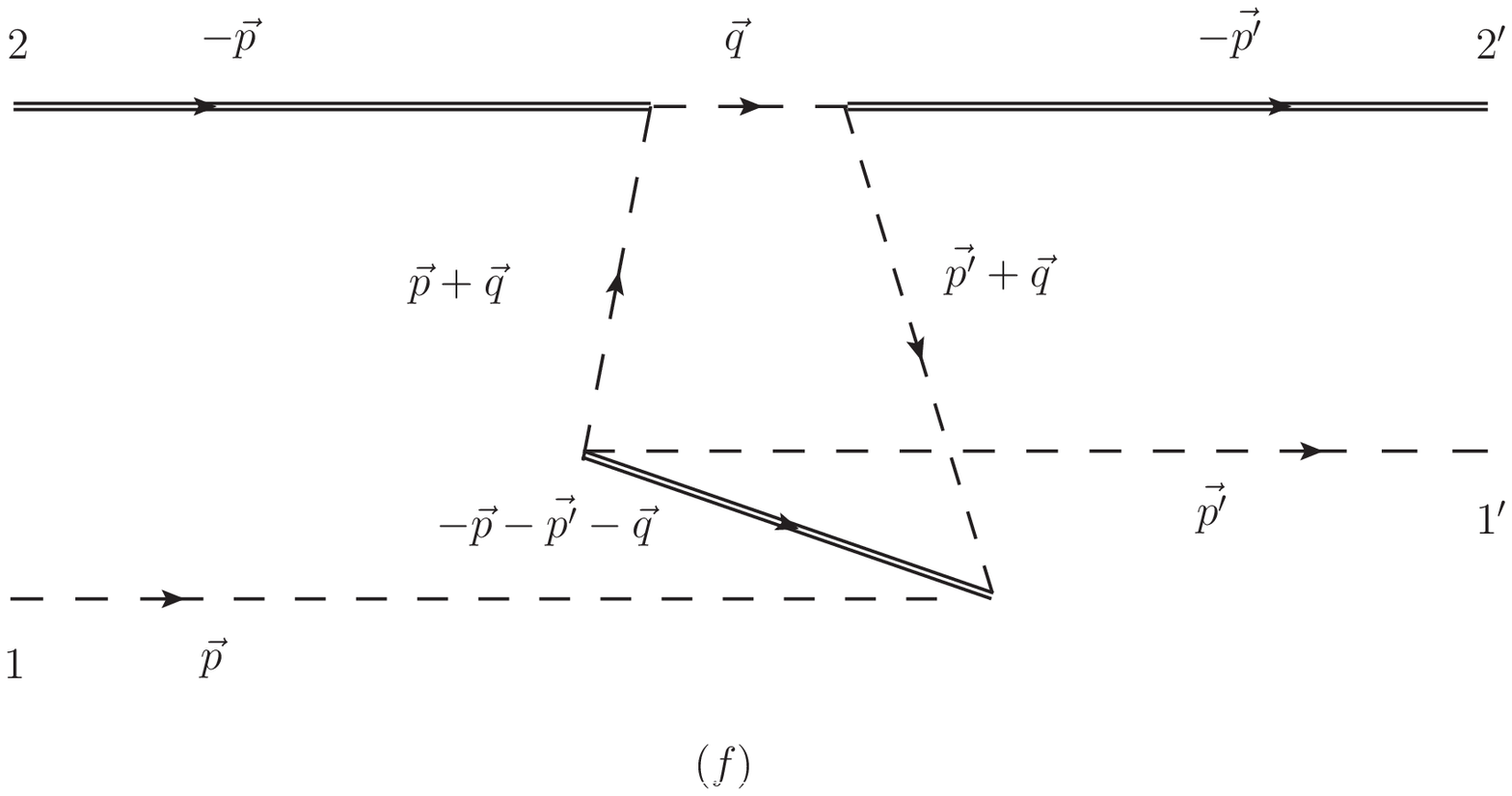}
\caption{The stretched boxes arise from $f_0(980)$ or $a_0(980)$ running backward
in time.}
\label{fig:stret-negs}
\end{center}
\end{figure*}

\begin{figure*}[ht]
\begin{center}
\includegraphics[scale=0.25]{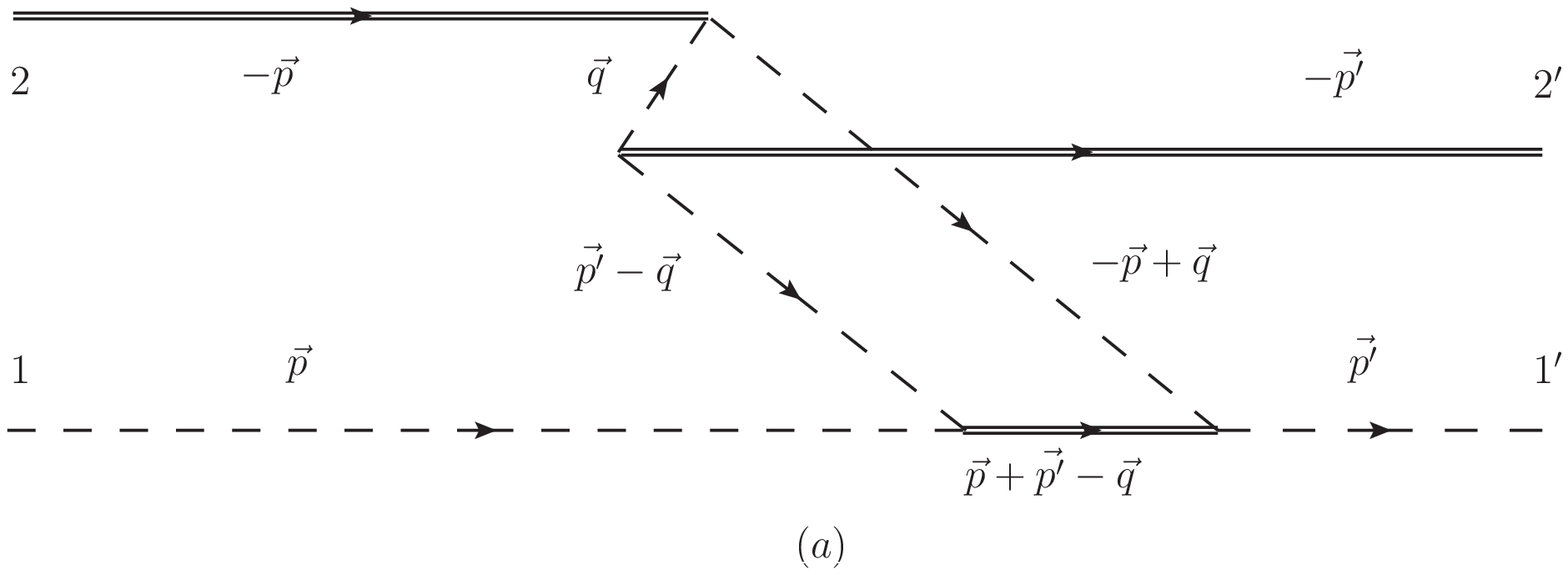}
\includegraphics[scale=0.25]{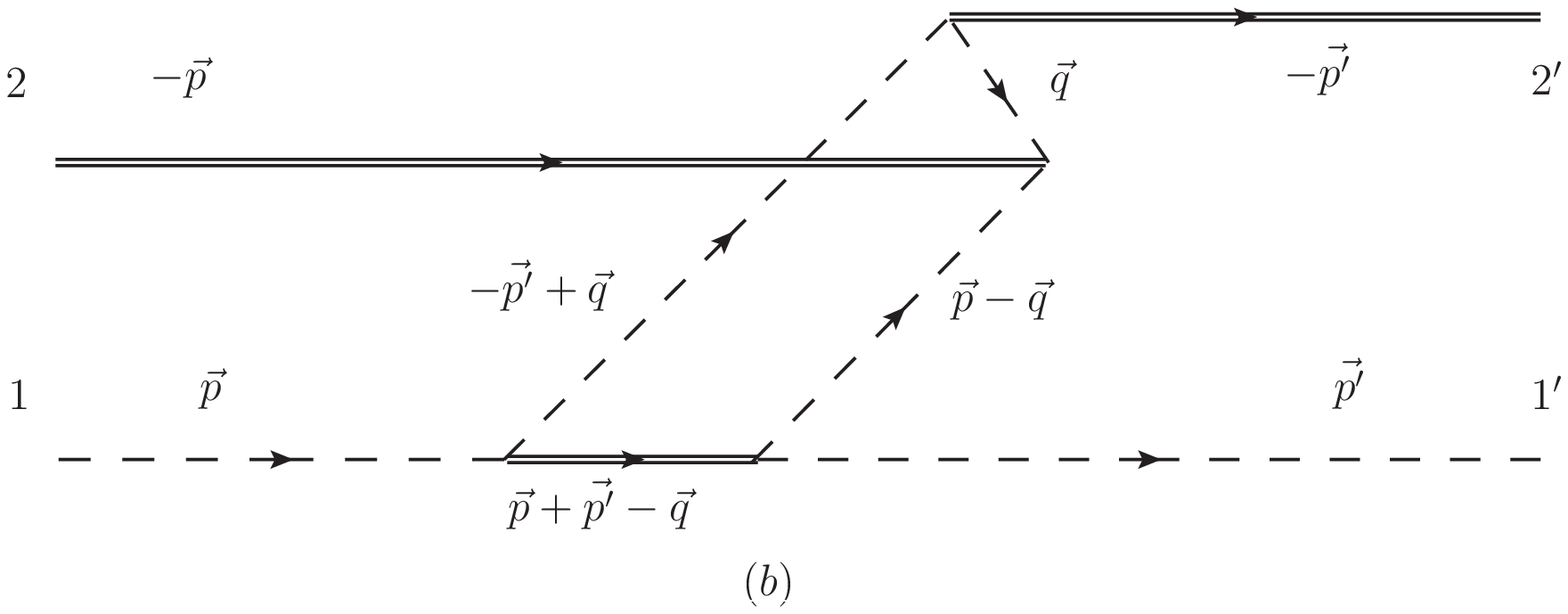}
\includegraphics[scale=0.25]{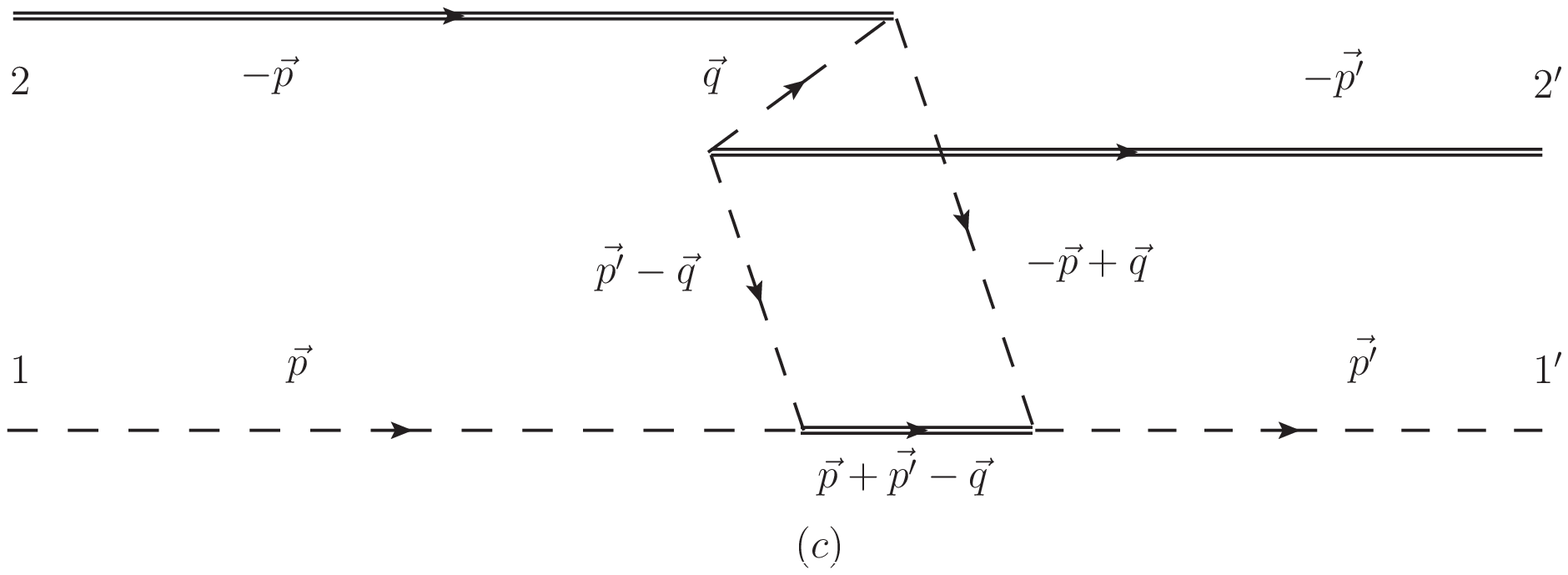}
\includegraphics[scale=0.25]{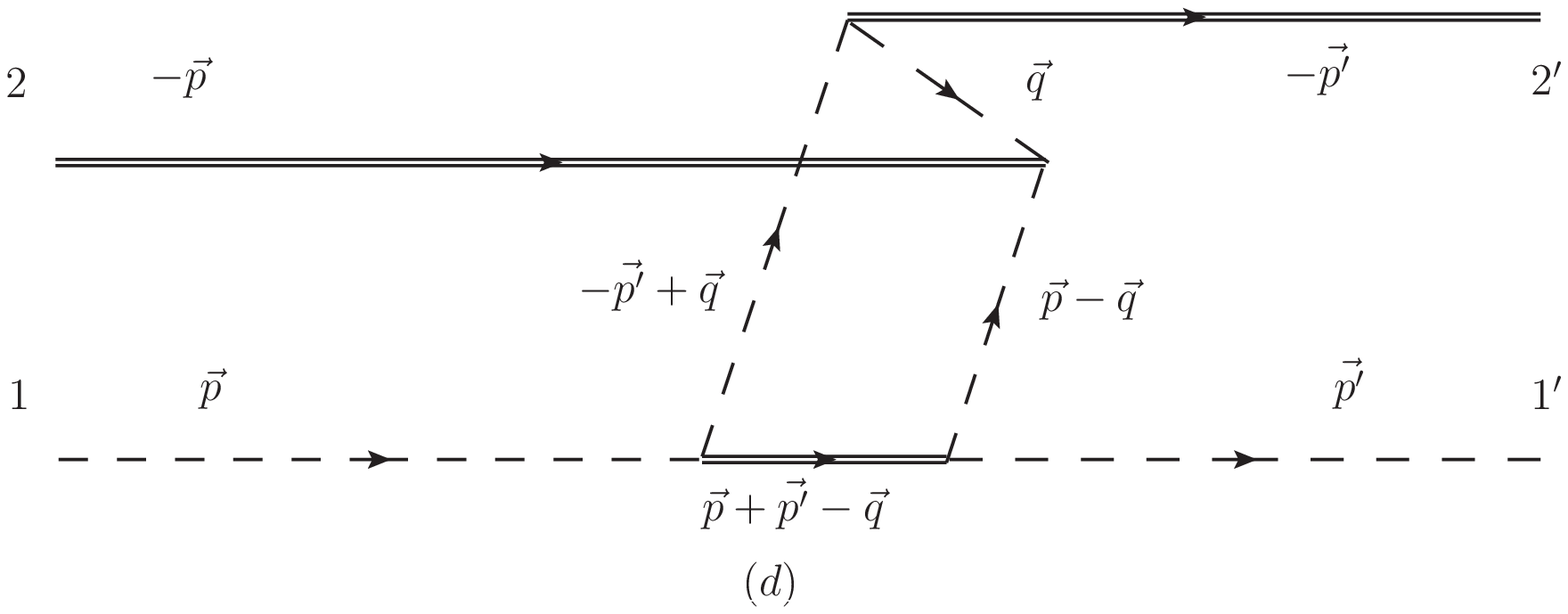}
\includegraphics[scale=0.25]{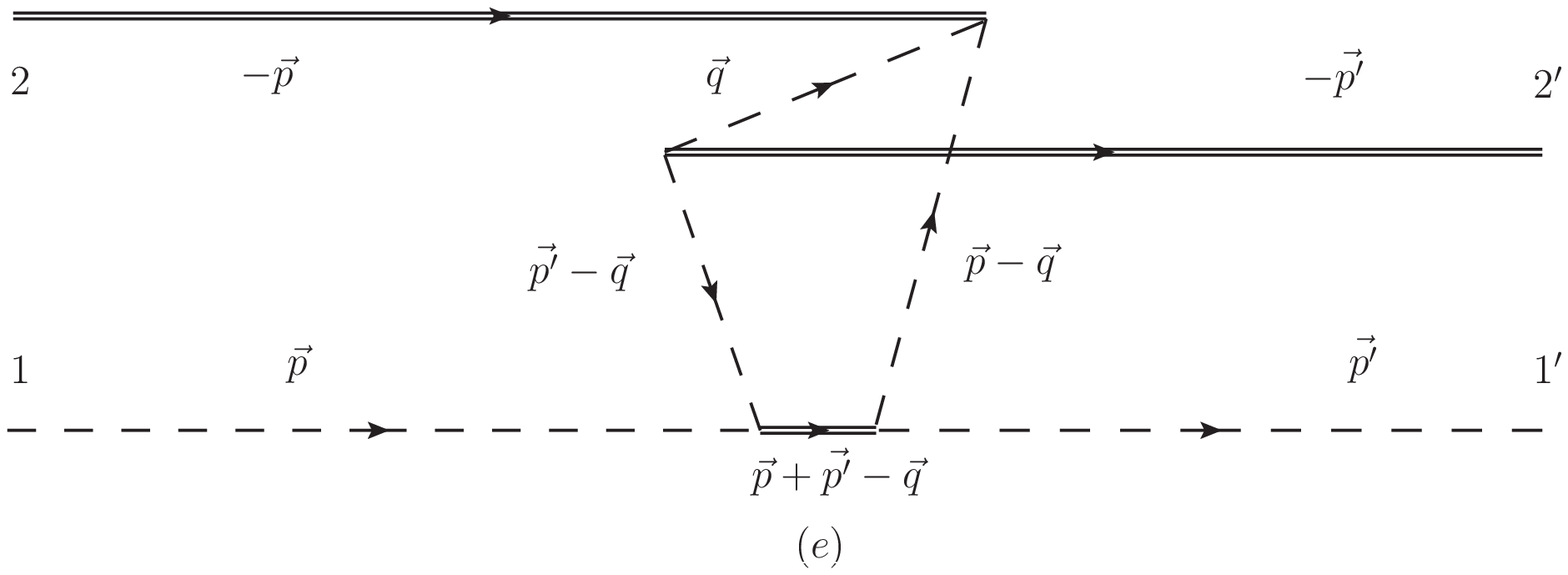}
\includegraphics[scale=0.25]{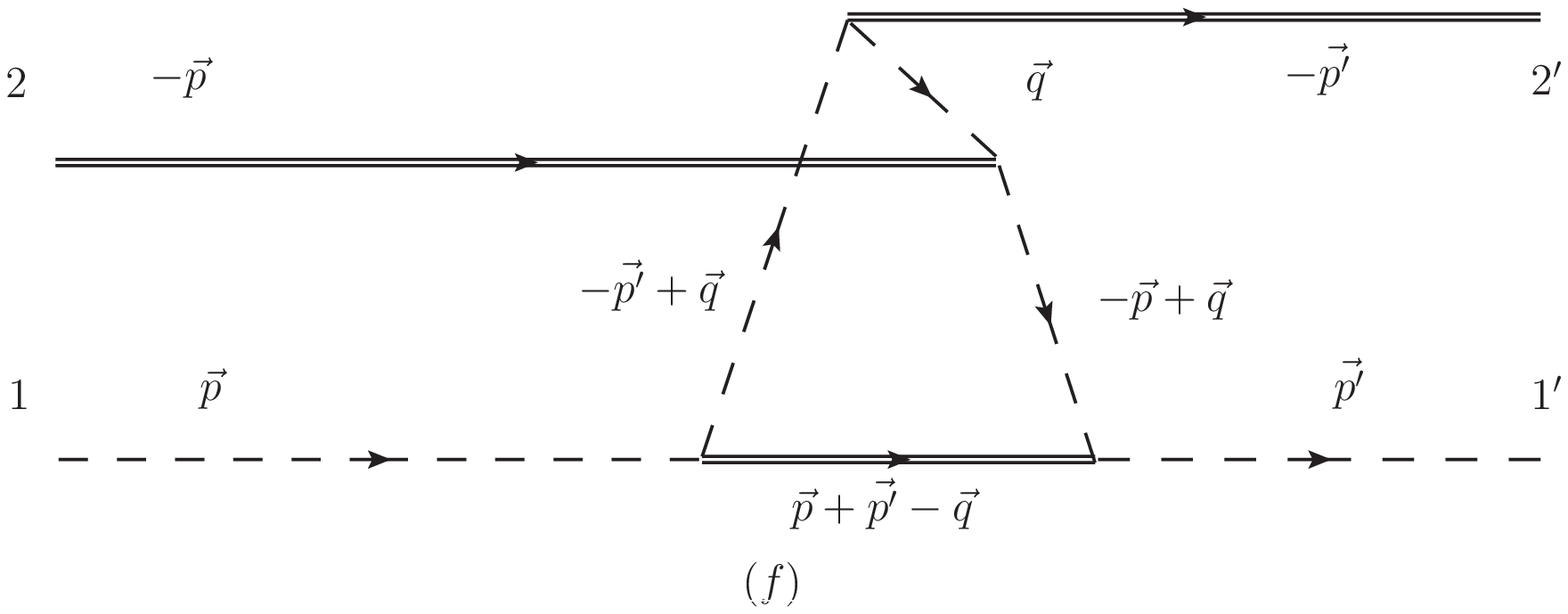}
\caption{The stretched boxes arise from kaons running backward
in time.}
\label{fig:stret-negk}
\end{center}
\end{figure*}

\begin{figure*}[ht]
\begin{center}
\includegraphics[scale=0.25]{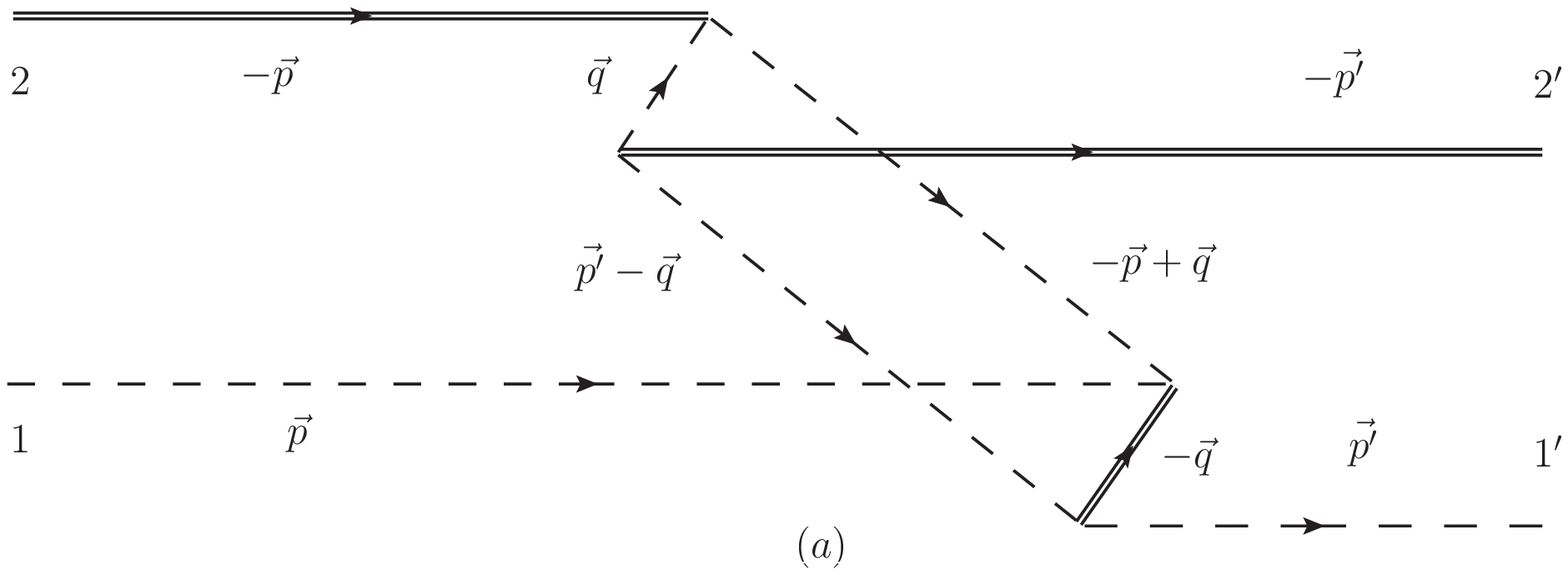}
\includegraphics[scale=0.25]{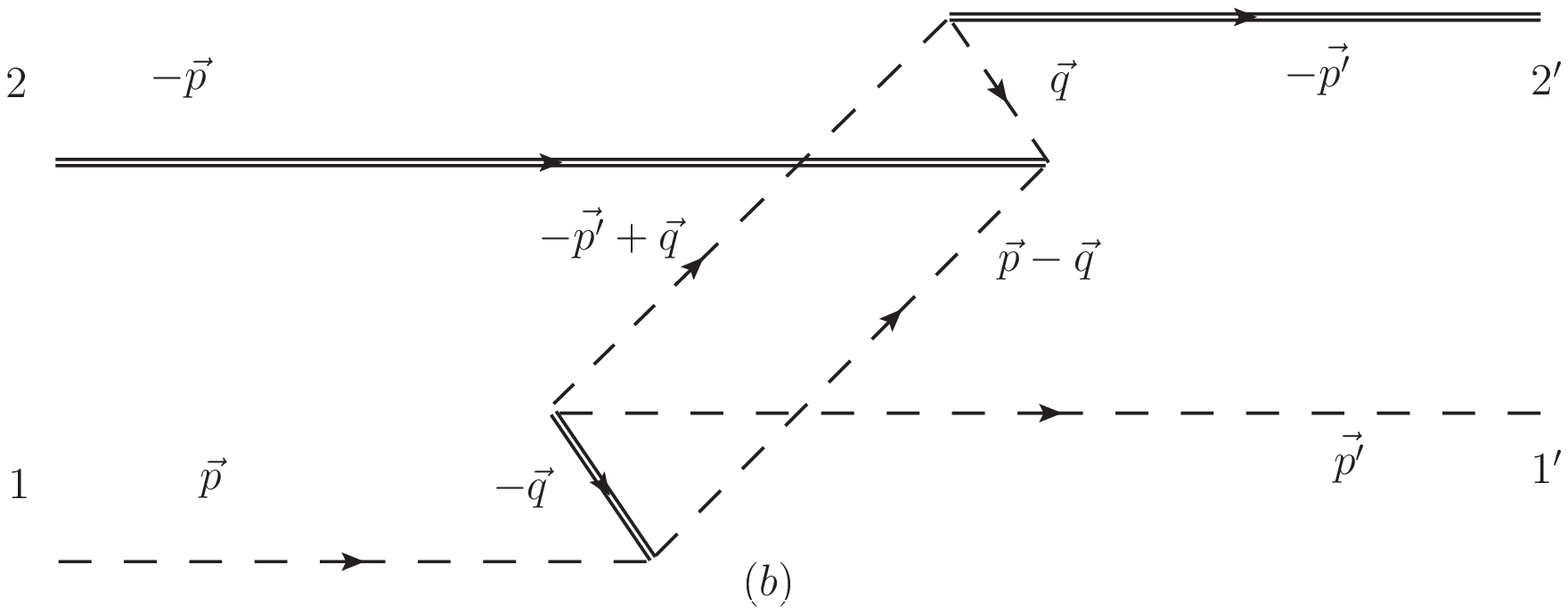}
\includegraphics[scale=0.25]{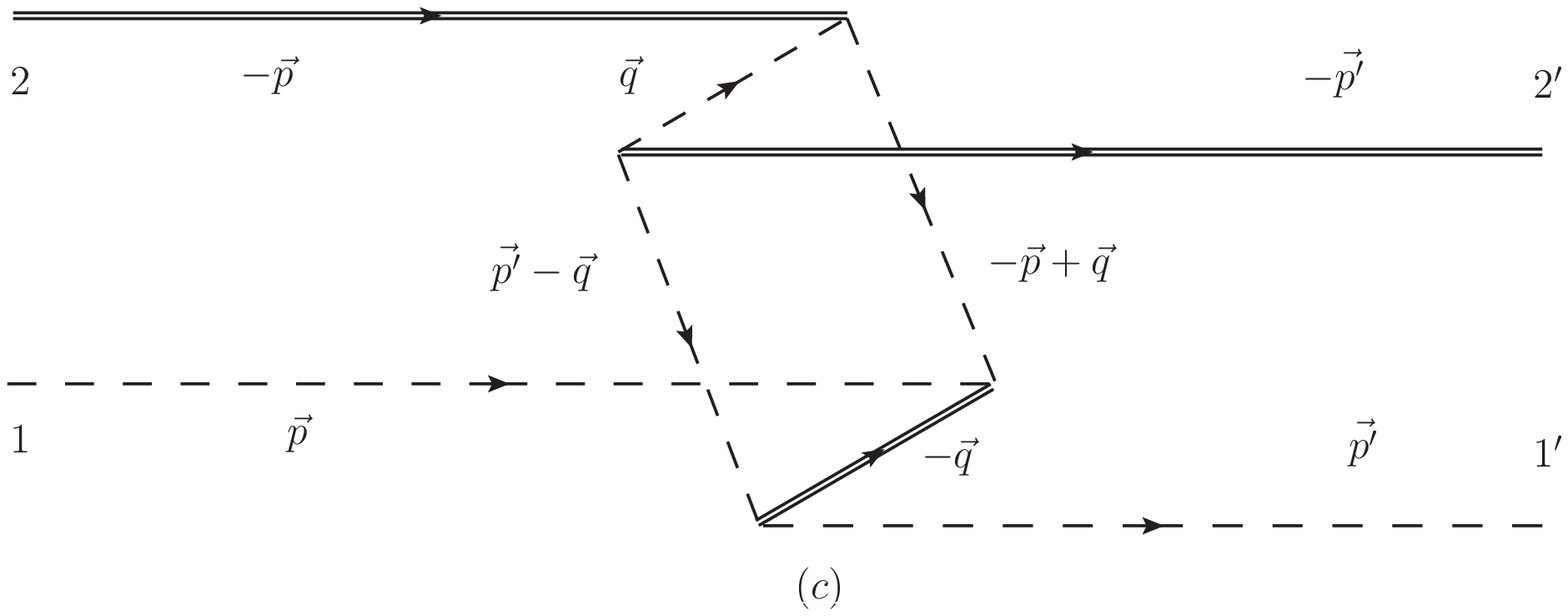}
\includegraphics[scale=0.25]{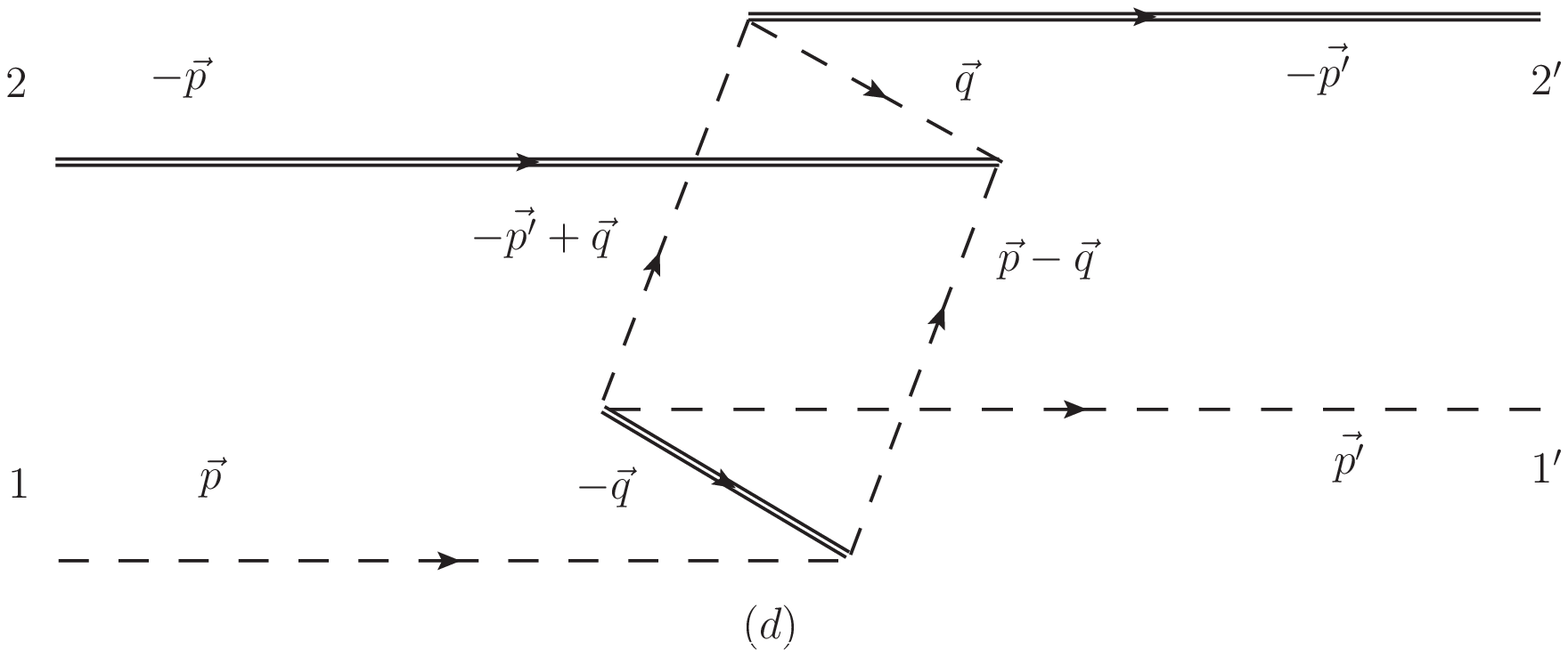}
\includegraphics[scale=0.25]{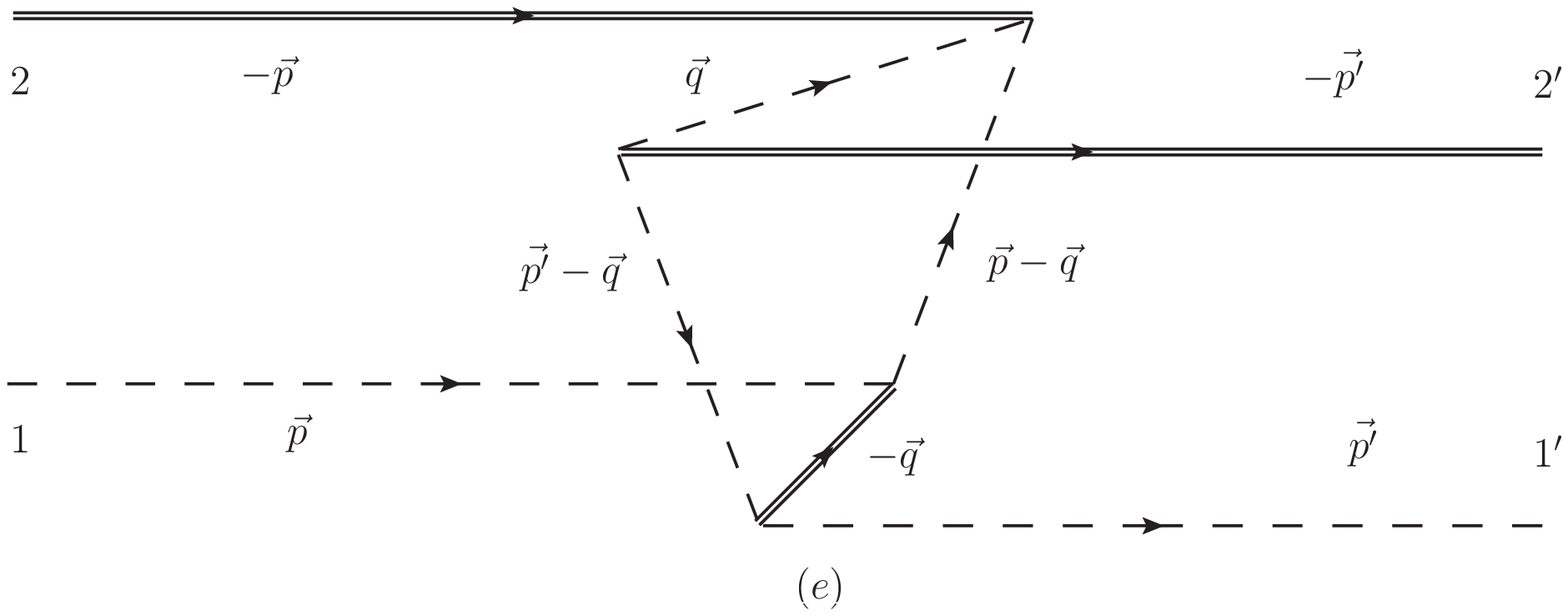}
\includegraphics[scale=0.25]{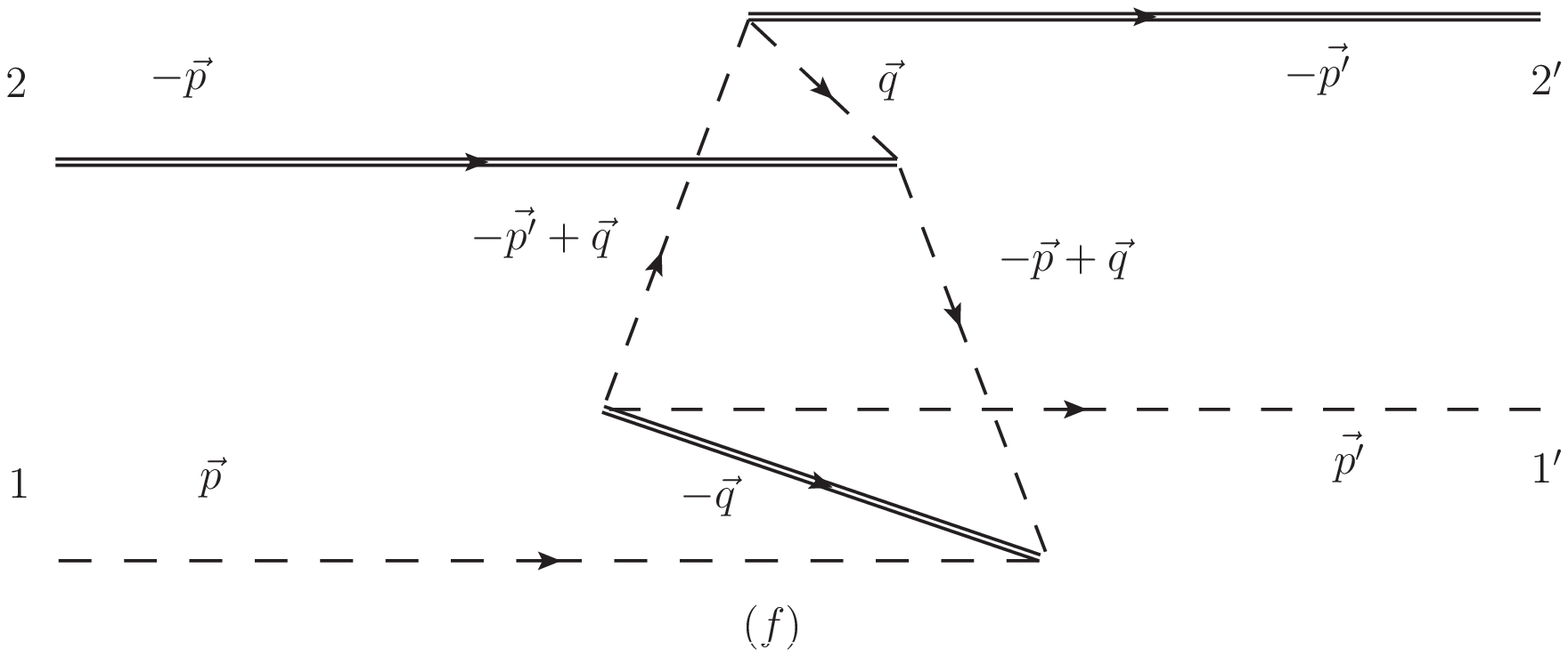}
\caption{The stretched boxes arise from both $f_0(980)$ or $a_0(980)$ and kaons running backward
in time.}
\label{fig:stret-negks}
\end{center}
\end{figure*}

To quantify the amount of violation of Lorentz invariance, we calculate explicitly the
contributions that restore it at the one loop level. This is achieved by 
the inclusion of the so-called streched boxes.  {The corresponding diagrams are
shown in Figs.~\ref{fig:stret},~\ref{fig:stret-negs},~\ref{fig:stret-negk} and~\ref{fig:stret-negks},
 the related amplitudes are given in Eqs.~(\ref{eq:stret-a}) to (\ref{eq:stret-negks-f})
in the appendix. The effect of the inclusion of these diagrams into the potential
on the resulting binding energies for the different
calculations is illustrated by the second to last column in Table~\ref{tab:result-three-body}.} The streched
boxes change the binding energies in all calculations
by about 20\% and we may regard this as a subleading 
contribution.
It should be noted that to restore covariance also at the two-loop level higher streched boxes
(contributing at order $f_S^6$) would need to be included. Those are even more suppressed
kinematically than the ones at order $f_S^4$, since more particles are included in the equal
time slices. We therefore conclude that the violation of Lorentz invariance for the equations
employed here is in the energy range studied indeed mild
and can be restored in a controlled way.
In contrast to this introducing the mentioned unphysical singularities generates an error
in the calculation that cannot be controlled quantitatively.

\begin{figure*}[ht]
\begin{center}
\includegraphics[scale=0.25]{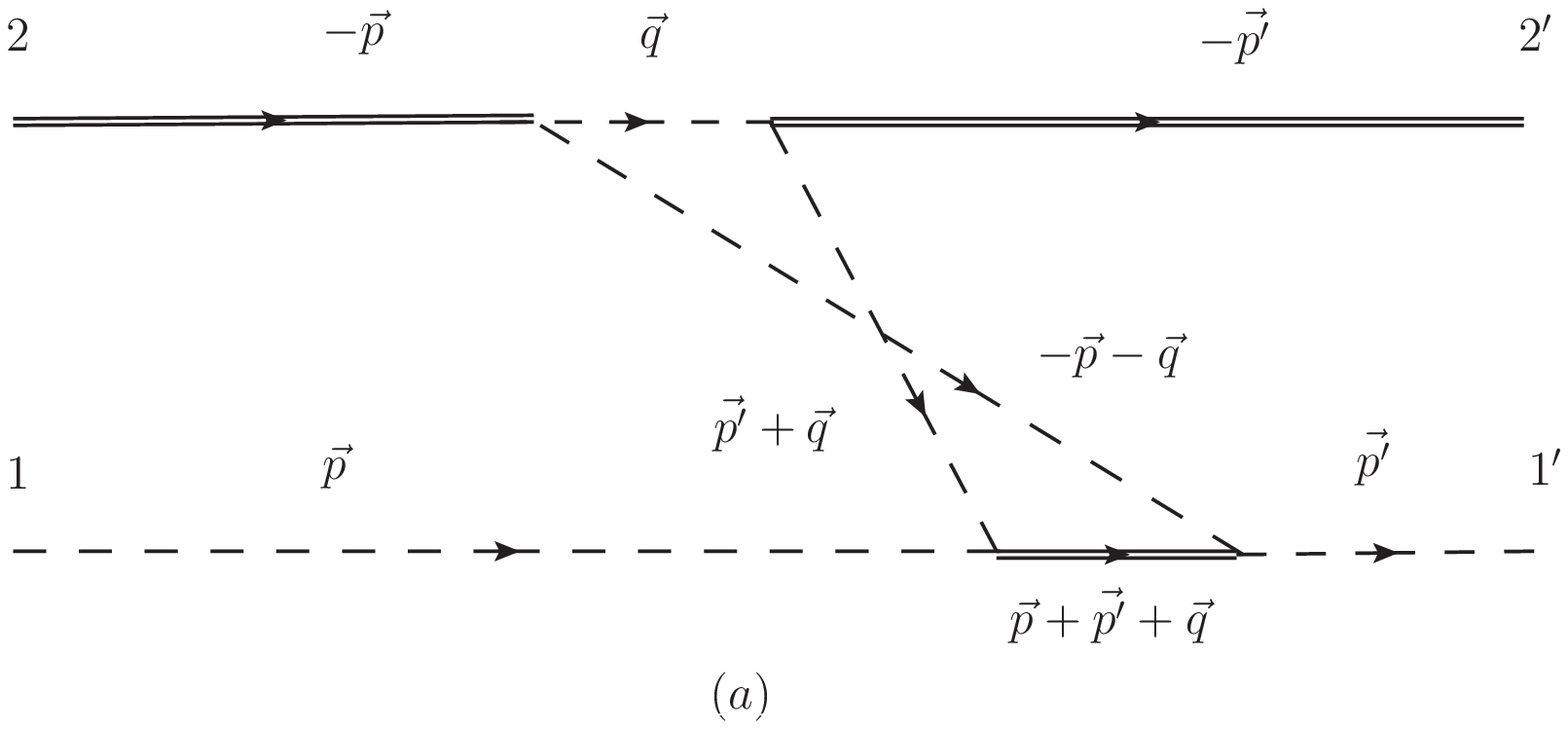}
\includegraphics[scale=0.25]{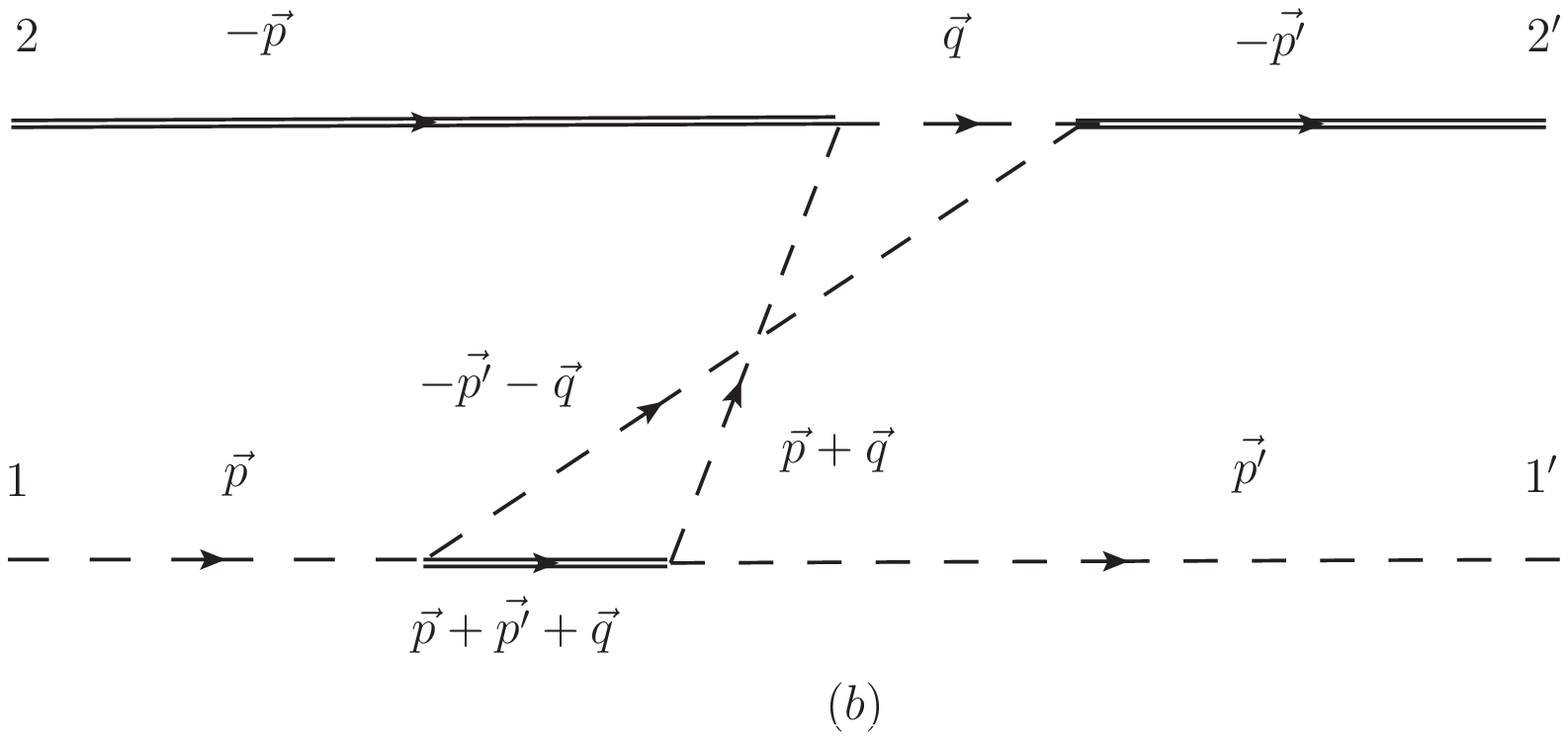} 
\includegraphics[scale=0.25]{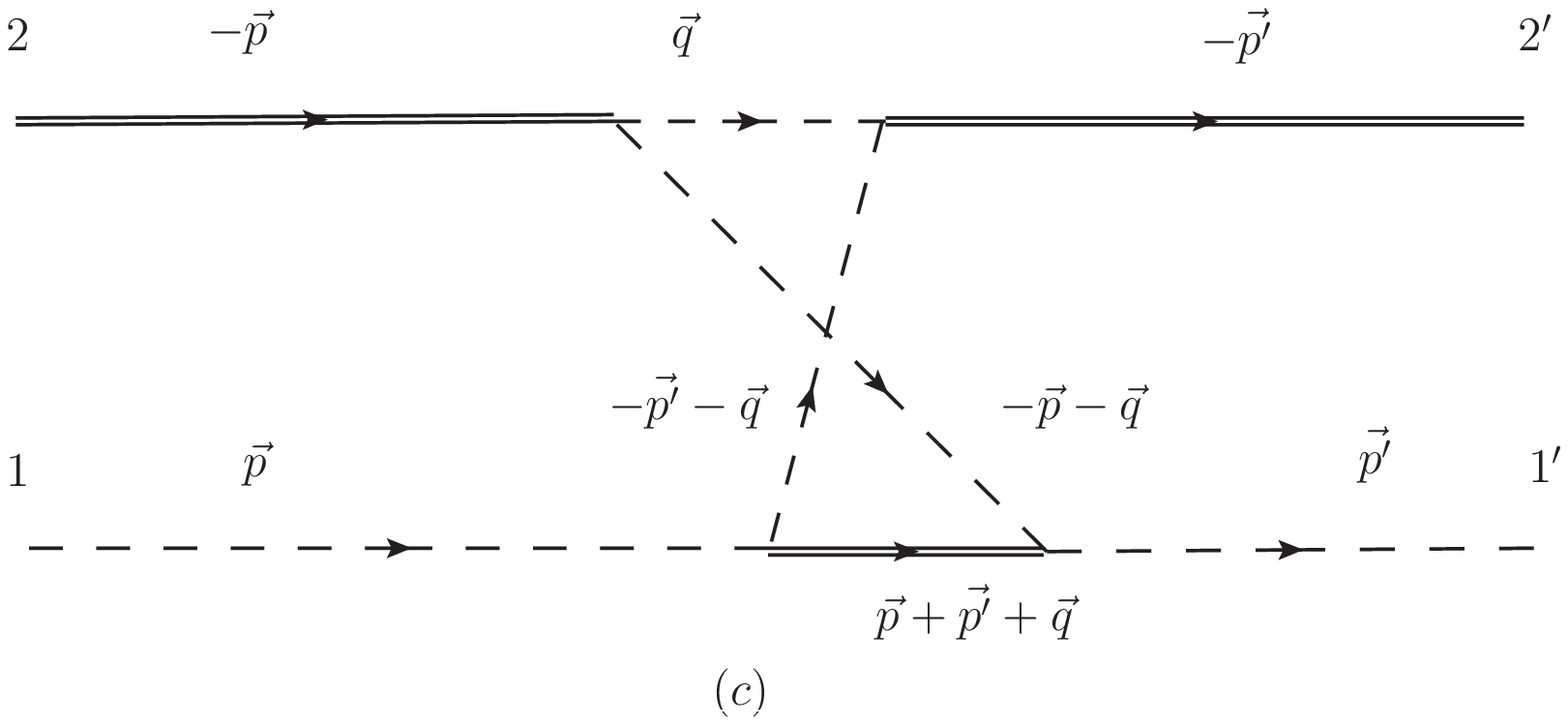}
\includegraphics[scale=0.25]{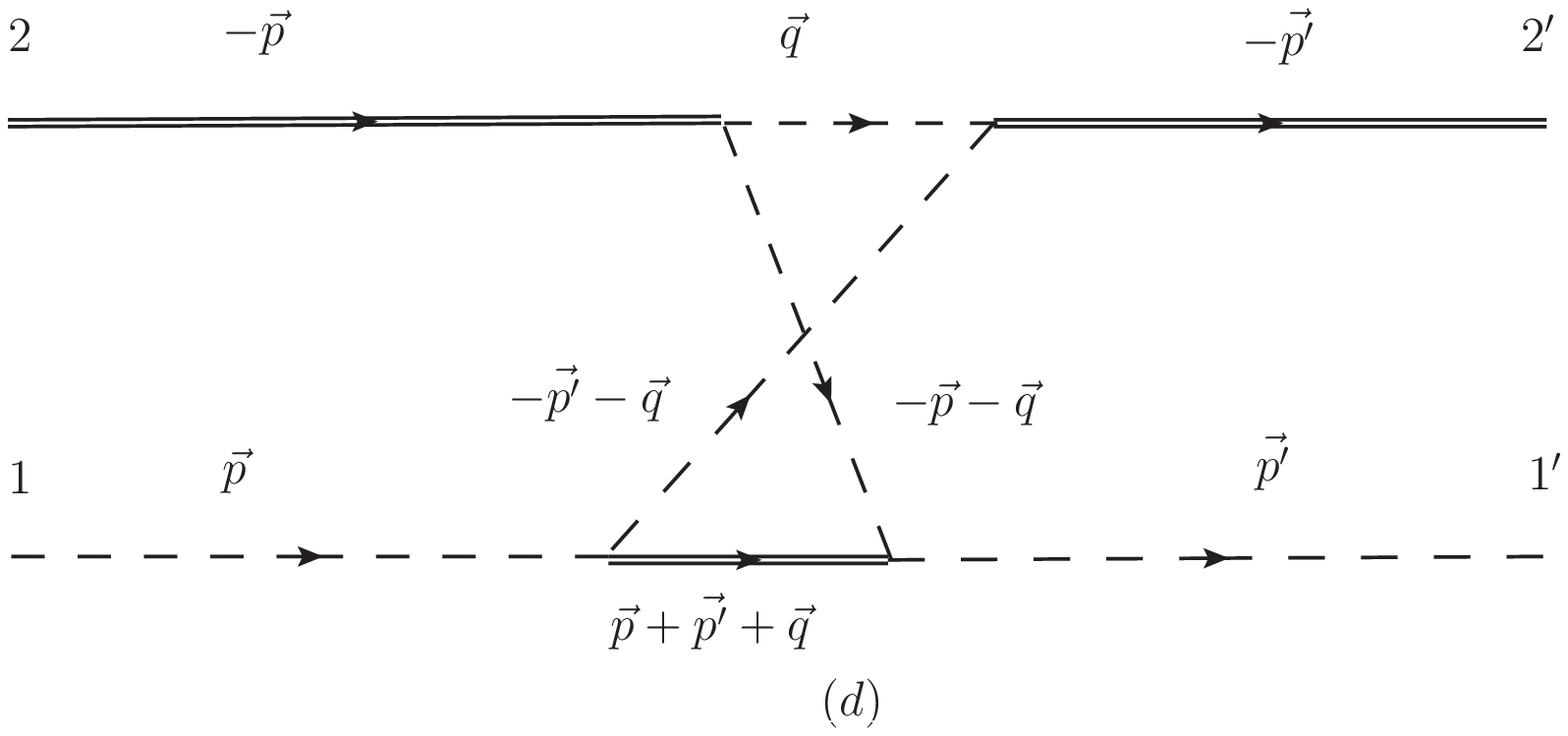} 
\includegraphics[scale=0.25]{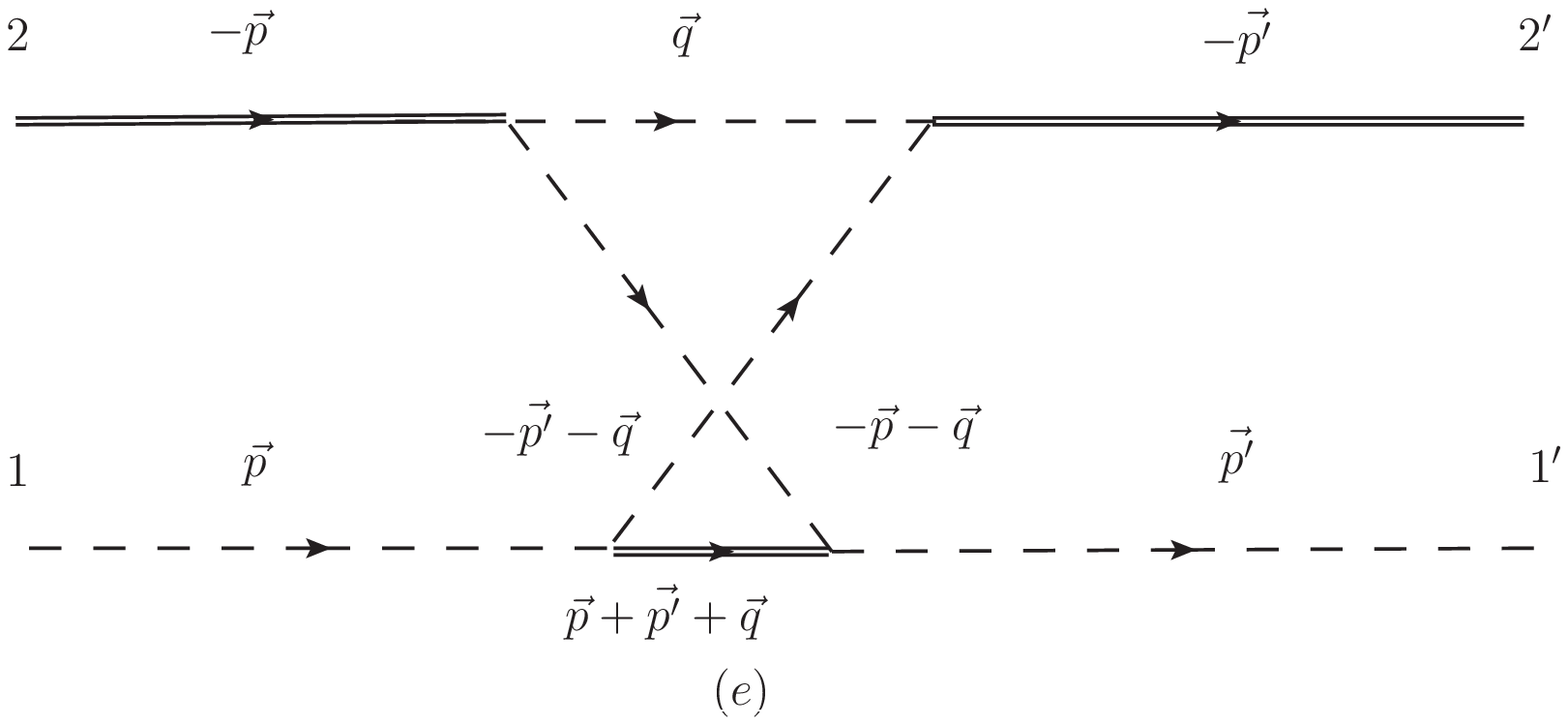}
\includegraphics[scale=0.25]{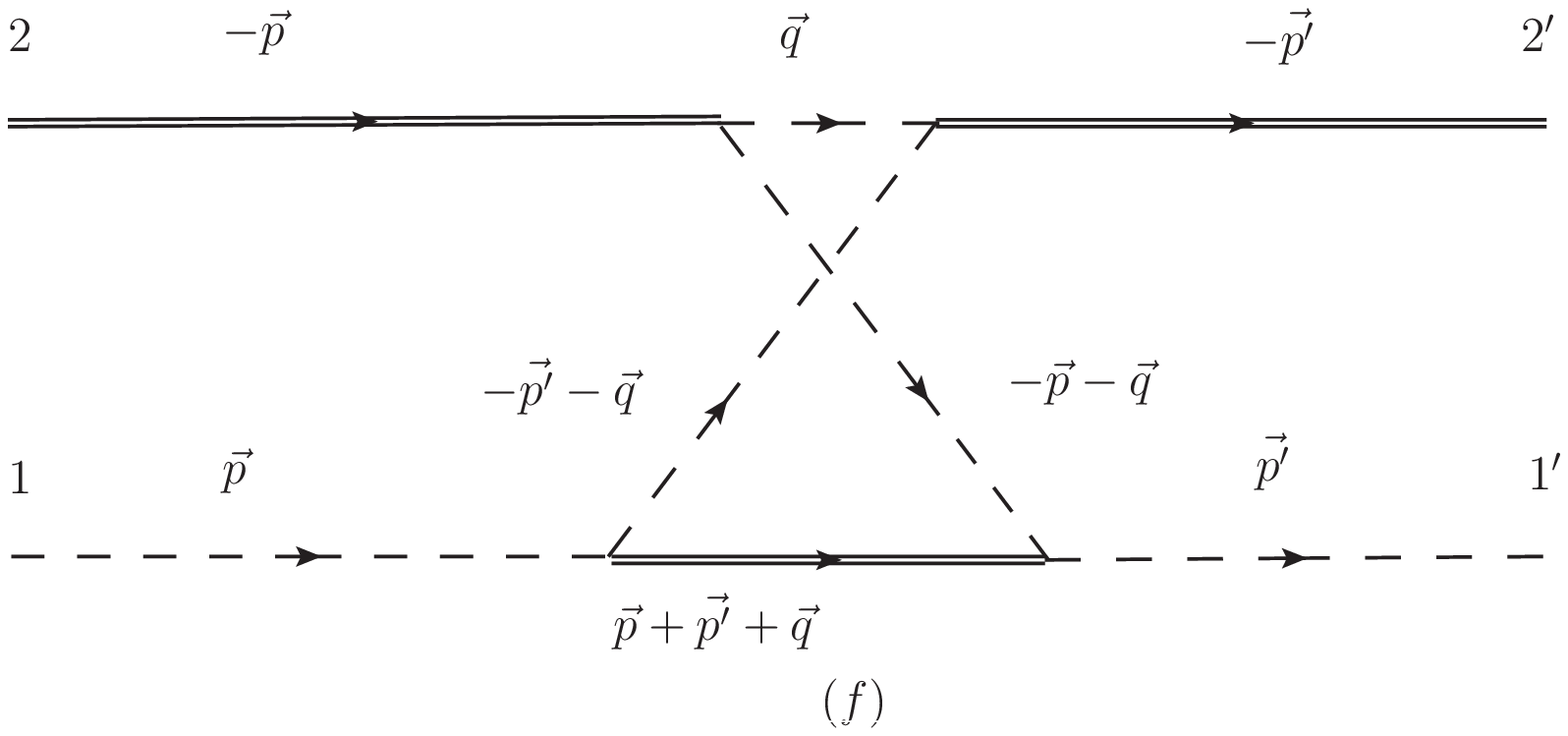}
\caption{The crossed boxes contribution.}
\label{fig:crossed}
\end{center}
\end{figure*}
\begin{figure*}[ht]
\begin{center}
\includegraphics[scale=0.25]{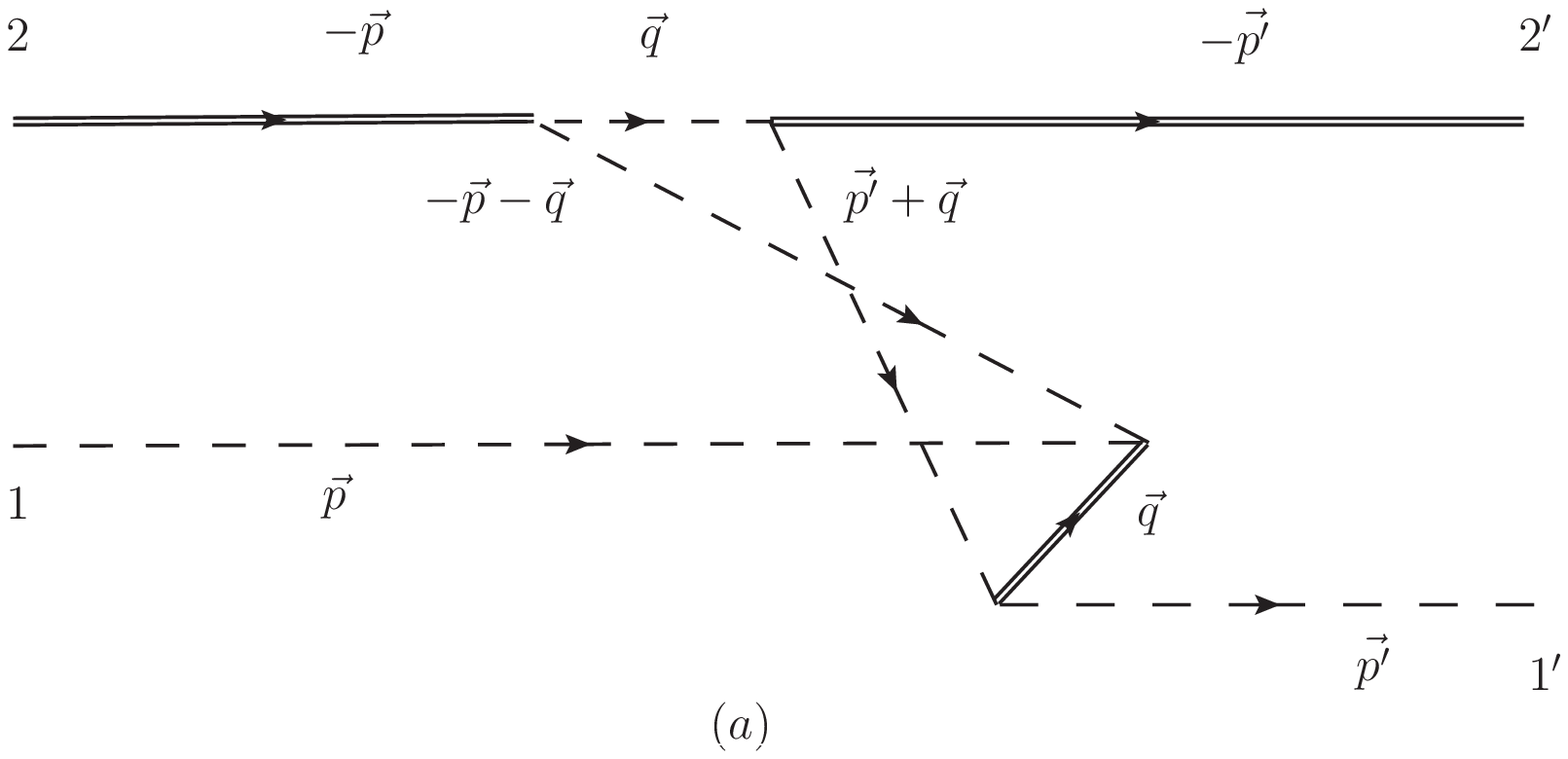}
\includegraphics[scale=0.25]{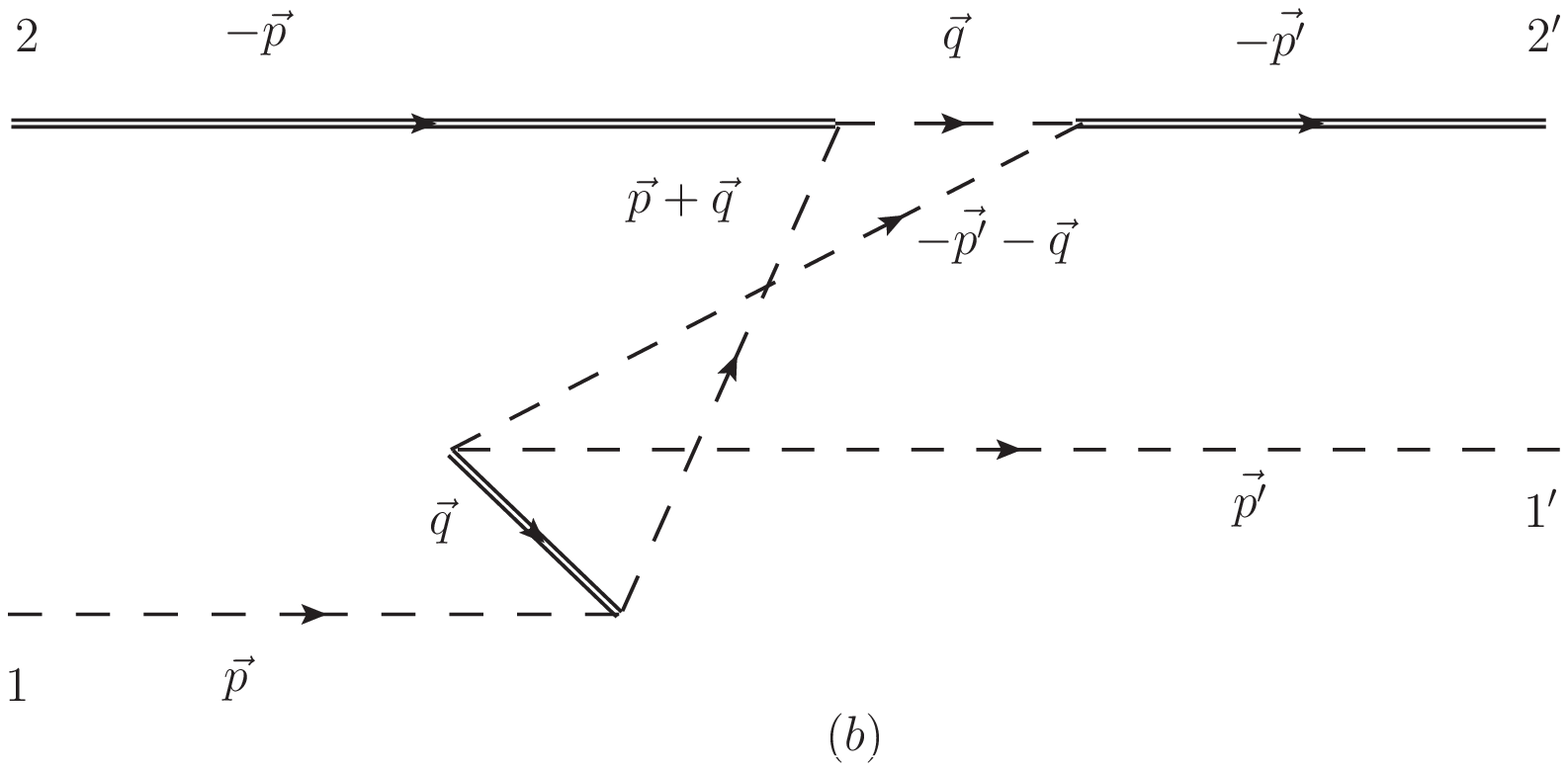} 
\includegraphics[scale=0.25]{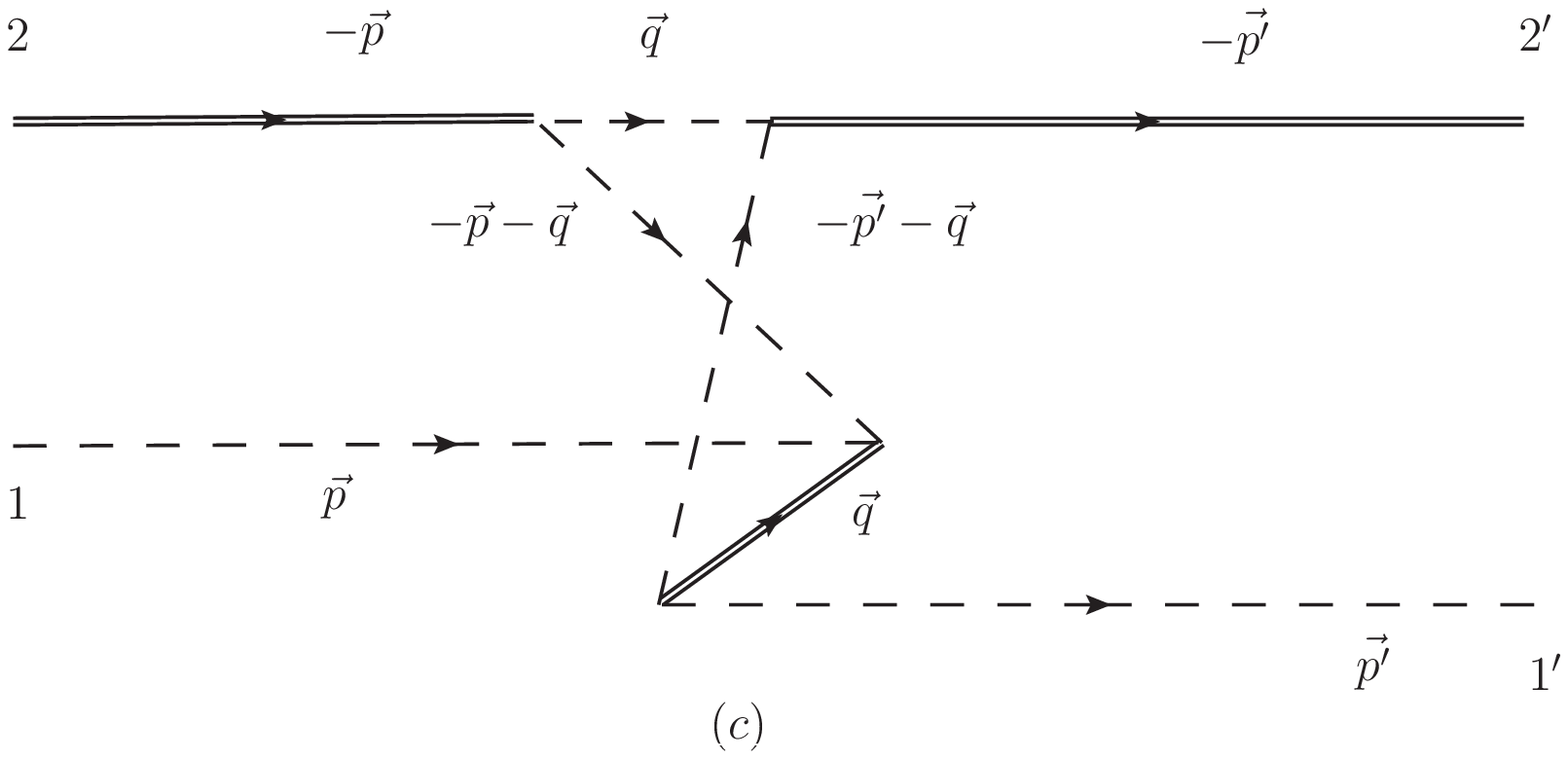}
\includegraphics[scale=0.25]{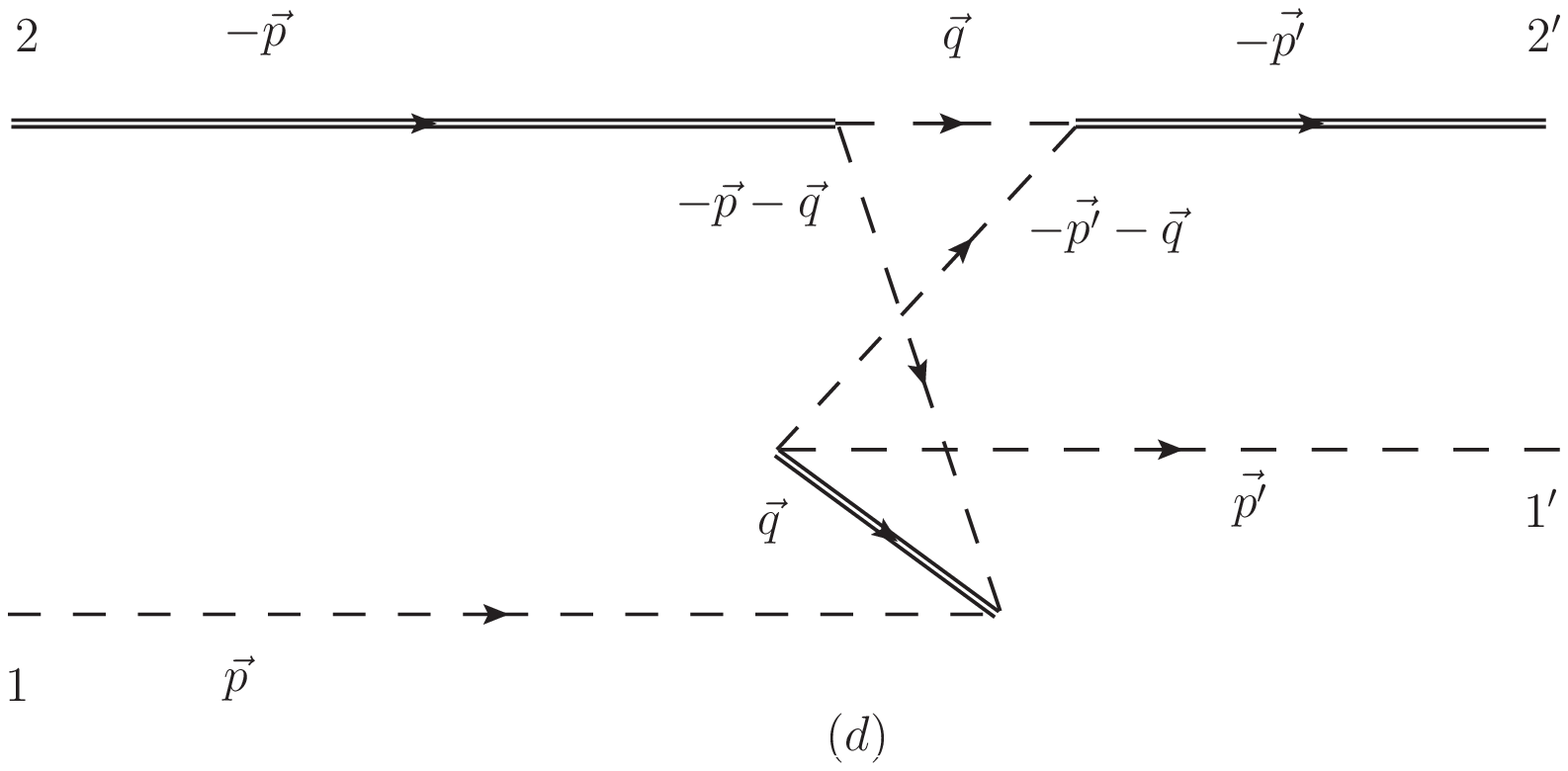} 
\includegraphics[scale=0.25]{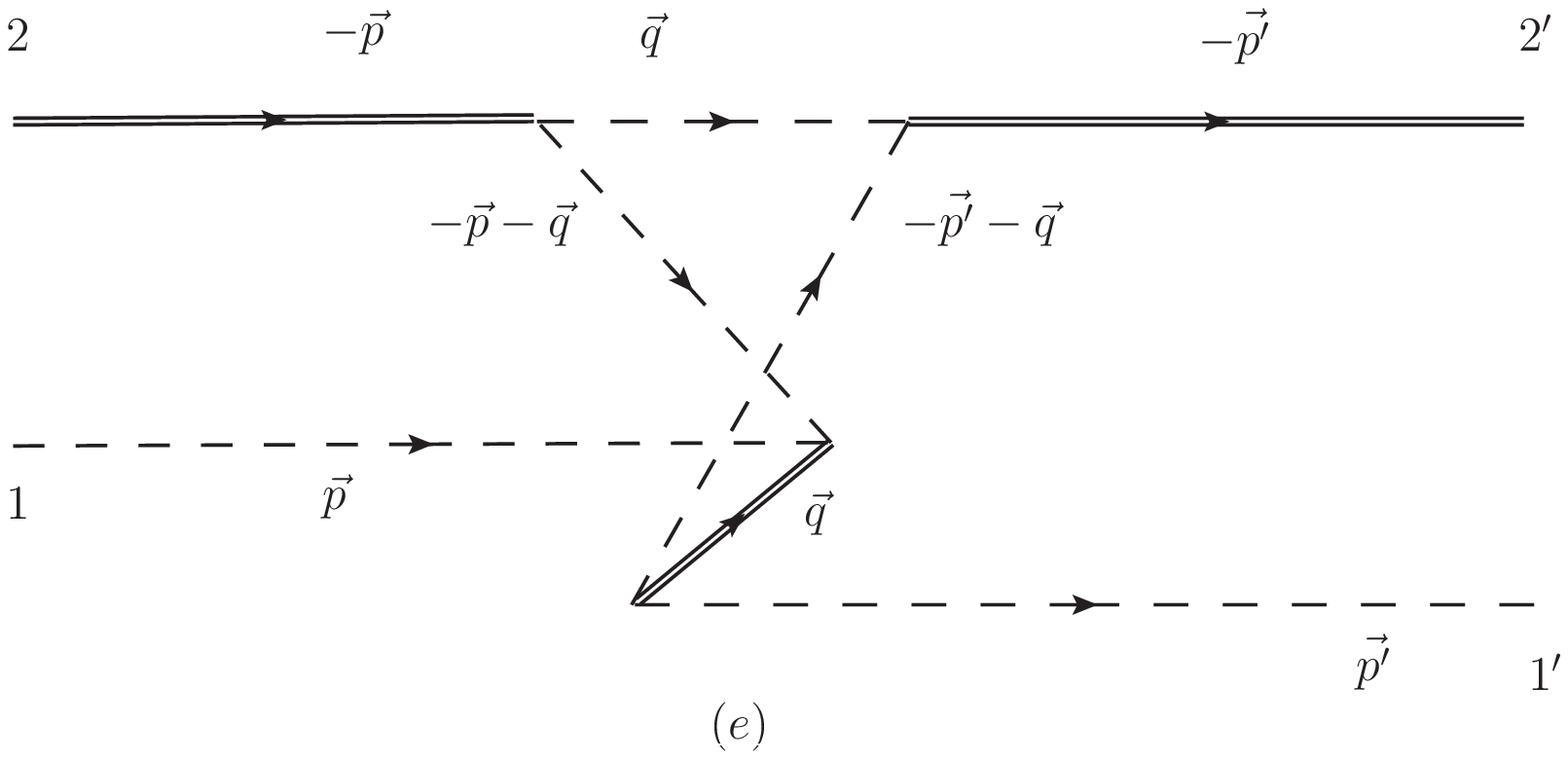}
\includegraphics[scale=0.25]{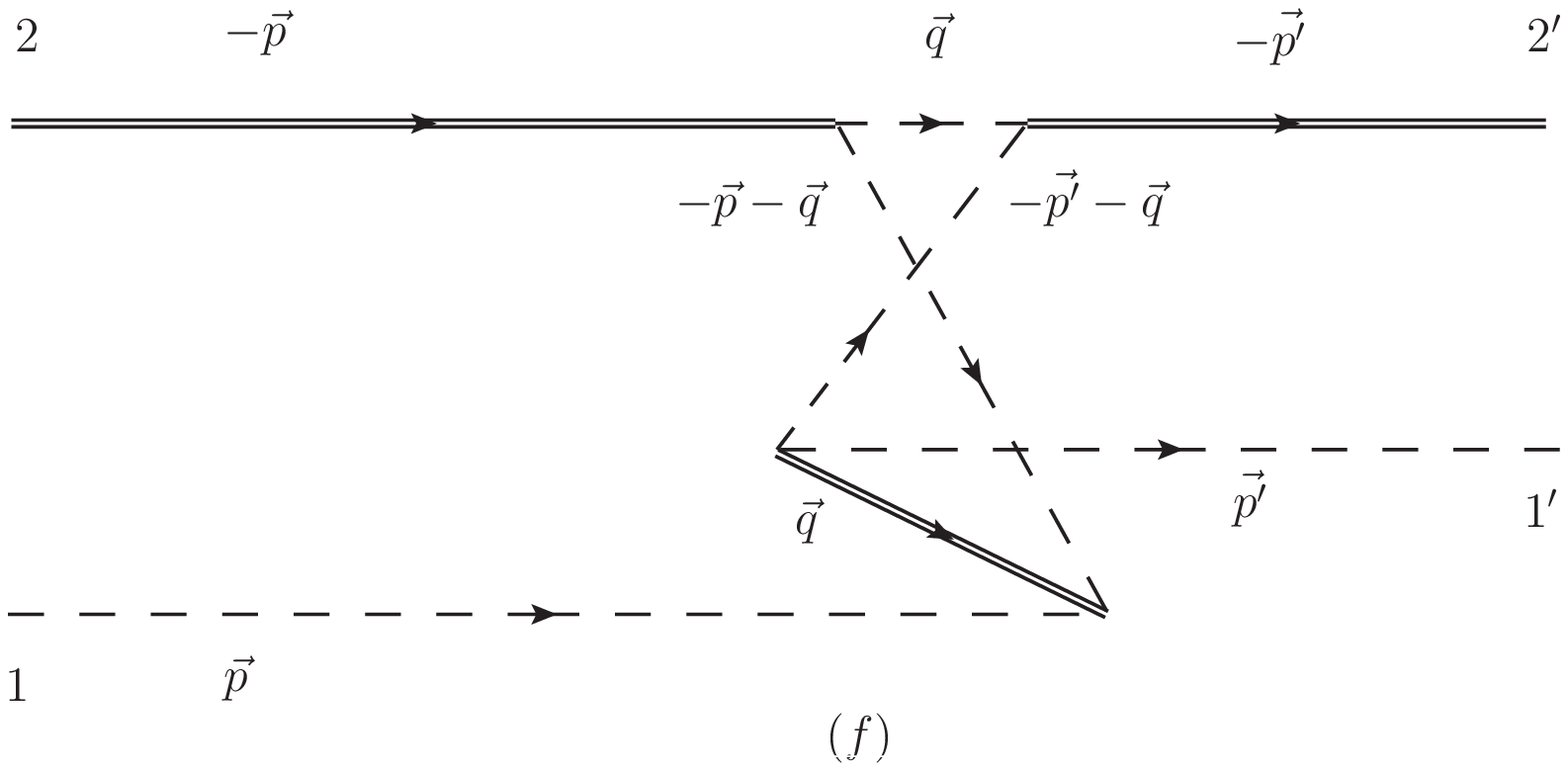}
\caption{The crossed boxes arise from $f_0(980)$ or $a_0(980)$ running backward
in time.}
\label{fig:crossed-negs}
\end{center}
\end{figure*}
\begin{figure*}[ht]
\begin{center}
\includegraphics[scale=0.25]{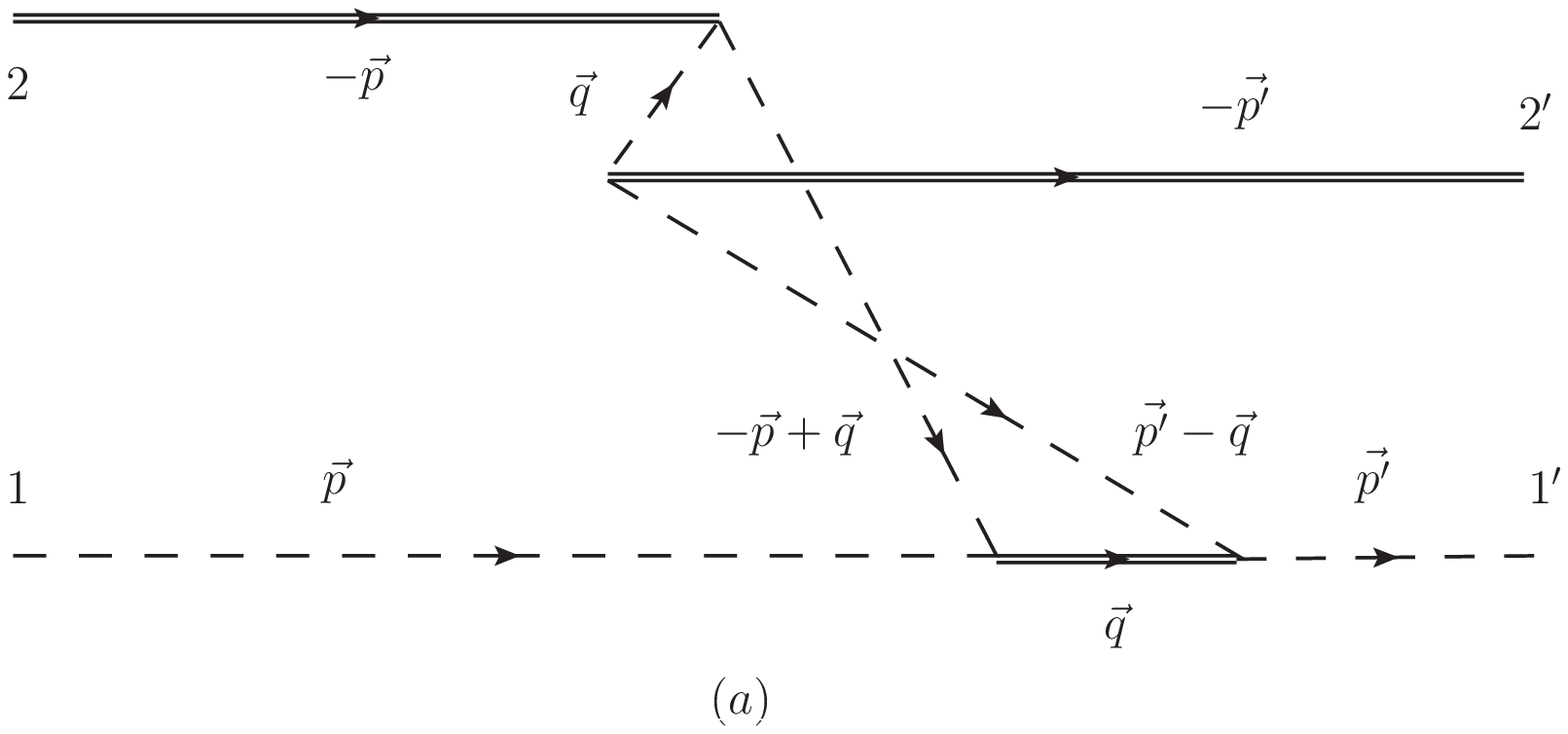}
\includegraphics[scale=0.25]{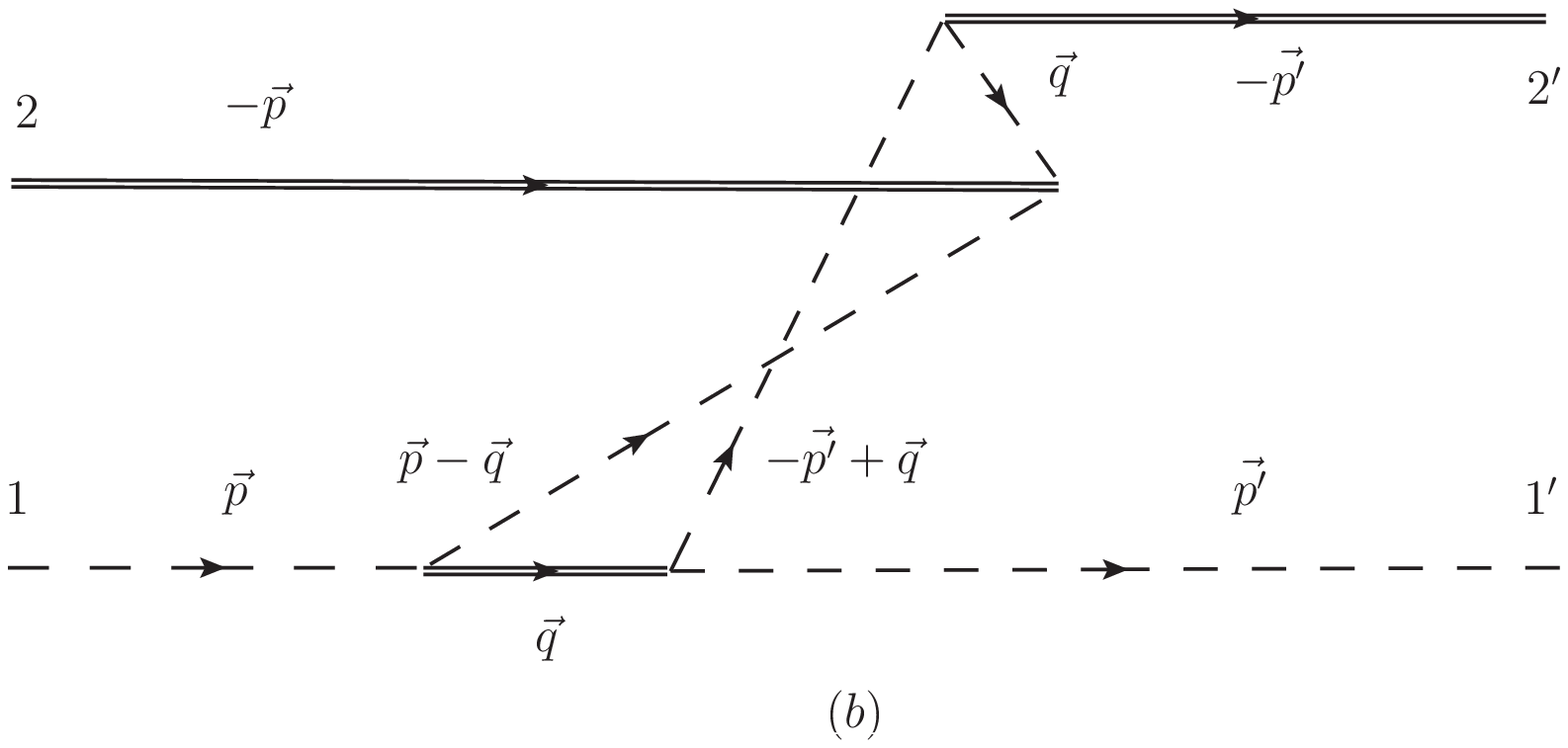} 
\includegraphics[scale=0.25]{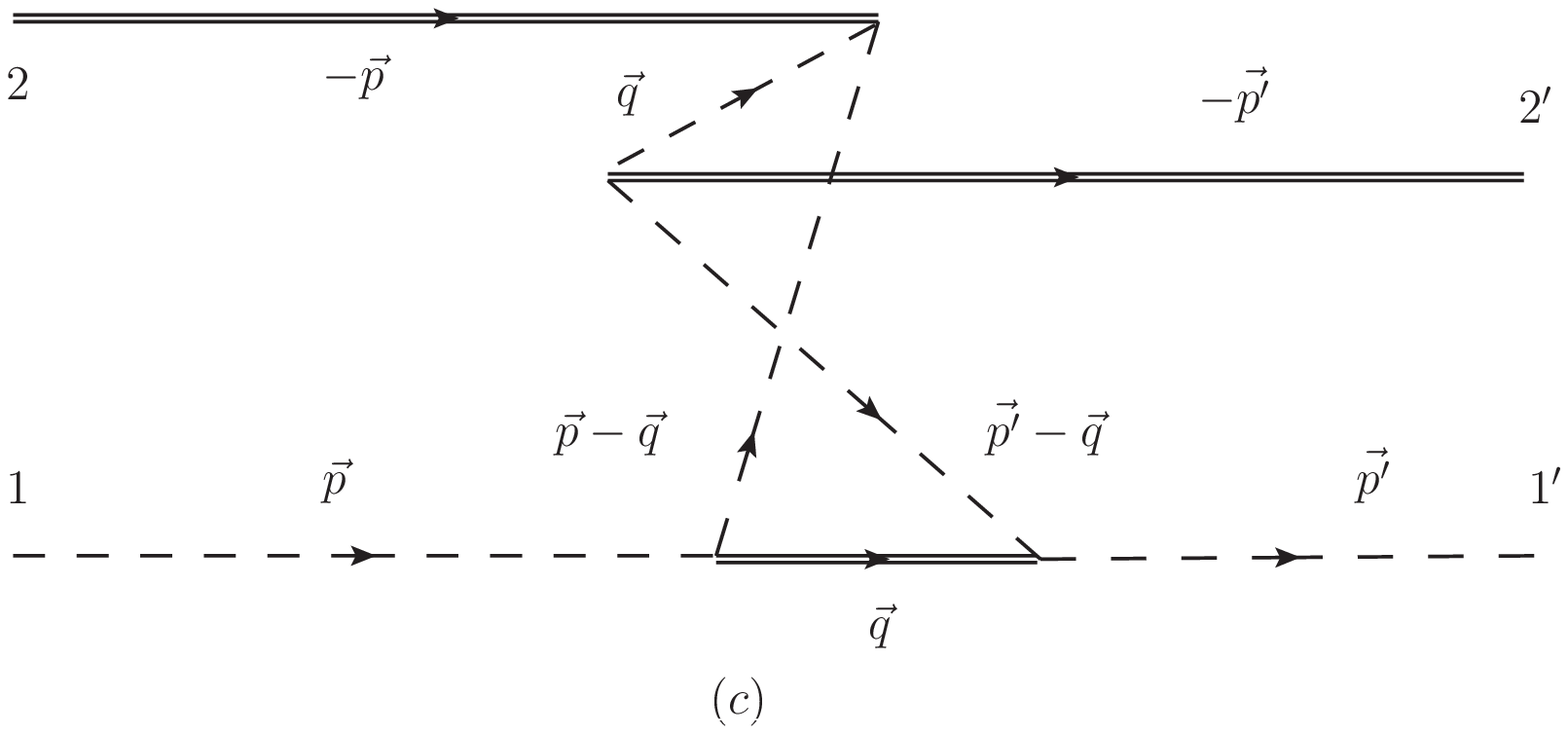}
\includegraphics[scale=0.25]{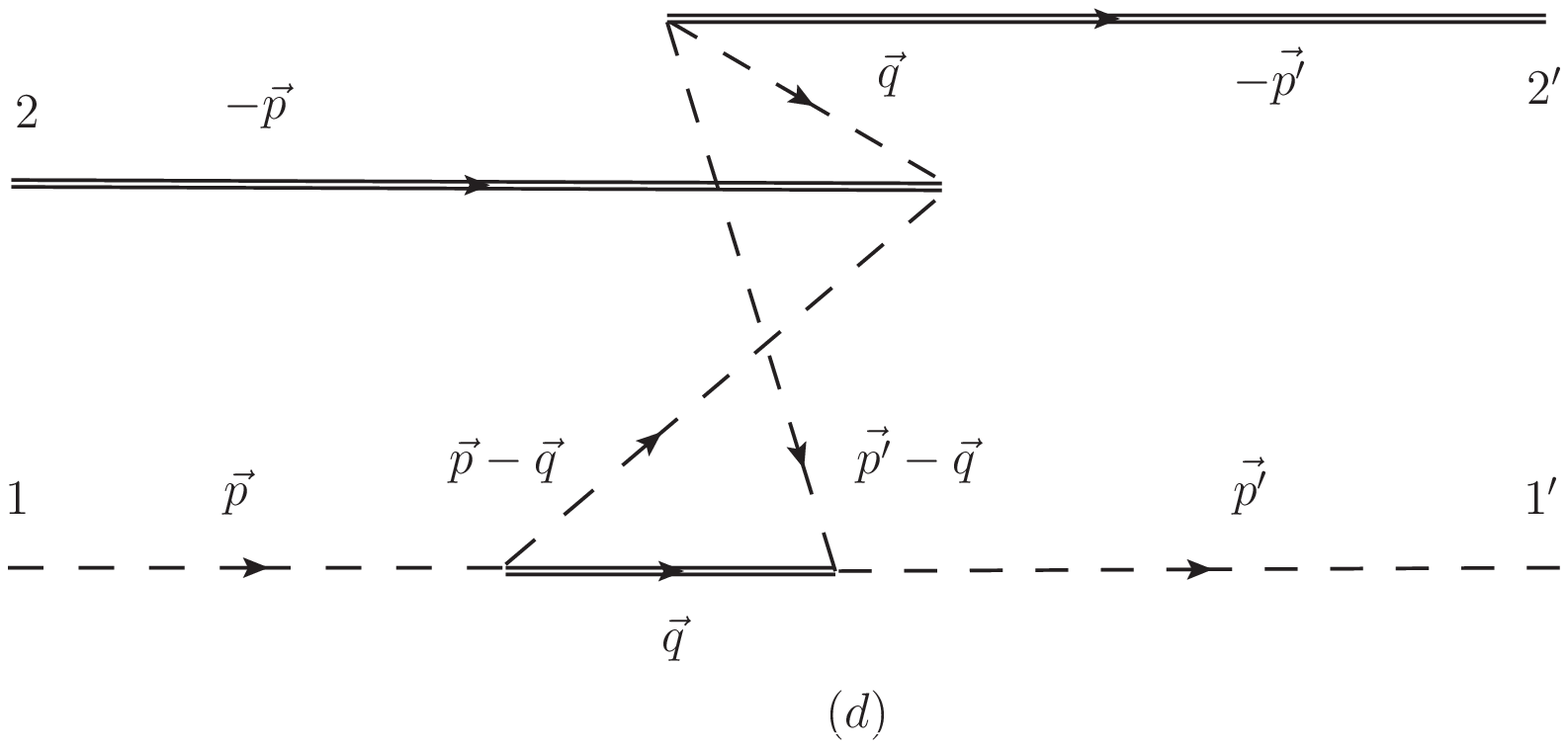} 
\includegraphics[scale=0.25]{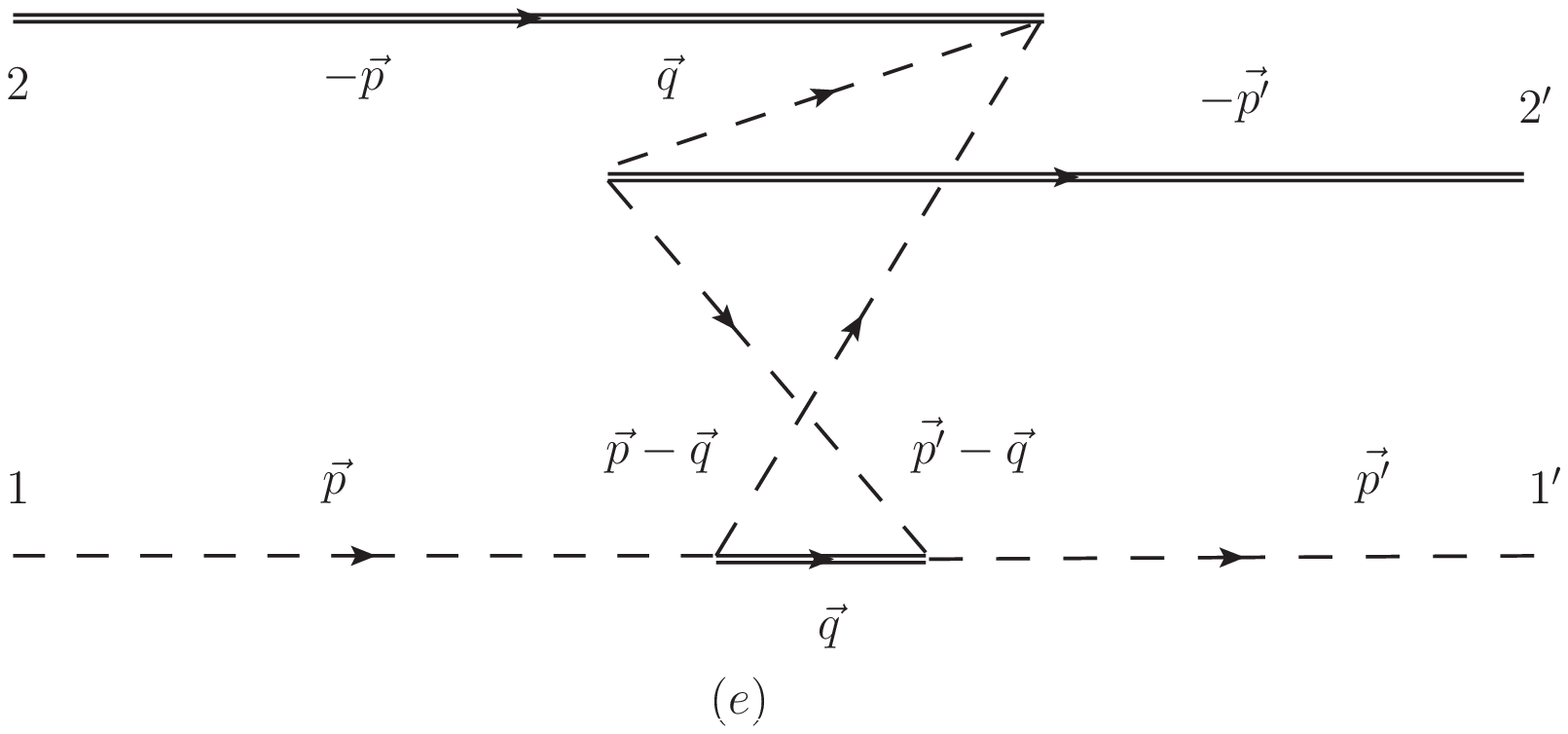}
\includegraphics[scale=0.25]{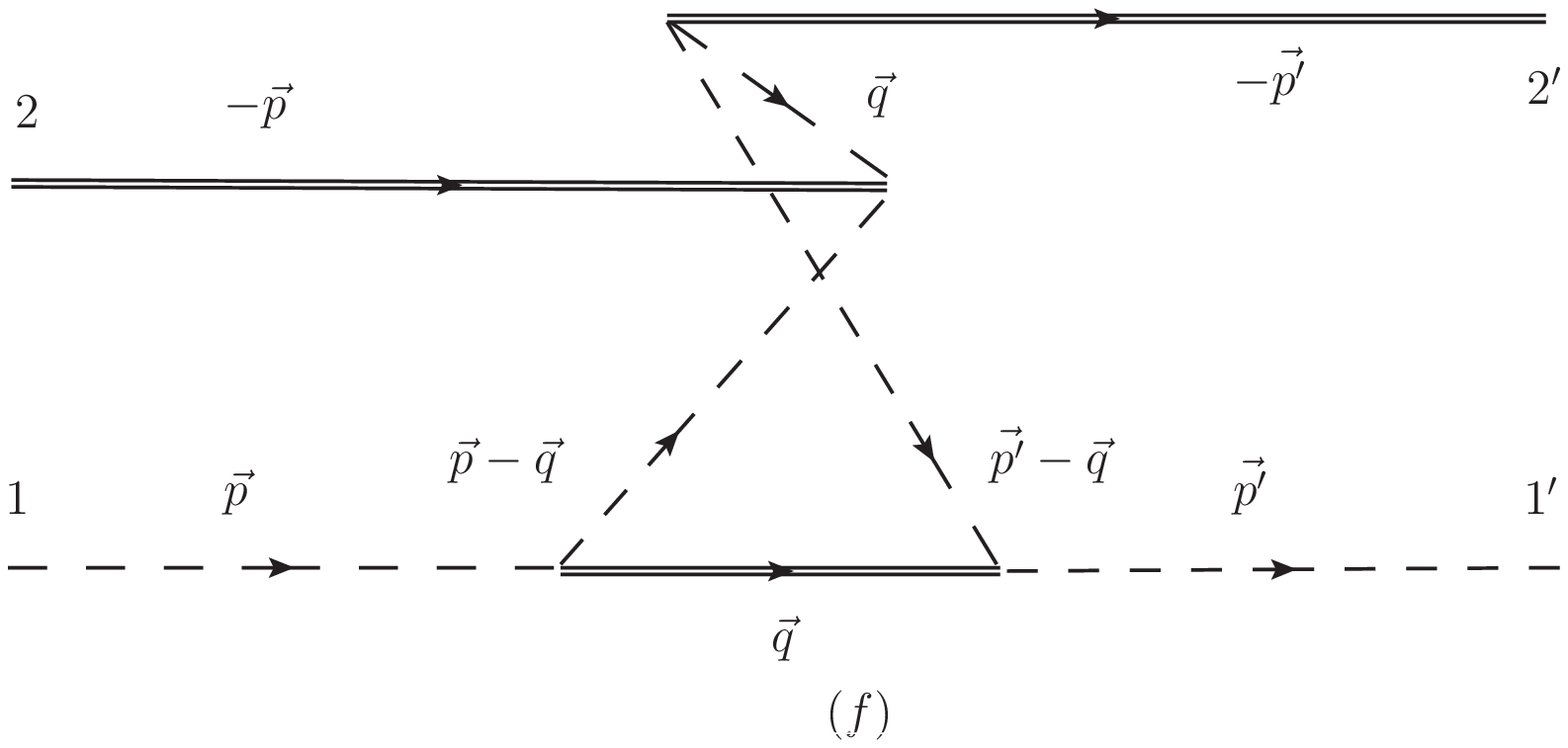}
\caption{The crossed boxes arise from kaons running backward
in time.}
\label{fig:crossed-negk}
\end{center}
\end{figure*}
\begin{figure*}[ht]
\begin{center}
\includegraphics[scale=0.25]{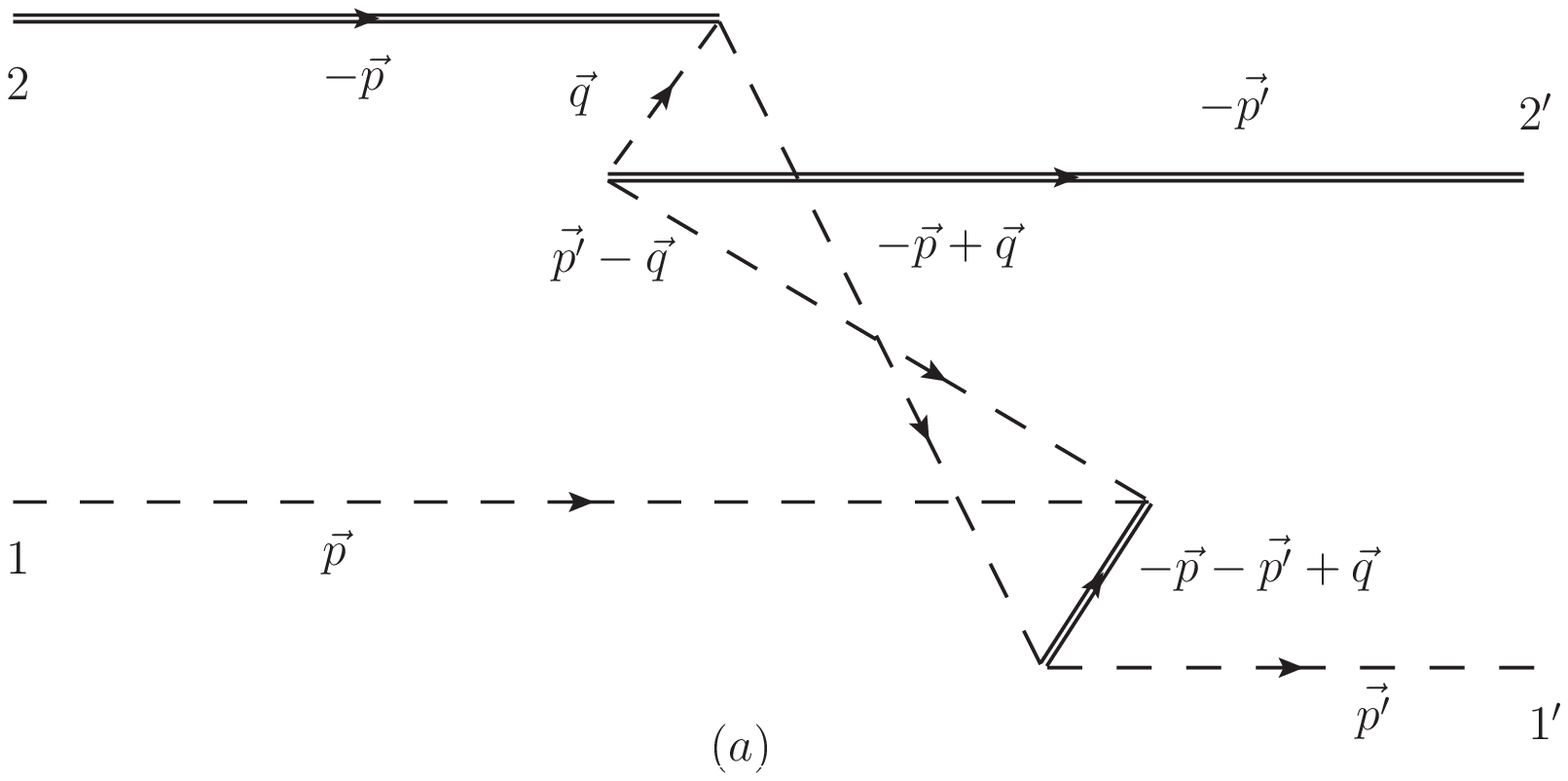}
\includegraphics[scale=0.25]{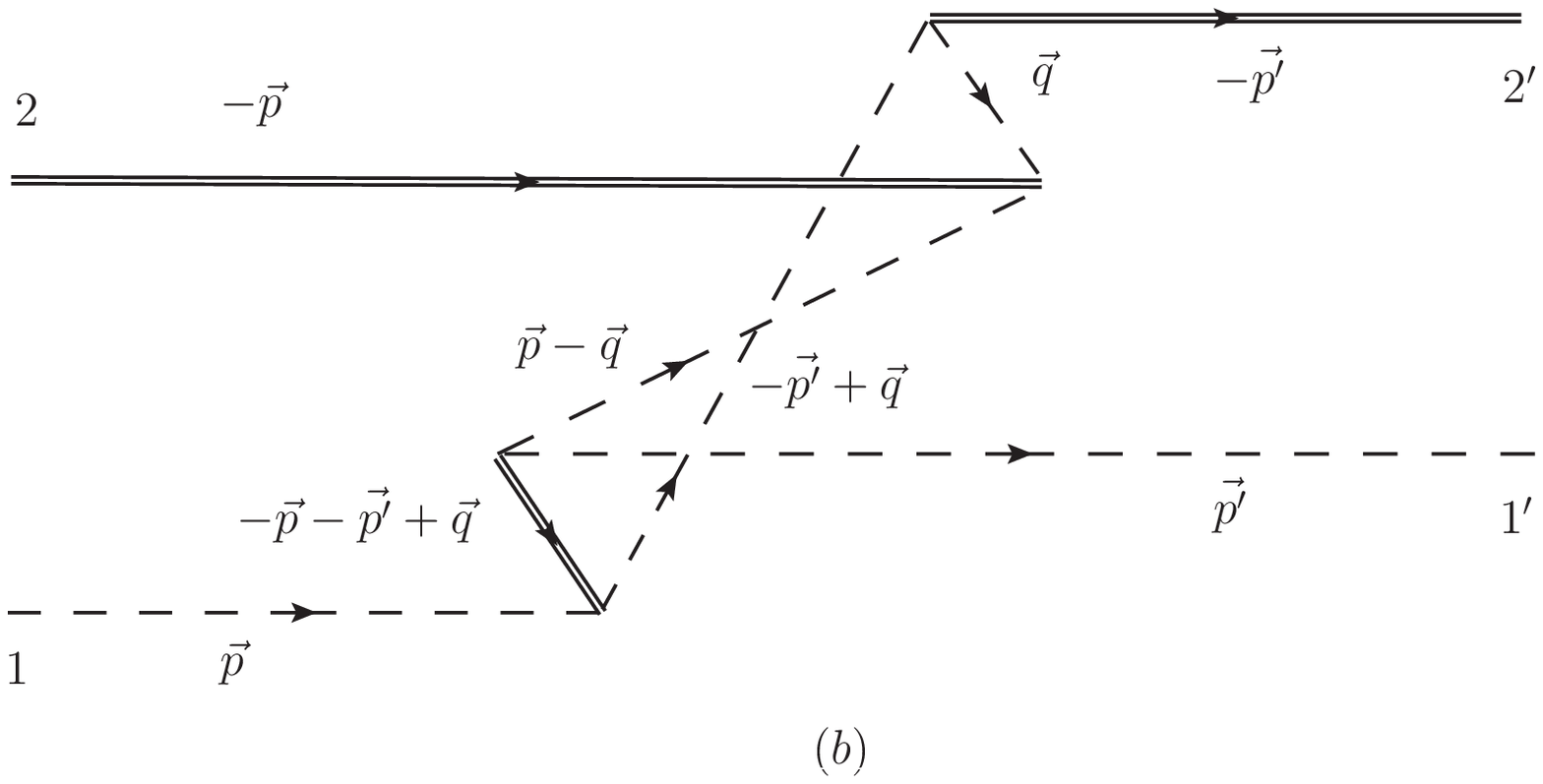} 
\includegraphics[scale=0.25]{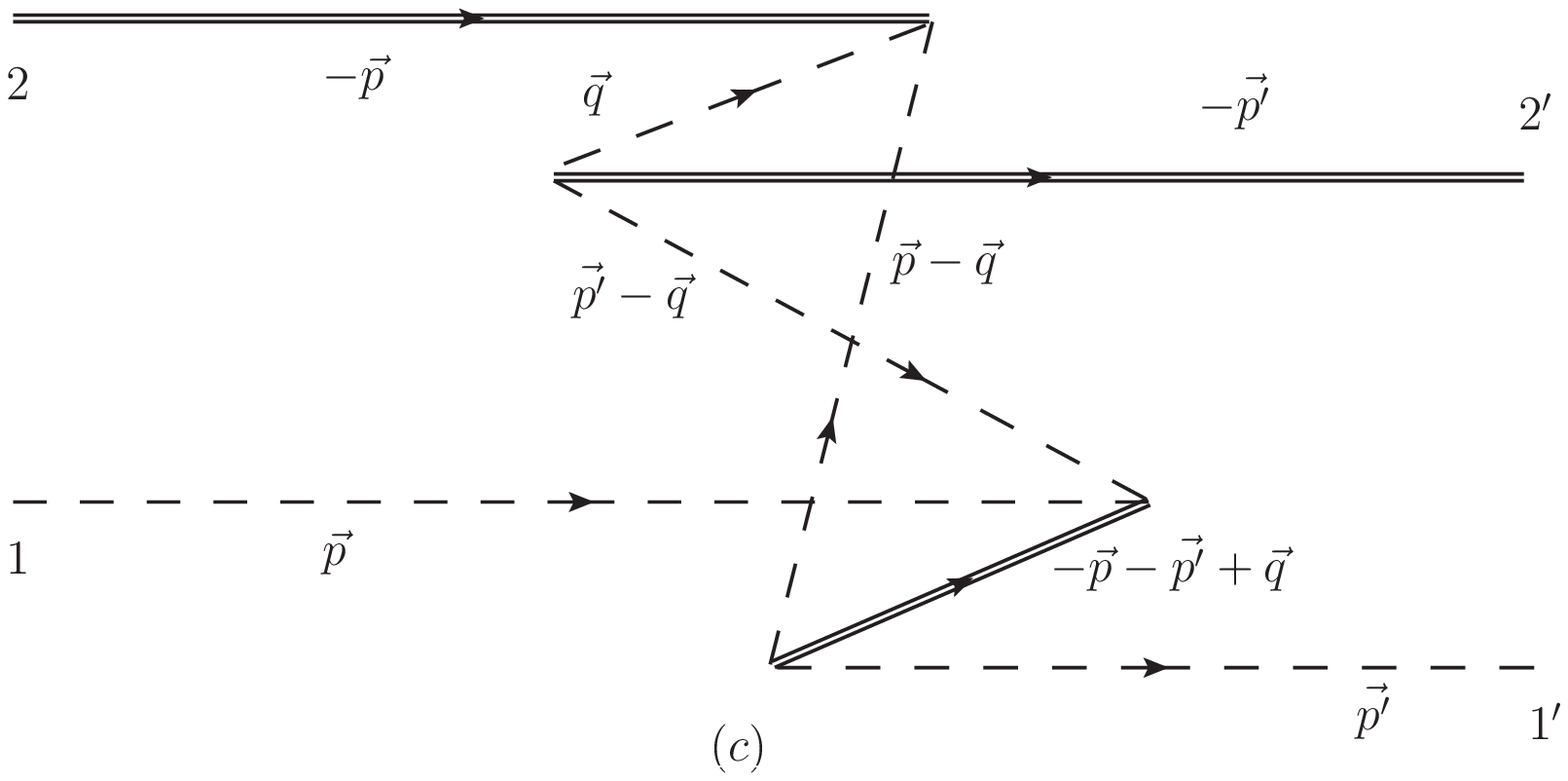}
\includegraphics[scale=0.25]{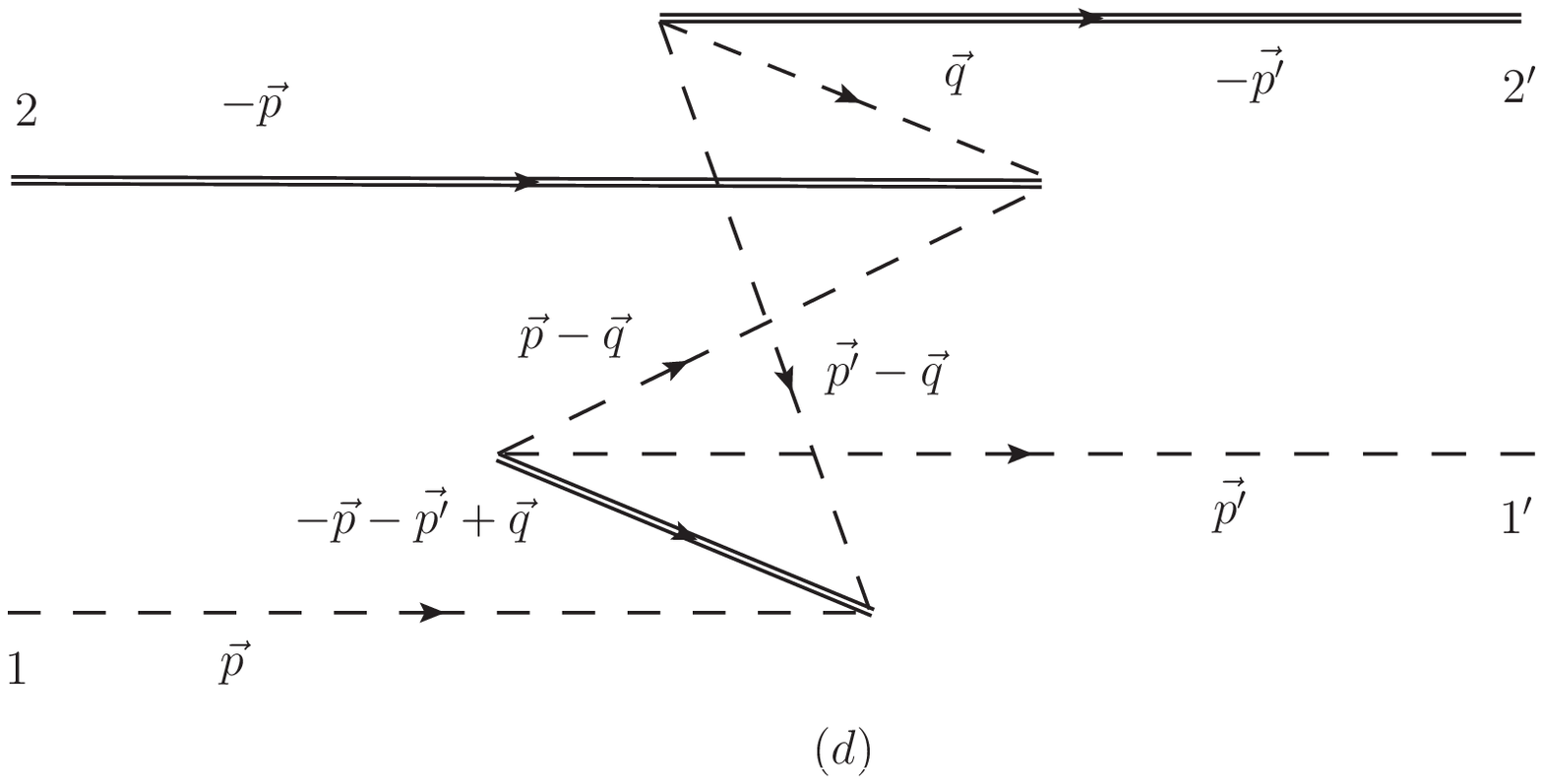} 
\includegraphics[scale=0.25]{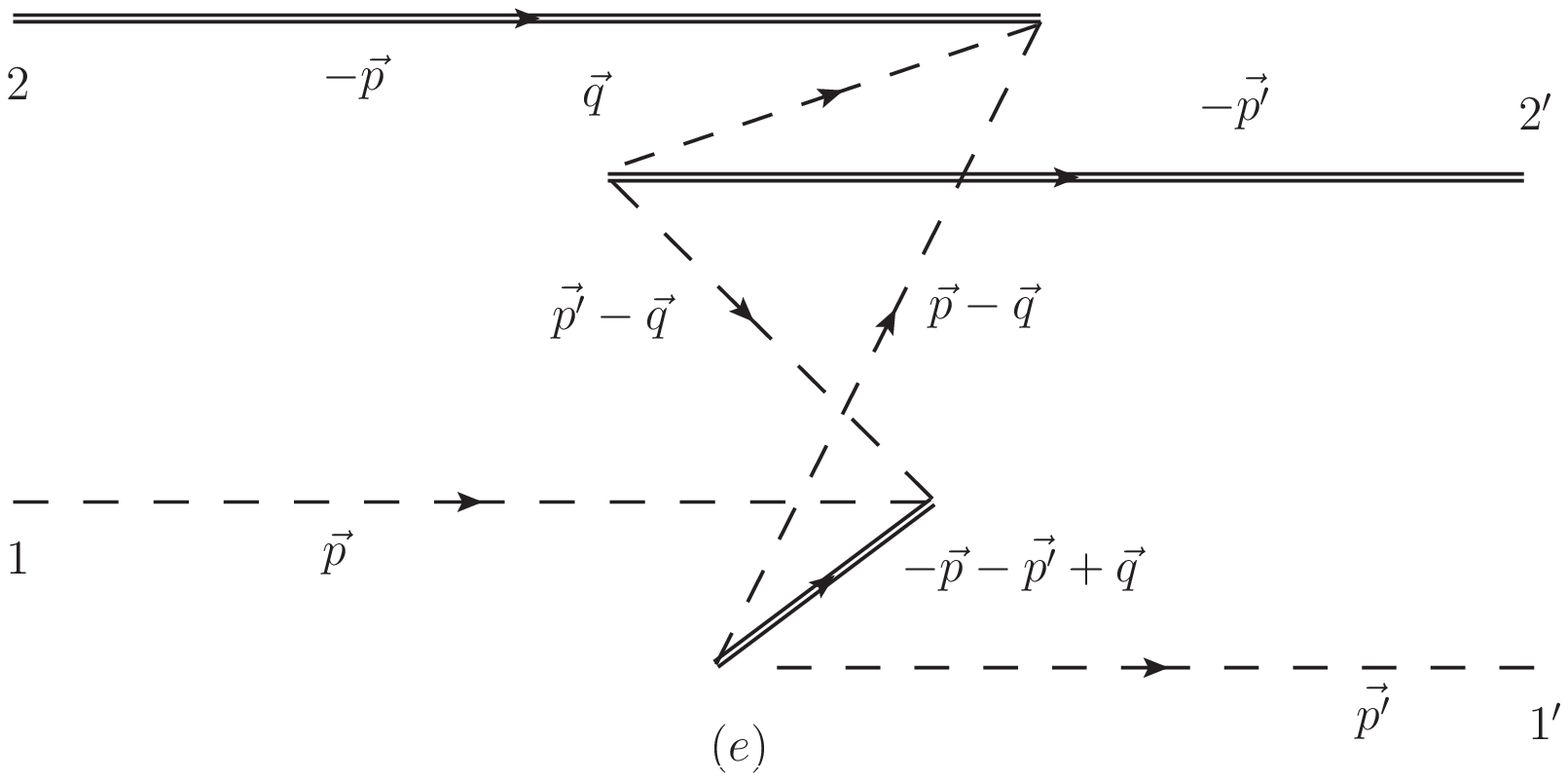}
\includegraphics[scale=0.25]{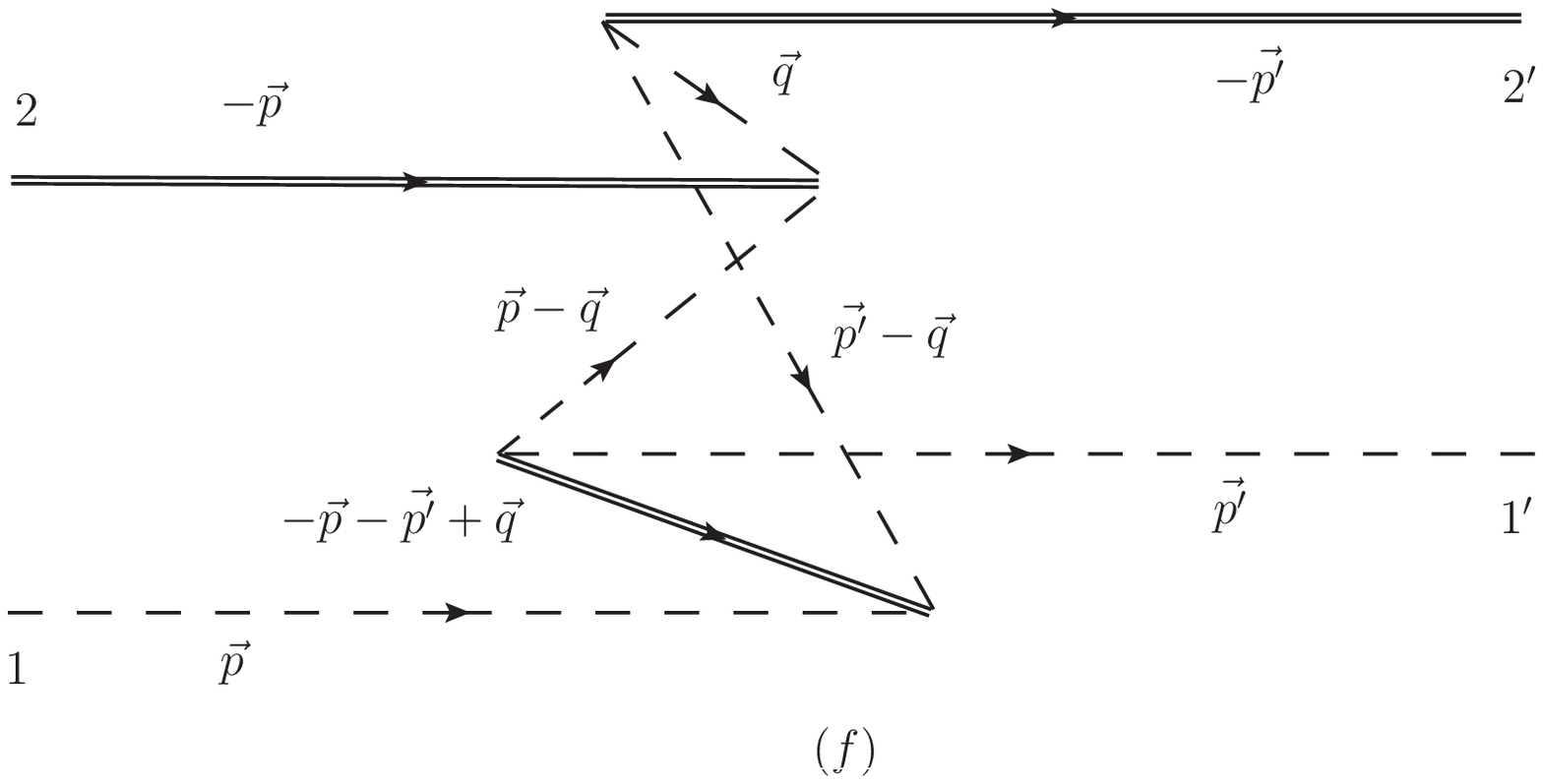}
\caption{The crossed boxes arise from both $f_0(980)$ or $a_0(980)$ and kaons running backward
in time.}
\label{fig:crossed-negks}
\end{center}
\end{figure*}

Moreover, there are additional diagrams at order $f_S^4$ that turn out to be of the same size
as the streched boxes, but are of a different
topology. {Those are the crossed boxes as shown in
Figs.~\ref{fig:crossed},~\ref{fig:crossed-negs},~\ref{fig:crossed-negk} and~\ref{fig:crossed-negks}.
The related amplitudes
 are given in Eqs.~\eqref{eq:cross-a} to~\eqref{eq:cross-negks-f} in the appendix.}
The effect of the mentioned contributions can again be
read off from the last column of
Table~\ref{tab:result-three-body}. 
It appears therefore not appropriate to employ a formalism
where the covariance of the ladder type diagrams is enforced
but diagrams of the crossed box type are ignored.




\begin{figure}[t]
\begin{center}
\includegraphics [scale=0.35] {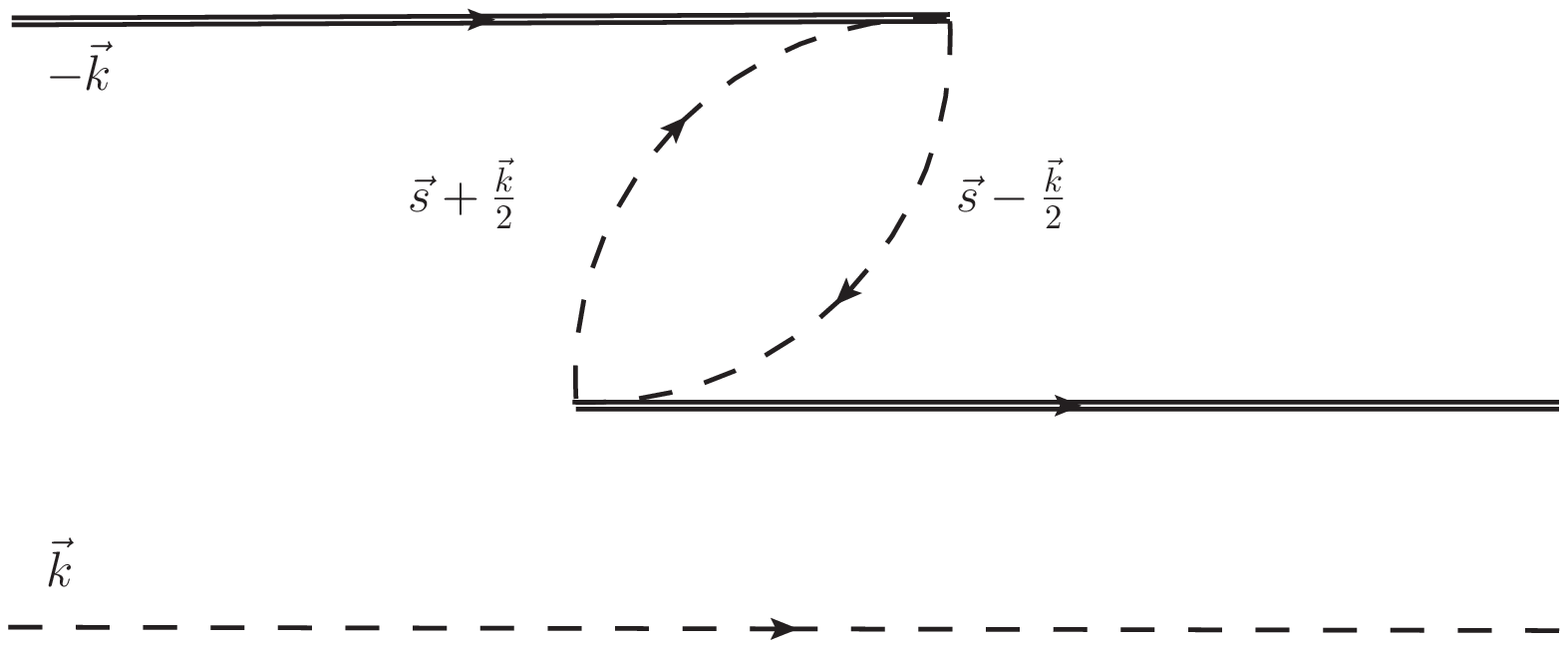}
\caption{Second time ordering for the self-energy correction of $f_0(980)$ and $a_0(980)$ mesons.
The double-solid line represents $f_0(980)$ or $a_0(980)$ meson, and the dashed line represents $K(\bar{K})$.
}
\label{fig:retardbubble}
\end{center}
\end{figure}

The final contribution necessary to restore covariance
at the one-loop level
is an additional one loop-correction to the self-energy
of the scalar mesons.
 {The corresponding diagram is shown
in Fig.~\ref{fig:retardbubble}. Since this contribution
is not an addition to the scattering potential, but
modifies the $f_0(980)/a_0(980) K$ propagator in the LS equation, it also modifies the
binding energies calculated from the scattering potential to order $f_S^2$. Therefore
we report its effect by additional lines 
in Table~\ref{tab:result-three-body}. Those are labeled
with $\alpha=1^\dagger$. The effect of this 
contribution is a moderate reduction of the calculated
binding energies, however, the effect is no larger than
the one of the crossed boxes.}
 This can be traced to the fact that the energy dependence
of the resulting self energy-contribution is quite weak and thus the threshold
subtraction to renormalize the self energy largely removes the contribution of the
second time ordering.

It should be noted that the final results now including the restoration of the covariance
to one loop shows a larger dependence of the regulator than the one without those corrections.
The reason for this is most probably that the
 added terms contain in the time slices more of the heavier
scalar fields that introduce larger momentum scales into the integrals. 
This observation indicates that a proper renormalization to this order might require a
three-kaon counter term. However, addressing this issue, which calls for a proper
power counting of the system, goes beyond the scope of this paper.

\section{Summary and Outlook}

In this work we provide additional support for the proposal
put forward in Ref.~\cite{Epelbaum:2016ffd}
to employ time ordered perturbation theory with relativistic
particle energies in calculations
of three-body scattering. To be concrete we demonstrate
on the example of isospin $1/2$ three-body $KK\bar K$ scattering
in the isobar formalism that the effect of the violation of
covariance on the binding energies for three-body bound
states is rather mild and can even be removed in a 
systematic way by inclusion of the streched boxes, although a restoration of covariance at the
two-loop level in practice would be a formidable task given
the large number of time orderings that would need to be included. 
Moreover, we demonstrate in addition that the so-called
crossed box contributions are of similar importance
than the streched boxes. This shows that a systematic calculation of three-body bound states
using time ordered perturbation theory could be set up by including at leading order just
the contributions to the potential at order $f_S^2$,
at next-to-leading order the two-particle
irreducible contributions (here two-particle
refers for the concrete example studied here to an intermediate state of an isobar and
a kaon) at order $f_S^4$ and so on, where the latter group should not only 
contain the new topologies that appear at this order but also the diagrams that
restore covariance at the one loop level. All two-particle reducible 
contributions are enhanced kinematically and 
are automatically generated via the LS equation.
Our results suggest that the binding energies of the three kaon system can be well estimated by
including the contribution from $t(a)$ only in the potential, although this violates covariance.

This kind of procedure avoids the need to
use covariant potentials in a three dimensional
formalism that can introduce
unphysical singularities~\cite{Mai:2017vot,Dawid:2020uhn} or to solve the very complicated 
four-dimensional scattering equations~\cite{Phillips:1996ed,Lahiff:2002wp}. 
Clearly, a further improved calculation needs to consider the contribution from inelastic
channels as well as  the Goldstone boson nature of the participating particles.

\begin{acknowledgements}
This work is supported in part by the National Natural Science Foundation of China (NSFC) and  the
Deutsche Forschungsgemeinschaft (DFG) through the funds provided to the Sino-German Collaborative Research
Center TRR110 ``Symmetries and the Emergence of Structure in QCD'' (NSFC Grant No. 12070131001,
DFG Project-ID 196253076),  by the Chinese Academy of Sciences (CAS) through a President's
International Fellowship Initiative (PIFI) (Grant No. 2018DM0034), by the VolkswagenStiftung
(Grant No. 93562), and by the EU Horizon 2020 research and innovation programme,  STRONG-2020 project
under grant agreement No. 824093. It is also supported by the National Natural Science Foundation
of China under Grant Nos. 12075288, 11735003, and 11961141012.
Furthermore, Xu Zhang acknowledges financial support from the China Scholarship Council.
\end{acknowledgements}

\begin{appendix}
\section{The renormalized propagator}
\label{app:f0renorm}
The unrenormalized or bare propagator in channel $\lambda$ reads 
\begin{align}
G_{b}^\lambda(E,k)=\frac{1}{E-\sqrt{{m_0^{(\lambda)}}^2+k^2}-\omega_K(k)-{f_S^0}^2\Sigma^{(\lambda)}(E,k)},
\end{align}
where $\sqrt{{m_0^{(\lambda)}}^2+k^2}$ is the bare free energy of the 
isobar, $f_S^0$ is the bare coupling 
for the isobar to $K\bar{K}$, and the unrenormalized self-energy ${f_S^0}^2\Sigma^{(\lambda)}(E,k)$
is given Eq.~(\ref{signonren}).

By requiring that $G_{b}^\lambda(E,k)$ has pole at
the $E_{on}$ defined in Eq.~(\ref{Eondef}), we may write
\begin{align}
G_{b}^\lambda(E,k)=\Big(&E-\omega^{(\lambda)}(k)-\omega_K(k)-{f_S^0}^2\Sigma^{(\lambda)}(E,k)\nonumber\\
&+{f_S^0}^2\Sigma^{(\lambda)}(E,k_{on})\Big)^{-1} ,
\end{align}
where we have fixed the bare energy via
\begin{align}
\sqrt{{m_0^{(\lambda)}}^2+k^2}=\omega^{(\lambda)}(k)-{f_S^0}^2\Sigma^{(\lambda)}(E,k_{on}).
\end{align}
 Then we may write 
 \begin{align}
G_{b}^\lambda&(E,k)=\nonumber \\
&\Big(E-\omega^{(\lambda)}(k)-\omega_K(k)-{f_S^0}^2\Sigma_R^{(\lambda)}(E,k)\Big)^{-1},
\end{align}
where 
\begin{align}
\Sigma_R^{(\lambda)}(E,k)= \Sigma^{(\lambda)}(E,k)-\Sigma^{(\lambda)}(E,k_{on}).
\end{align}
Note that for $E<\sqrt{{m^{(\lambda)}}^2-m_K^2}$, the continuation $\omega_K(k_{on}) \to -\omega_K(k_{on})$ needs to be employed.

 The renormalized propagator, $G_r^\lambda(E,k)$ is related to
 the bare propagator, $G_{b}^\lambda(E,k)$, by $G_r^\lambda(E,k)=G_{b}^\lambda(E,k)Z^{-1}$. 
 Its residue at the pole is 
 \begin{align}
&\lim_{E  \to E_{on} }(E - \omega^{(\lambda)}(k) - \omega_K(k))Z^{-1}G_{b}(E,k)=\nonumber \\
& \qquad Z^{-1}\left(\frac{1}{1-\frac{d}{dE}{f_S^0}^2\Sigma^{(\lambda)}(E,k_{on}) \big |_{E=E_{on} }}\right).
\end{align}
By definition this residue needs to be one. Using the
definition of the renormalized coupling $f_S^2={f_S^0}^2Z$, we get
\begin{align}
\frac{{f_S^0}^2}{f_S^2}=1-\frac{d}{dE}{f_S^0}^2\Sigma^{(\lambda)}(E,k_{on}) \big |_{E=E_{on} }.
\end{align}
Hence, we have 
\begin{align}
{f_S^0}^2=\frac{f_S^2}{1+\frac{d}{dE}f_S^2\Sigma^{(\lambda)}(E,k_{on}) \big |_{E=E_{on}}},
\end{align}
and 
\begin{align}
Z={1+\frac{d}{dE}f_S^2\Sigma^{(\lambda)}(E,k_{on}) \big |_{E=E_{on} }}.
\end{align}

Accordingly the expression for the renormalized propagator reads
\begin{align}
G_r^\lambda(E,k)=\Big[&Z\Big(E-\sqrt{{m^{(\lambda)}}^2+k^2}-\omega_K(k)\Big) \nonumber\\
&-\alpha f_S^2\Sigma_R^{(\lambda)}(E,k)\Big]^{-1} .
\end{align}

\section{Renormalization of the kaon pole contribution}
\label{app:Krenorm}

In this appendix we outline the renormalization procedure for the kaon pole contribution following the procedure used in  Ref.~\cite{Krehl:1999km} to renormalize the nucleon pole 
in $\pi N$ scattering. 
Clearly, when the kaon $s$-channel pole is included into the potential of the LS-equation
both its coupling to the scalar fields and a kaon as well as its mass get renormalized. 
Thus, to have in the final amplitude kaon pole and residue correct we need to employ
a potential that is formulated in terms of bare parameters ($c.f.$ Eq.~(\ref{eq:Vsch})).
In general one has 
\begin{align}
\label{eq:ccbfo}
\frac{1}{2}f_1f_1=&(\Gamma_{11}f_1^{(0)} + \Gamma_{12}f_2^{(0)} ) (1-{f_1^{(0)}}^2{\Sigma_{11}^{(3)}}'-2f_1^{(0)}f_2^{(0)}{\Sigma_{12}^{(3)}}' \nonumber\\
&-{f_2^{(0)}}^2{\Sigma_{22}^{(3)}}')^{-1}(\Gamma_{11}^{T}f_1^{(0)} + \Gamma_{21}^{T}f_2^{(0)} ),
\end{align}
\begin{align}
\label{eq:ccbft}
\frac{\sqrt{3}}{2}f_1f_2=&(\Gamma_{11}f_1^{(0)} + \Gamma_{12}f_2^{(0)} ) (1-{f_1^{(0)}}^2{\Sigma_{11}^{(3)}}'-2f_1^{(0)}f_2^{(0)}{\Sigma_{12}^{(3)}}' \nonumber\\
&-{f_2^{(0)}}^2{\Sigma_{22}^{(3)}}')^{-1}(\Gamma_{12}^{T}f_1^{(0)} + \Gamma_{22}^{T}f_2^{(0)} ),
\end{align}
\begin{align}
\label{eq:ccbfh}
\frac{3}{2}f_2f_2=&(\Gamma_{21}f_1^{(0)} + \Gamma_{22}f_2^{(0)} ) (1-{f_1^{(0)}}^2{\Sigma_{11}^{(3)}}'-2f_1^{(0)}f_2^{(0)}{\Sigma_{12}^{(3)}}' \nonumber\\
&-{f_2^{(0)}}^2{\Sigma_{22}^{(3)}}')^{-1}(\Gamma_{12}^{T}f_1^{(0)} + \Gamma_{22}^{T}f_2^{(0)} ).
\end{align}
In these expressions the kaon self energy $f_{\lambda'}^{{0}}f_{\lambda}^{{0}}\Sigma_{\lambda'\lambda}^{(3)}(E)$ 
and dressed vertex function $\Gamma_{\lambda'\lambda}(E,k)$ get
generated by the LS-equation. Explicitly the $\Sigma^{(3)}_{\lambda'\lambda}(E)$ can be written as
\begin{align}
&\Sigma^{(3)}_{\lambda'\lambda}(E)=\sum_{\lambda\lambda'}({\cal I}^{\lambda})({\cal I}^{\lambda'})\frac{1}{2m_K} \nonumber \\
\times &\bigg ( \delta_{\lambda\lambda'} \int \frac{dq^3}{(2\pi)^3 2\omega^{(\lambda)}(q)2\omega_K(q)}G_{r}(E,q)   \nonumber \\
&+ \int \frac{d^3qd^3q'}{(2\pi)^62\omega^{(\lambda)}(q)2\omega_K(q)2\omega^{(\lambda')}(q')2\omega_K(q')} \nonumber \\
& \times G_{r}(E,q') {\sqrt{2\omega^{(\lambda')}(q')2\omega_K(q')}}{\sqrt{2\omega^{(\lambda)}(q)2\omega_K(q)}}  \nonumber \\
& \times \widetilde{T}^{\lambda'\lambda} (E,{q'},{q}) G_{r}(E,q)   \bigg),     
\end{align}
where $ \widetilde{T}^{\lambda'\lambda} (E,{q'},{q})$ is
the solution of the LS-equation employing the non-pole part of the potential, namley the $t$-channel contribution in Fig.~\ref{fig:tchannel}(a)~and~(b) and $s$-channel contribution in Fig.~\ref{fig:schannel}(a).
Isospin factors ${\cal I}^{\lambda}{\cal I}^{\lambda'}$ are ${\cal I}^{1}{\cal I}^{1}=\frac{1}{2}$, ${\cal I}^{1}{\cal I}^{2}=\frac{\sqrt{3}}{2}$ and ${\cal I}^{2} {\cal I}^{2}=\frac{3}{2}$.
Clearly the procedure can be straightforwardly generalised to include also higher orders in the potential.

The vertex functions $\Gamma_{\lambda'\lambda}(E,k)$ can be written as
\begin{align}
&\Gamma_{\lambda'\lambda}(E,k)={\cal I}^{\lambda}\Big( \delta_{\lambda'\lambda}+ \nonumber \\
& \int \frac{dq^3}{(2\pi)^3 2\omega^{(\lambda)}(q)2\omega_K(q)}{\sqrt{2\omega^{(\lambda)}(k)2\omega_K(k)}}\nonumber \\
&\times {\sqrt{2\omega^{(\lambda)}(q)2\omega_K(q)}}\widetilde{T}^{\lambda'\lambda} (E,k,q) G_{r}(E,q)   \Big),
\end{align}
where the isospin factors are ${\cal I}^1=\sqrt{\frac{1}{2}}$ and ${\cal I}^2=\sqrt{\frac{3}{2}}$.

As explained in the main text, based on general considerations, the physical coupling $f_S$ is known when we assume that both $a_0(980)$ and $f_0(980)$ are 
$\bar KK$ bound states.
Moreover, the physical kaon mass is known, while the bare parameters are unknown. Thus we need
to express the latter in terms of the former. Solving Eq.~(\ref{eq:ccbft}) and Eq.~(\ref{eq:ccbfh}) for $f_1^{(0)}$ and $f_2^{(0)}$ and taking $f_1=f_2=f_S$ gives
\begin{align}
{f_1^{(0)}}^2=\frac{1}{2}f_S^2\Big/\Big[&(\Gamma_{11}+ R\Gamma_{12} ) (\Gamma_{11}^{T}+ R\Gamma_{21}^{T} )+\frac{1}{2}f_S^2 \nonumber \\
&\times({\Sigma_{11}^{(3)}}'+2R{\Sigma_{12}^{(3)}}' +R^2{\Sigma_{22}^{(3)}}')\Big],
\end{align}
\begin{align}
f_2^{(0)}=f_1^{(0)}R,
\end{align}
which agrees to Eq.~(\ref{eq:Kwffrenorm}) and Eq.~(\ref{eq:Kwfarenorm}). With the bare coupling determined we can now also calculate
the bare mass from
\begin{align}
m_K^{(0)}= m_K -\Big(&{f_1^{(0)}}^2{\Sigma_{11}^{(3)}}+2f_1^{(0)}f_2^{(0)}{\Sigma_{12}^{(3)}}+{f_2^{(0)}}^2{\Sigma_{22}^{(3)}}\Big) .
\end{align}

\section{Contributions proportional to $f_S^4$}

In the framework of TOPT, the expressions corresponding to the diagrams of Fig.~\ref{fig:stret} may be written in
the following form
\begin{align} 
 \label{eq:stret-a}
V&_{stret-a}^{\lambda' \lambda}(E,\vec{p'},\vec{p}\ )=f_S^4 {\cal I} N\int_0^{\Lambda} \frac{d^3{q}}{(2\pi)^3} \nonumber \\
&\times \frac{1}{16\omega_K({p}+{q})\omega_K({q})\omega^{(1)}({q})\omega_K(p'+{q})}\nonumber\\
&\times \frac{1}{E-\omega_1(p)-\omega_K({p}+{q})-\omega_K({q})+i \varepsilon} \nonumber \\
& \times \frac{1}{E-\omega_1(p)-\omega_{2'}(p')-\omega_K(p'+q)-\omega_K(p+q)+i \varepsilon}\nonumber \\
&\times \frac{1}{E-\omega_{2'}(p')-\omega_K(p'+{q})-\omega^{(1)}({q})+i \varepsilon},
\end{align}
 \begin{align} 
 \label{eq:stret-b}
V&_{stret-b}^{\lambda' \lambda}(E,\vec{p'},\vec{p}\ )=f_S^4 {\cal I} N\int_0^{\Lambda} \frac{d^3{q}}{(2\pi)^3}\nonumber \\
&\times \frac{1}{16\omega_K({p}+{q})\omega_K({q})\omega^{(1)}({q})\omega_K(p'+{q})}\nonumber\\
&\times \frac{1}{E-\omega_2(p)-\omega_K(p+q)-\omega^{(1)}({q})+i \varepsilon} \nonumber \\
& \times \frac{1}{E-\omega_2(p)-\omega_K(p+q)-\omega_K(p'+q)-\omega_{1'}(p')+i \varepsilon}\nonumber \\
&\times \frac{1}{E-\omega_{1'}(p')-\omega_K(p'+{q})-\omega_K({q})+i \varepsilon}.
\end{align}

In the framework of TOPT, the expressions corresponding to the diagrams of Fig.~\ref{fig:stret-negs} may be written in
the following form
\begin{align} 
\label{eq:stret-negs-a}
V&_{stret-negs-a}^{\lambda' \lambda}(E,\vec{p'},\vec{p}\ )=f_S^4 {\cal I} N\int_0^{\Lambda} \frac{d^3{q}}{(2\pi)^3}\nonumber\\
&\times \frac{1}{16\omega_K({p}+{q})\omega_K({q})\omega^{(1)}({p+p'+q})\omega_K(p'+{q})}\nonumber\\
&\times \frac{1}{E-\omega_1(p)-\omega_K(p+q)-\omega_K({q})+i \varepsilon} \nonumber \\
&\times \frac{1}{E-\omega_1(p)-\omega_{2'}(p')-\omega_K({p}+{q})-\omega_K(p'+{q})+i \varepsilon}\nonumber \\
&\times 1/[E-\omega_{1}(p)-\omega_{1'}(p')-\omega_{2'}(p')\nonumber \\
&\qquad  \qquad  -\omega_K(p'+q)-\omega^{(1)}({p+p'+q})+i \varepsilon],
\end{align}
\begin{align} 
 \label{eq:stret-negs-b}
V&_{stret-negs-b}^{\lambda' \lambda}(E,\vec{p'},\vec{p}\ )=f_S^4 {\cal I} N\int_0^{\Lambda} \frac{d^3{q}}{(2\pi)^3}\nonumber\\
&\times \frac{1}{16\omega_K({p}+{q})\omega_K({q})\omega^{(1)}({p+p'+q})\omega_K(p'+{q})}\nonumber\\
&\times 1/ [E-\omega_{1}(p)-\omega_{2}(p)-\omega_{1'}(p')-\omega_K(p+q)\nonumber\\
&\qquad  \qquad -\omega^{(1)}({p+p'+q})+i \varepsilon ] \nonumber \\
&\times \frac{1}{E-\omega_2(p)-\omega_{1'}(p')-\omega_K({p}+{q})-\omega_K(p'+{q})+i \varepsilon}  \nonumber \\
&\times \frac{1}{E-\omega_{1'}(p')-\omega_K(p'+q)-\omega_K({q})+i \varepsilon},
\end{align}
\begin{align} 
 \label{eq:stret-negs-c}
V&_{stret-negs-c}^{\lambda' \lambda}(E,\vec{p'},\vec{p}\ )=f_S^4 {\cal I} N\int_0^{\Lambda} \frac{d^3{q}}{(2\pi)^3}\nonumber\\
&\times \frac{1}{16\omega_K({p}+{q})\omega_K({q})\omega^{(1)}({p+p'+q})\omega_K(p'+{q})}\nonumber\\
&\times \frac{1}{E-\omega_1(p)-\omega_K(p+q)-\omega_K({q})+i \varepsilon} \nonumber\\
&\times \frac{1}{E-\omega_1(p)-\omega_{1'}(p')-\omega_K(q)-\omega^{(1)}({p+p'+q})+i \varepsilon}\nonumber \\
&\times 1/ [E-\omega_{1}(p)-\omega_{1'}(p')-\omega_{2'}(p')\nonumber \\
&\qquad  \qquad  -\omega_K(p'+q)-\omega^{(1)}({p+p'+q})+i \varepsilon],
\end{align}
\begin{align} 
 \label{eq:stret-negs-d}
V&_{stret-negs-d}^{\lambda' \lambda}(E,\vec{p'},\vec{p}\ )=f_S^4 {\cal I} N\int_0^{\Lambda} \frac{d^3{q}}{(2\pi)^3}\nonumber\\
&\times \frac{1}{16\omega_K({p}+{q})\omega_K({q})\omega^{(1)}({p+p'+q})\omega_K(p'+{q})}\nonumber\\
&\times 1/[E-\omega_{1}(p)-\omega_{2}(p)-\omega_{1'}(p')-\omega_K(p+q)\nonumber\\
&\qquad  \qquad  -\omega^{(1)}({p+p'+q})+i \varepsilon ] \nonumber \\
&\times \frac{1}{E-\omega_1(p)-\omega_{1'}(p')-\omega_K(q)-\omega^{(1)}({p+p'+q})+i \varepsilon} \nonumber \\
&\times \frac{1}{E-\omega_{1'}(p')-\omega_K(p'+q)-\omega_K({q})+i \varepsilon},
\end{align}
\begin{align} 
 \label{eq:stret-negs-e}
V&_{stret-negs-e}^{\lambda' \lambda}(E,\vec{p'},\vec{p}\ )=f_S^4 {\cal I} N\int_0^{\Lambda} \frac{d^3{q}}{(2\pi)^3}\nonumber\\
&\times \frac{1}{16\omega_K({p}+{q})\omega_K({q})\omega^{(1)}({p+p'+q})\omega_K(p'+{q})}\nonumber\\
&\times \frac{1}{E-\omega_1(p)-\omega_K(p+q)-\omega_K({q})+i \varepsilon} \nonumber \\
&\times \frac{1}{E-\omega_1(p)-\omega_{1'}(p')-\omega_K(q)-\omega^{(1)}(p+p'+q)+i \varepsilon}\nonumber \\
&\times \frac{1}{E-\omega_{1'}(p')-\omega_K(p'+q)-\omega_K(q)+i \varepsilon},
\end{align}
\begin{align} 
 \label{eq:stret-negs-f}
V&_{stret-negs-f}^{\lambda' \lambda}(E,\vec{p'},\vec{p}\ )=f_S^4 {\cal I} N\int_0^{\Lambda} \frac{d^3{q}}{(2\pi)^3}\nonumber\\
&\times \frac{1}{16\omega_K({p}+{q})\omega_K({q})\omega^{(1)}({p+p'+q})\omega_K(p'+{q})}\nonumber\\
&\times 1/[E-\omega_{1}(p)-\omega_{2}(p)-\omega_{1'}(p') \nonumber\\
&\qquad  \qquad -\omega_K(p+q)-\omega^{(1)}({p+p'+q})+i \varepsilon] \nonumber \\
&\times \frac{1}{E-\omega_1(p)-\omega_{1'}(p')-\omega_K(q)-\omega^{(1)}(p+p'+q)+i \varepsilon}\nonumber \\
&\times 1/[E-\omega_{1}(p)-\omega_{1'}(p')-\omega_{2'}(p')\nonumber\\
&\qquad  \qquad -\omega_K(p'+q)-\omega^{(1)}({p+p'+q})+i \varepsilon].
\end{align}

In the framework of TOPT, the expressions corresponding to the diagrams of Fig.~\ref{fig:stret-negk} may be written in
the following form
\begin{align} 
\label{eq:stret-negk-a}
V&_{stret-negk-a}^{\lambda' \lambda}(E,\vec{p'},\vec{p}\ )=f_S^4 {\cal I} N\int_0^{\Lambda} \frac{d^3{q}}{(2\pi)^3}\nonumber\\
&\times \frac{1}{16\omega_K(p'-{q})\omega_K({q})\omega^{(1)}({p+p'-q})\omega_K({p}-{q})}\nonumber\\
&\times 1/[E-\omega_1(p)-\omega_2(p)-\omega_{2'}(p')-\omega_K(p'-q)\nonumber\\
&\qquad \qquad  -\omega_K({q})+i \varepsilon] \nonumber \\
&\times \frac{1}{E-\omega_1(p)-\omega_{2'}(p')-\omega_K(p-{q})-\omega_K(p'-{q})+i \varepsilon}\nonumber \\
&\times \frac{1}{E-\omega_{2'}(p')-\omega_K(p-q)-\omega^{(1)}({p+p'-q})+i\varepsilon},
\end{align}
\begin{align} 
\label{eq:stret-negk-b}
V&_{stret-negk-b}^{\lambda' \lambda}(E,\vec{p'},\vec{p}\ )=f_S^4 {\cal I} N\int_0^{\Lambda} \frac{d^3{q}}{(2\pi)^3}\nonumber\\
&\times \frac{1}{16\omega_K(p'-{q})\omega_K({q})\omega^{(1)}({p+p'-q})\omega_K({p}-{q})}\nonumber\\
&\times \frac{1}{E-\omega_{2}(p)-\omega_K(p'-q)-\omega^{(1)}({p+p'-q})+i\varepsilon}\nonumber\\
&\times \frac{1}{E-\omega_2(p)-\omega_{1'}(p')-\omega_K(p-{q})-\omega_K(p'-{q})+i \varepsilon}\nonumber \\
&\times 1/[E-\omega_2(p)-\omega_{1'}(p')-\omega_{2'}(p')-\omega_K(p-q)\nonumber \\
&\qquad \qquad -\omega_K({q})+i \varepsilon],
\end{align}
\begin{align} 
\label{eq:stret-negk-c}
V&_{stret-negk-c}^{\lambda' \lambda}(E,\vec{p'},\vec{p}\ )=f_S^4 {\cal I} N\int_0^{\Lambda} \frac{d^3{q}}{(2\pi)^3}\nonumber\\
&\times \frac{1}{16\omega_K(p'-{q})\omega_K({q})\omega^{(1)}({p+p'-q})\omega_K({p}-{q})}\nonumber\\
&\times 1/[E-\omega_1(p)-\omega_2(p)-\omega_{2'}(p')-\omega_K(p'-q)\nonumber \\
&\qquad \qquad -\omega_K({q})+i \varepsilon] \nonumber \\
&\times \frac{1}{E-\omega_2(p)-\omega_{2'}(p')-\omega_K(q)-\omega^{(1)}({p+p'-q})+i \varepsilon}\nonumber \\
&\times \frac{1}{E-\omega_{2'}(p')-\omega_K(p-q)-\omega^{(1)}({p+p'-q})+i\varepsilon},
\end{align}
\begin{align} 
\label{eq:stret-negk-d}
V&_{stret-negk-d}^{\lambda' \lambda}(E,\vec{p'},\vec{p}\ )=f_S^4 {\cal I} N\int_0^{\Lambda} \frac{d^3{q}}{(2\pi)^3}\nonumber\\
&\times \frac{1}{16\omega_K(p'-{q})\omega_K({q})\omega^{(1)}({p+p'-q})\omega_K({p}-{q})}\nonumber\\
&\times \frac{1}{E-\omega_{2}(p)-\omega_K(p'-q)-\omega^{(1)}({p+p'-q})+i\varepsilon}\nonumber\\
&\times \frac{1}{E-\omega_2(p)-\omega_{2'}(p')-\omega_K(q)-\omega^{(1)}({p+p'-q})+i \varepsilon}\nonumber \\
&\times 1/[E-\omega_2(p)-\omega_{1'}(p')-\omega_{2'}(p')-\omega_K(p-q)\nonumber \\
&\qquad\qquad-\omega_K({q})+i \varepsilon],
\end{align}
\begin{align} 
\label{eq:stret-negk-e}
V&_{stret-negk-e}^{\lambda' \lambda}(E,\vec{p'},\vec{p}\ )=f_S^4 {\cal I} N\int_0^{\Lambda} \frac{d^3{q}}{(2\pi)^3}\nonumber\\
&\times \frac{1}{16\omega_K(p'-{q})\omega_K({q})\omega^{(1)}({p+p'-q})\omega_K({p}-{q})}\nonumber\\
&\times 1/[E-\omega_1(p)-\omega_2(p)-\omega_{2'}(p')-\omega_K(p'-q) \nonumber \\
& \qquad\qquad -\omega_K({q})+i \varepsilon] \nonumber \\
&\times \frac{1}{E-\omega_2(p)-\omega_{2'}(p')-\omega_K(q)-\omega^{(1)}({p+p'-q})+i \varepsilon}\nonumber \\
&\times 1/[E-\omega_2(p)-\omega_{1'}(p')-\omega_{2'}(p')-\omega_K(p-q) \nonumber \\
&\qquad \qquad -\omega_K(q)+i\varepsilon],
\end{align}
\begin{align} 
\label{eq:stret-negk-f}
V&_{stret-negk-f}^{\lambda' \lambda}(E,\vec{p'},\vec{p}\ )=f_S^4 {\cal I} N\int_0^{\Lambda} \frac{d^3{q}}{(2\pi)^3}\nonumber\\
&\times \frac{1}{16\omega_K(p'-{q})\omega_K({q})\omega^{(1)}({p+p'-q})\omega_K({p}-{q})}\nonumber\\
&\times \frac{1}{E-\omega_{2}(p)-\omega_K(p'-q)-\omega^{(1)}({p+p'-q})+i\varepsilon}\nonumber\\
&\times \frac{1}{E-\omega_2(p)-\omega_{2'}(p')-\omega_K(q)-\omega^{(1)}({p+p'-q})+i \varepsilon}\nonumber \\
&\times \frac{1}{E-\omega_{2'}(p')-\omega_K(p-q)-\omega^{(1)}({p+p'-q})+i\varepsilon}.
\end{align}

In the framework of TOPT, the expression corresponding to Fig.~\ref{fig:stret-negks} may be written in
the following form
\begin{align} 
\label{eq:stret-negks-a}
V&_{stret-negks-a}^{\lambda' \lambda}(E,\vec{p'},\vec{p}\ )=f_S^4 {\cal I} N\int_0^{\Lambda} \frac{d^3{q}}{(2\pi)^3}\nonumber\\
&\times \frac{1}{16\omega_K(p'-{q})\omega_K({q})\omega^{(1)}({q})\omega_K({p}-{q})}\nonumber\\
&\times 1/[E-\omega_1(p)-\omega_2(p)-\omega_{2'}(p')-\omega_K(p'-q) \nonumber\\
&\qquad\qquad -\omega_K({q})+i \varepsilon] \nonumber \\
&\times \frac{1}{E-\omega_1(p)-\omega_{2'}(p')-\omega_K(p-{q})-\omega_K(p'-{q})+i \varepsilon}\nonumber \\
&\times 1/[E-\omega_1(p)-\omega_{1'}(p')-\omega_{2'}(p')-\omega_K(p-q)\nonumber\\
&\qquad\qquad-\omega^{(1)}({q})+i \varepsilon],
\end{align}
\begin{align} 
\label{eq:stret-negks-b}
V&_{stret-negks-b}^{\lambda' \lambda}(E,\vec{p'},\vec{p}\ )=f_S^4 {\cal I} N\int_0^{\Lambda} \frac{d^3{q}}{(2\pi)^3}\nonumber\\
&\times \frac{1}{16\omega_K(p'-{q})\omega_K({q})\omega^{(1)}({q})\omega_K({p}-{q})}\nonumber\\
&\times1/[E-\omega_1(p)-\omega_2(p)-\omega_{1'}(p')-\omega_K(p'-q)\nonumber\\
&\qquad\qquad -\omega^{(1)}({q})+i \varepsilon] \nonumber \\
&\times \frac{1}{E-\omega_2(p)-\omega_{1'}(p')-\omega_K(p-{q})-\omega_K(p'-{q})+i \varepsilon}\nonumber \\
&\times 1/[E-\omega_2(p)-\omega_{1'}(p')-\omega_{2'}(p')-\omega_K(p-q)\nonumber\\
&\qquad\qquad-\omega_K({q})+i \varepsilon],
\end{align}
\begin{align} 
\label{eq:stret-negks-c}
V&_{stret-negks-c}^{\lambda' \lambda}(E,\vec{p'},\vec{p}\ )=f_S^4 {\cal I} N\int_0^{\Lambda} \frac{d^3{q}}{(2\pi)^3}\nonumber\\
&\times \frac{1}{16\omega_K(p'-{q})\omega_K({q})\omega^{(1)}({q})\omega_K({p}-{q})}\nonumber\\
& \times 1/[E-\omega_1(p)-\omega_2(p)-\omega_{2'}(p')-\omega_K(p'-q)\nonumber \\
&\qquad  \qquad -\omega_K({q})+i \varepsilon] \nonumber \\
&\times 1/[E-\omega_1(p)-\omega_2(p)-\omega_{1'}(p')-\omega_{2'}(p')\nonumber \\
&\qquad  \qquad -\omega_K(q)-\omega^{(1)}({q})+i \varepsilon]\nonumber \\
&\times 1/ [E-\omega_1(p)-\omega_{1'}(p')-\omega_{2'}(p')-\omega_K(p-q)\nonumber \\
&\qquad  \qquad -\omega^{(1)}({q})+i \varepsilon] ,
\end{align}
\begin{align} 
\label{eq:stret-negks-d}
V&_{stret-negks-d}^{\lambda' \lambda}(E,\vec{p'},\vec{p}\ )=f_S^4 {\cal I} N\int_0^{\Lambda} \frac{d^3{q}}{(2\pi)^3}\nonumber\\
&\times \frac{1}{16\omega_K(p'-{q})\omega_K({q})\omega^{(1)}({q})\omega_K({p}-{q})}\nonumber\\
&\times 1/[E-\omega_1(p)-\omega_2(p)-\omega_{1'}(p')-\omega_K(p'-q) \nonumber \\
&\qquad  \qquad -\omega^{(1)}({q})+i \varepsilon] \nonumber \\
&\times 1/ [E-\omega_1(p)-\omega_2(p)-\omega_{1'}(p')-\omega_{2'}(p') \nonumber \\
&\qquad  \qquad -\omega_K(q)-\omega^{(1)}({q})+i \varepsilon ]\nonumber \\
&\times 1/[E-\omega_2(p)-\omega_{1'}(p')-\omega_{2'}(p')-\omega_K(p-q)\nonumber \\
&\qquad  \qquad -\omega_K({q})+i \varepsilon],
\end{align}
\begin{align} 
\label{eq:stret-negks-e}
V&_{stret-negks-e}^{\lambda' \lambda}(E,\vec{p'},\vec{p}\ )=f_S^4 {\cal I} N\int_0^{\Lambda} \frac{d^3{q}}{(2\pi)^3}\nonumber\\
&\times \frac{1}{16\omega_K(p'-{q})\omega_K({q})\omega^{(1)}({q})\omega_K({p}-{q})}\nonumber\\
&\times 1/[E-\omega_1(p)-\omega_2(p)-\omega_{2'}(p')-\omega_K(p'-q)\nonumber \\
&\qquad  \qquad-\omega_K({q})+i \varepsilon] \nonumber \\
&\times 1/[E-\omega_1(p)-\omega_2(p)-\omega_{1'}(p')-\omega_{2'}(p')\nonumber \\
&\qquad  \qquad -\omega_K(q)-\omega^{(1)}({q})+i \varepsilon]\nonumber \\
&\times 1/[E-\omega_2(p)-\omega_{1'}(p')-\omega_{2'}(p')-\omega_K(p-q)\nonumber \\
&\qquad  \qquad -\omega_K({q})+i \varepsilon],
\end{align}
\begin{align} 
\label{eq:stret-negks-f}
V&_{stret-negks-f}^{\lambda' \lambda}(E,\vec{p'},\vec{p}\ )=f_S^4 {\cal I} N\int_0^{\Lambda} \frac{d^3{q}}{(2\pi)^3}\nonumber\\
&\times \frac{1}{16\omega_K(p'-{q})\omega_K({q})\omega^{(1)}({q})\omega_K({p}-{q})}\nonumber\\
&\times 1/[E-\omega_1(p)-\omega_2(p)-\omega_{1'}(p')-\omega_K(p'-q) \nonumber \\
&\qquad\qquad-\omega^{(1)}({q})+i \varepsilon] \nonumber \\
&\times 1/[E-\omega_1(p)-\omega_2(p)-\omega_{1'}(p')-\omega_{2'}(p')\nonumber \\
&\qquad  \qquad -\omega_K(q)-\omega^{(1)}({q})+i \varepsilon]\nonumber \\
&\times 1/[E-\omega_1(p)-\omega_{1'}(p')-\omega_{2'}(p')-\omega_K(p-q)\nonumber \\
&\qquad  \qquad -\omega^{(1)}({q})+i \varepsilon].
\end{align}

In the framework of TOPT, the expressions corresponding to the diagrams of Fig.~\ref{fig:crossed} may be written in
the following form
\begin{align} 
 \label{eq:cross-a}
V&_{cros-a}^{\lambda' \lambda}(E,\vec{p'},\vec{p}\ )=f_S^4 {\cal I} N\int_0^{\Lambda} \frac{d^3{q}}{(2\pi)^3}\nonumber \\
&\times \frac{1}{16\omega_K({p}+{q})\omega_K({q})\omega^{(1)}({p+p'+q})\omega_K(p'+{q})}\nonumber\\
&\times \frac{1}{E-\omega_1(p)-\omega_K(p+q)-\omega_K({q})+i \varepsilon} \nonumber \\
&\times \frac{1}{E-\omega_1(p)-\omega_K(p+q)-\omega_K(p'+q)-\omega_{2'}(p')+i \varepsilon}\nonumber \\
&\times \frac{1}{E-\omega_{2'}(p')-\omega_K(p+q)-\omega^{(1)}(p+p'+q)+i \varepsilon},
\end{align}
 \begin{align} 
  \label{eq:cross-b}
V&_{cros-b}^{\lambda' \lambda}(E,\vec{p'},\vec{p}\ )=f_S^4 {\cal I} N\int_0^{\Lambda} \frac{d^3{q}}{(2\pi)^3}\nonumber \\
&\times \frac{1}{16\omega_K({p}+{q})\omega_K({q})\omega^{(1)}({p+p'+q})\omega_K(p'+{q})}\nonumber\\
&\times \frac{1}{E-\omega_2(p)-\omega_K(p'+q)-\omega^{(1)}(p+p'+q)+i \varepsilon} \nonumber \\
&\times \frac{1}{E-\omega_{2}(p)-\omega_K(p+q)-\omega_K(p'+q)-\omega_{1'}(p')+i \varepsilon}\nonumber \\
&\times \frac{1}{E-\omega_{1'}(p')-\omega_K(p'+q)-\omega_{K}({q})+i \varepsilon},
\end{align}
 \begin{align} 
  \label{eq:cross-c}
V&_{cros-c}^{\lambda' \lambda}(E,\vec{p'},\vec{p}\ )=f_S^4 {\cal I} N\int_0^{\Lambda} \frac{d^3{q}}{(2\pi)^3}\nonumber \\
&\times \frac{1}{16\omega_K({p}+{q})\omega_K({q})\omega^{(1)}({p+p'+q})\omega_K(p'+{q})}\nonumber\\
&\times \frac{1}{E-\omega_1(p)-\omega_K(p+q)-\omega_K({q})+i \varepsilon} \nonumber \\
&\times 1/[E-\omega_K(q)-\omega_K(p+q)-\omega_K(p'+q)\nonumber \\
&\qquad  \qquad -\omega^{(1)}(p+p'+q)+i \varepsilon]\nonumber \\
&\times \frac{1}{E-\omega_{2'}(p')-\omega_K(p+q)-\omega^{(1)}(p+p'+q)+i \varepsilon},
\end{align}
 \begin{align} 
  \label{eq:cross-d}
V&_{cros-d}^{\lambda' \lambda}(E,\vec{p'},\vec{p}\ )=f_S^4 {\cal I} N\int_0^{\Lambda} \frac{d^3{q}}{(2\pi)^3}\nonumber \\
&\times \frac{1}{16\omega_K({p}+{q})\omega_K({q})\omega^{(1)}({p+p'+q})\omega_K(p'+{q})}\nonumber\\
&\times \frac{1}{E-\omega_2(p)-\omega_K(p'+q)-\omega^{(1)}(p+p'+q)+i \varepsilon} \nonumber \\
&\times 1/[E-\omega_K(q)-\omega_K(p+q)-\omega_K(p'+q)\nonumber \\
&\qquad  \qquad -\omega^{(1)}(p+p'+q)+i \varepsilon]\nonumber \\
&\times \frac{1}{E-\omega_{1'}(p')-\omega_K(p'+q)-\omega_{K}({q})+i \varepsilon},
\end{align}
 \begin{align} 
  \label{eq:cross-e}
V&_{cros-e}^{\lambda' \lambda}(E,\vec{p'},\vec{p}\ )=f_S^4 {\cal I} N\int_0^{\Lambda} \frac{d^3{q}}{(2\pi)^3}\nonumber \\
&\times \frac{1}{16\omega_K({p}+{q})\omega_K({q})\omega^{(1)}({p+p'+q})\omega_K(p'+{q})}\nonumber\\
&\times \frac{1}{E-\omega_1(p)-\omega_K(p+q)-\omega_K({q})+i \varepsilon} \nonumber \\
&\times 1/[E-\omega_K(q)-\omega_K(p+q)-\omega_K(p'+q)\nonumber \\
&\qquad  \qquad -\omega^{(1)}(p+p'+q)+i \varepsilon]\nonumber \\
&\times \frac{1}{E-\omega_{1'}(p')-\omega_K(p'+q)-\omega_{K}({q})+i \varepsilon},
\end{align}
 \begin{align} 
  \label{eq:cross-f}
V&_{cros-f}^{\lambda' \lambda}(E,\vec{p'},\vec{p}\ )=f_S^4 {\cal I} N\int_0^{\Lambda} \frac{d^3{q}}{(2\pi)^3}\nonumber \\
&\times \frac{1}{16\omega_K({p}+{q})\omega_K({q})\omega^{(1)}({p+p'+q})\omega_K(p'+{q})}\nonumber\\
&\times \frac{1}{E-\omega_2(p)-\omega_K(p'+q)-\omega^{(1)}(p+p'+q)+i \varepsilon} \nonumber \\
&\times 1/[E-\omega_K(q)-\omega_K(p+q)-\omega_K(p'+q)\nonumber \\
&\qquad  \qquad -\omega^{(1)}(p+p'+q)+i \varepsilon]\nonumber \\
&\times \frac{1}{E-\omega_{2'}(p')-\omega_K(p+q)-\omega^{(1)}(p+p'+q)+i \varepsilon}.
\end{align}

In the framework of TOPT, the expressions corresponding to the diagrams of Fig.~\ref{fig:crossed-negs} may be written in
the following form
\begin{align} 
 \label{eq:cross-negs-a}
V&_{cros-negs-a}^{\lambda' \lambda}(E,\vec{p'},\vec{p}\ )=f_S^4 {\cal I} N\int_0^{\Lambda} \frac{d^3{q}}{(2\pi)^3}\nonumber \\
&\times \frac{1}{16\omega_K({p}+{q})\omega_K({q})\omega^{(1)}({q})\omega_K(p'+{q})}\nonumber\\
&\times \frac{1}{E-\omega_1(p)-\omega_K(p+q)-\omega_K({q})+i \varepsilon} \nonumber \\
&\times \frac{1}{E-\omega_1(p)-\omega_{2'}(p')-\omega_K(p+q)-\omega_K(p'+q)+i \varepsilon}\nonumber \\
&\times 1/[E-\omega_1(p)-\omega_{1'}(p')-\omega_{2'}(p')-\omega_K(p+q)\nonumber \\
&\qquad\qquad -\omega^{(1)}(q)+i \varepsilon],
\end{align}
\begin{align} 
\label{eq:cross-negs-b}
V&_{cros-negs-b}^{\lambda' \lambda}(E,\vec{p'},\vec{p}\ )=f_S^4 {\cal I} N\int_0^{\Lambda} \frac{d^3{q}}{(2\pi)^3}\nonumber \\
&\times \frac{1}{16\omega_K({p}+{q})\omega_K({q})\omega^{(1)}({q})\omega_K(p'+{q})}\nonumber\\
&\times 1/[E-\omega_1(p)-\omega_2(p)-\omega_{1'}(p')-\omega_K(p'+{q})\nonumber\\
&\qquad\qquad-\omega^{(1)}(q)+i \varepsilon] \nonumber \\
&\times \frac{1}{E-\omega_{2}(p)-\omega_{1'}(p')-\omega_K(p+q)-\omega_K(p'+q)+i \varepsilon}\nonumber \\
&\times \frac{1}{E-\omega_{1'}(p')-\omega_K(p'+q)-\omega_{K}({q})+i \varepsilon},
\end{align}
\begin{align} 
 \label{eq:cross-negs-c}
V&_{cros-negs-c}^{\lambda' \lambda}(E,\vec{p'},\vec{p}\ )=f_S^4 {\cal I} N\int_0^{\Lambda} \frac{d^3{q}}{(2\pi)^3}\nonumber \\
&\times \frac{1}{16\omega_K({p}+{q})\omega_K({q})\omega^{(1)}({q})\omega_K(p'+{q})}\nonumber\\
&\times \frac{1}{E-\omega_1(p)-\omega_K(p+q)-\omega_K({q})+i \varepsilon} \nonumber \\
&\times 1/[E-\omega_1(p)-\omega_{1'}(p')-\omega_K(p+q)-\omega_K(p'+q)\nonumber \\
&\qquad  \qquad -\omega_K({q})-\omega^{(1)}({q})+i \varepsilon]\nonumber \\
&\times 1/[E-\omega_1(p)-\omega_{1'}(p')-\omega_{2'}(p')-\omega_K(p+q) \nonumber\\
&\qquad\qquad-\omega^{(1)}(q)+i \varepsilon],
\end{align}
\begin{align} 
\label{eq:cross-negs-d}
V&_{cros-negs-d}^{\lambda' \lambda}(E,\vec{p'},\vec{p}\ )=f_S^4 {\cal I} N\int_0^{\Lambda} \frac{d^3{q}}{(2\pi)^3}\nonumber \\
&\times \frac{1}{16\omega_K({p}+{q})\omega_K({q})\omega^{(1)}({q})\omega_K(p'+{q})}\nonumber\\
&\times 1/[E-\omega_1(p)-\omega_2(p)-\omega_{1'}(p')-\omega_K(p'+{q})\nonumber\\
&\qquad\qquad -\omega^{(1)}(q)+i \varepsilon] \nonumber \\
&\times 1/[E-\omega_1(p)-\omega_{1'}(p')-\omega_K(p+q)-\omega_K(p'+q)\nonumber \\
&\qquad  \qquad -\omega_K({q})-\omega^{(1)}({q})+i \varepsilon]\nonumber \\
&\times \frac{1}{E-\omega_{1'}(p')-\omega_K(p'+q)-\omega_{K}({q})+i \varepsilon},
\end{align}
\begin{align} 
 \label{eq:cross-negs-e}
V&_{cros-negs-e}^{\lambda' \lambda}(E,\vec{p'},\vec{p}\ )=f_S^4 {\cal I} N\int_0^{\Lambda} \frac{d^3{q}}{(2\pi)^3}\nonumber \\
&\times \frac{1}{16\omega_K({p}+{q})\omega_K({q})\omega^{(1)}({q})\omega_K(p'+{q})}\nonumber\\
&\times \frac{1}{E-\omega_1(p)-\omega_K(p+q)-\omega_K({q})+i \varepsilon} \nonumber \\
&\times 1/[E-\omega_1(p)-\omega_{1'}(p')-\omega_K(p+q)-\omega_K(p'+q)\nonumber \\
&\qquad  \qquad -\omega_K({q})-\omega^{(1)}({q})+i \varepsilon]\nonumber \\
&\times \frac{1}{E-\omega_{1'}(p')-\omega_K(p'+q)-\omega_{K}({q})+i \varepsilon},
\end{align}
\begin{align} 
\label{eq:cross-negs-f}
V&_{cros-negs-f}^{\lambda' \lambda}(E,\vec{p'},\vec{p}\ )=f_S^4 {\cal I} N\int_0^{\Lambda} \frac{d^3{q}}{(2\pi)^3}\nonumber \\
&\times \frac{1}{16\omega_K({p}+{q})\omega_K({q})\omega^{(1)}({q})\omega_K(p'+{q})}\nonumber\\
&\times 1/[E-\omega_1(p)-\omega_2(p)-\omega_{1'}(p')-\omega_K(p'+{q})\nonumber\\
&\qquad\qquad -\omega^{(1)}(q)+i \varepsilon] \nonumber \\
&\times 1/[E-\omega_1(p)-\omega_{1'}(p')-\omega_K(p+q)-\omega_K(p'+q)\nonumber \\
&\qquad  \qquad -\omega_K({q})-\omega^{(1)}({q})+i \varepsilon]\nonumber \\
&\times 1/[E-\omega_1(p)-\omega_{1'}(p')-\omega_{2'}(p')-\omega_K(p+q)\nonumber\\
&\qquad\qquad -\omega^{(1)}(q)+i \varepsilon].
\end{align}

In the framework of TOPT, the expressions corresponding to the diagrams of Fig.~\ref{fig:crossed-negk} may be written in
the following form
\begin{align} 
\label{eq:cross-negk-a}
V&_{cros-negk-a}^{\lambda' \lambda}(E,\vec{p'},\vec{p}\ )=f_S^4 {\cal I} N\int_0^{\Lambda} \frac{d^3{q}}{(2\pi)^3}\nonumber\\
&\times \frac{1}{16\omega_K(p'-{q})\omega_K({q})\omega^{(1)}({q})\omega_K({p}-{q})}\nonumber\\
&\times 1/[E-\omega_1(p)-\omega_2(p)-\omega_{2'}(p')-\omega_K(p'-q)\nonumber \\
&\qquad \qquad -\omega_K({q})+i \varepsilon] \nonumber \\
&\times \frac{1}{E-\omega_1(p)-\omega_{2'}(p')-\omega_K(p-{q})-\omega_K(p'-{q})+i \varepsilon}\nonumber \\
&\times \frac{1}{E-\omega_{2'}(p')-\omega_K(p'-q)-\omega^{(1)}({q})+i \varepsilon} ,
\end{align}
\begin{align} 
\label{eq:cross-negk-b}
V&_{cros-negk-b}^{\lambda' \lambda}(E,\vec{p'},\vec{p}\ )=f_S^4 {\cal I} N\int_0^{\Lambda} \frac{d^3{q}}{(2\pi)^3}\nonumber\\
&\times \frac{1}{16\omega_K(p'-{q})\omega_K({q})\omega^{(1)}({q})\omega_K({p}-{q})}\nonumber\\
&\times \frac{1}{E-\omega_2(p)-\omega_K(p-q)-\omega^{(1)}({q})+i \varepsilon} \nonumber \\
&\times \frac{1}{E-\omega_2(p)-\omega_{1'}(p')-\omega_K(p-{q})-\omega_K(p'-{q})+i \varepsilon}\nonumber \\
&\times 1/[E-\omega_2(p)-\omega_{1'}(p')-\omega_{2'}(p')-\omega_K(p-q) \nonumber \\
&\qquad \qquad -\omega_K({q})+i \varepsilon],
\end{align}
\begin{align} 
\label{eq:cross-negk-c}
V&_{cros-negk-c}^{\lambda' \lambda}(E,\vec{p'},\vec{p}\ )=f_S^4 {\cal I} N\int_0^{\Lambda} \frac{d^3{q}}{(2\pi)^3}\nonumber\\
&\times \frac{1}{16\omega_K(p'-{q})\omega_K({q})\omega^{(1)}({q})\omega_K({p}-{q})}\nonumber\\
&\times 1/[E-\omega_1(p)-\omega_2(p)-\omega_{2'}(p')-\omega_K(p'-q)\nonumber\\
&\qquad \qquad -\omega_K({q})+i \varepsilon] \nonumber \\
&\times  1/[E-\omega_2(p)-\omega_{2'}(p')-\omega_K(p-{q})-\omega_K(p'-{q}) \nonumber \\
&\qquad  \qquad -\omega_K({q})-\omega^{(1)}({q})+i \varepsilon]\nonumber \\
&\times \frac{1}{E-\omega_{2'}(p')-\omega_K(p'-q)-\omega^{(1)}({q})+i \varepsilon},
\end{align}
\begin{align} 
\label{eq:cross-negk-d}
V&_{cros-negk-d}^{\lambda' \lambda}(E,\vec{p'},\vec{p}\ )=f_S^4 {\cal I} N\int_0^{\Lambda} \frac{d^3{q}}{(2\pi)^3}\nonumber\\
&\times \frac{1}{16\omega_K(p'-{q})\omega_K({q})\omega^{(1)}({q})\omega_K({p}-{q})}\nonumber\\
&\times \frac{1}{E-\omega_2(p)-\omega_K(p-q)-\omega^{(1)}({q})+i \varepsilon} \nonumber \\
&\times  1/[E-\omega_2(p)-\omega_{2'}(p')-\omega_K(p-{q})-\omega_K(p'-{q}) \nonumber \\
&\qquad  \qquad -\omega_K({q})-\omega^{(1)}({q})+i \varepsilon]\nonumber \\
&\times 1/[E-\omega_2(p)-\omega_{1'}(p')-\omega_{2'}(p')-\omega_K(p-q)\nonumber\\
&\qquad \qquad -\omega_K({q})+i \varepsilon],
\end{align}
\begin{align} 
\label{eq:cross-negk-e}
V&_{cros-negk-e}^{\lambda' \lambda}(E,\vec{p'},\vec{p}\ )=f_S^4 {\cal I} N\int_0^{\Lambda} \frac{d^3{q}}{(2\pi)^3}\nonumber\\
&\times \frac{1}{16\omega_K(p'-{q})\omega_K({q})\omega^{(1)}({q})\omega_K({p}-{q})}\nonumber\\
&\times 1/[E-\omega_1(p)-\omega_2(p)-\omega_{2'}(p')-\omega_K(p'-q)\nonumber\\
&\qquad\qquad -\omega_K({q})+i \varepsilon] \nonumber \\
&\times  1/[E-\omega_2(p)-\omega_{2'}(p')-\omega_K(p-{q})-\omega_K(p'-{q}) \nonumber \\
&\qquad  \qquad -\omega_K({q})-\omega^{(1)}({q})+i \varepsilon]\nonumber \\
&\times 1/[E-\omega_2(p)-\omega_{1'}(p')-\omega_{2'}(p')-\omega_K(p-q)\nonumber\\
&\qquad\qquad -\omega_K({q})+i \varepsilon],
\end{align}
\begin{align} 
\label{eq:cross-negk-f}
V&_{cros-negk-f}^{\lambda' \lambda}(E,\vec{p'},\vec{p}\ )=f_S^4 {\cal I} N\int_0^{\Lambda} \frac{d^3{q}}{(2\pi)^3}\nonumber\\
&\times \frac{1}{16\omega_K(p'-{q})\omega_K({q})\omega^{(1)}({q})\omega_K({p}-{q})}\nonumber\\
&\times \frac{1}{E-\omega_2(p)-\omega_K(p-q)-\omega^{(1)}({q})+i \varepsilon} \nonumber \\
&\times  1/[E-\omega_2(p)-\omega_{2'}(p')-\omega_K(p-{q})-\omega_K(p'-{q}) \nonumber \\
&\qquad  \qquad -\omega_K({q})-\omega^{(1)}({q})+i \varepsilon]\nonumber \\
&\times \frac{1}{E-\omega_{2'}(p')-\omega_K(p'-q)-\omega^{(1)}({q})+i \varepsilon}.
\end{align}

In the framework of TOPT, the expressions corresponding to the diagrams of Fig.~\ref{fig:crossed-negks} may be written in
the following form
\begin{align} 
\label{eq:cross-negks-a}
V&_{cros-negks-a}^{\lambda' \lambda}(E,\vec{p'},\vec{p}\ )=f_S^4 {\cal I} N\int_0^{\Lambda} \frac{d^3{q}}{(2\pi)^3}\nonumber\\
&\times \frac{1}{16\omega_K(p'-{q})\omega_K({q})\omega^{(1)}(p+p'-q)\omega_K({p}-{q})}\nonumber\\
&\times 1/[E-\omega_1(p)-\omega_2(p)-\omega_{2'}(p')-\omega_K(p'-q)\nonumber\\
&\qquad\qquad-\omega_K({q})+i \varepsilon] \nonumber \\
&\times \frac{1}{E-\omega_1(p)-\omega_{2'}(p')-\omega_K(p-{q})-\omega_K(p'-{q})+i \varepsilon}\nonumber \\
&\times  1/[E-\omega_1(p)-\omega_{1'}(p')-\omega_{2'}(p')-\omega_K(p'-q) \nonumber \\
&\qquad  \qquad -\omega^{(1)}(p+p'-q)+i \varepsilon],
\end{align}
\begin{align} 
\label{eq:cross-negks-b}
V&_{cros-negks-b}^{\lambda' \lambda}(E,\vec{p'},\vec{p}\ )=f_S^4 {\cal I} N\int_0^{\Lambda} \frac{d^3{q}}{(2\pi)^3}\nonumber\\
&\times \frac{1}{16\omega_K(p'-{q})\omega_K({q})\omega^{(1)}(p+p'-q)\omega_K({p}-{q})}\nonumber\\
&\times  1/[E-\omega_1(p)-\omega_2(p)-\omega_{1'}(p')-\omega_K(p-{q})\nonumber\\
&\qquad  \qquad-\omega^{(1)}(p+p'-q)+i \varepsilon] \nonumber \\
&\times \frac{1}{E-\omega_2(p)-\omega_{1'}(p')-\omega_K(p-{q})-\omega_K(p'-{q})+i \varepsilon}\nonumber \\
&\times 1/[E-\omega_2(p)-\omega_{1'}(p')-\omega_{2'}(p')-\omega_K(p-q)\nonumber\\
&\qquad\qquad-\omega_K({q})+i \varepsilon],
\end{align}
\begin{align} 
\label{eq:cross-negks-c}
V&_{cros-negks-c}^{\lambda' \lambda}(E,\vec{p'},\vec{p}\ )=f_S^4 {\cal I} N\int_0^{\Lambda} \frac{d^3{q}}{(2\pi)^3}\nonumber\\
&\times \frac{1}{16\omega_K(p'-{q})\omega_K({q})\omega^{(1)}(p+p'-q)\omega_K({p}-{q})}\nonumber\\
&\times 1/[E-\omega_1(p)-\omega_2(p)-\omega_{2'}(p')-\omega_K(p'-q)\nonumber\\
&\qquad\qquad-\omega_K({q})+i \varepsilon] \nonumber \\
&\times  1/[E-\omega_1(p)-\omega_{1'}(p)-\omega_2(p)-\omega_{2'}(p')-\omega_K(p-{q})\nonumber \\
&\qquad  \qquad -\omega_K(p'-{q})-\omega_K({q})-\omega^{(1)}(p+p'-q)+i \varepsilon]\nonumber \\
&\times  1/[E-\omega_1(p)-\omega_{1'}(p')-\omega_{2'}(p')-\omega_K(p'-q) \nonumber \\
&\qquad  \qquad -\omega^{(1)}(p+p'-q)+i \varepsilon],
\end{align}
\begin{align} 
\label{eq:cross-negks-d}
V&_{cros-negks-d}^{\lambda' \lambda}(E,\vec{p'},\vec{p}\ )=f_S^4 {\cal I} N\int_0^{\Lambda} \frac{d^3{q}}{(2\pi)^3}\nonumber\\
&\times \frac{1}{16\omega_K(p'-{q})\omega_K({q})\omega^{(1)}(p+p'-q)\omega_K({p}-{q})}\nonumber\\
&\times  1/[E-\omega_1(p)-\omega_2(p)-\omega_{1'}(p')-\omega_K(p-{q})\nonumber\\
&\qquad  \qquad-\omega^{(1)}(p+p'-q)+i \varepsilon] \nonumber \\
&\times  1/[E-\omega_1(p)-\omega_{1'}(p)-\omega_2(p)-\omega_{2'}(p')-\omega_K(p-{q})\nonumber \\
&\qquad  \qquad -\omega_K(p'-{q})-\omega_K({q})-\omega^{(1)}(p+p'-q)+i \varepsilon]\nonumber \\
&\times 1/[E-\omega_2(p)-\omega_{1'}(p')-\omega_{2'}(p')-\omega_K(p-q) \nonumber\\
&\qquad\qquad -\omega_K({q})+i \varepsilon],
\end{align}
\begin{align} 
\label{eq:cross-negks-e}
V&_{cros-negks-e}^{\lambda' \lambda}(E,\vec{p'},\vec{p}\ )=f_S^4 {\cal I} N\int_0^{\Lambda} \frac{d^3{q}}{(2\pi)^3}\nonumber\\
&\times \frac{1}{16\omega_K(p'-{q})\omega_K({q})\omega^{(1)}(p+p'-q)\omega_K({p}-{q})}\nonumber\\
&\times 1/[E-\omega_1(p)-\omega_2(p)-\omega_{2'}(p')-\omega_K(p'-q)\nonumber\\
&\qquad\qquad-\omega_K({q})+i \varepsilon] \nonumber \\
&\times  1/[E-\omega_1(p)-\omega_{1'}(p)-\omega_2(p)-\omega_{2'}(p')-\omega_K(p-{q})\nonumber \\
&\qquad  \qquad -\omega_K(p'-{q})-\omega_K({q})-\omega^{(1)}(p+p'-q)+i \varepsilon]\nonumber \\
&\times  1/[E-\omega_2(p)-\omega_{1'}(p')-\omega_{2'}(p')-\omega_K(p-q) \nonumber \\
&\qquad  \qquad -\omega_K({q})+i \varepsilon],
\end{align}
\begin{align} 
\label{eq:cross-negks-f}
V&_{cros-negks-f}^{\lambda' \lambda}(E,\vec{p'},\vec{p}\ )=f_S^4 {\cal I} N\int_0^{\Lambda} \frac{d^3{q}}{(2\pi)^3}\nonumber\\
&\times \frac{1}{16\omega_K(p'-{q})\omega_K({q})\omega^{(1)}(p+p'-q)\omega_K({p}-{q})}\nonumber\\
&\times  1/[E-\omega_1(p)-\omega_2(p)-\omega_{1'}(p')-\omega_K(p-{q})\nonumber\\
&\qquad  \qquad-\omega^{(1)}(p+p'-q)+i \varepsilon] \nonumber \\
&\times  1/[E-\omega_1(p)-\omega_{1'}(p)-\omega_2(p)-\omega_{2'}(p')-\omega_K(p-{q})\nonumber \\
&\qquad  \qquad -\omega_K(p'-{q})-\omega_K({q})-\omega^{(1)}(p+p'-q)+i \varepsilon]\nonumber \\
&\times  1/[E-\omega_1(p)-\omega_{1'}(p')-\omega_{2'}(p')-\omega_K(p'-q)\nonumber \\
&\qquad  \qquad -\omega^{(1)}(p+p'-q)+i \varepsilon].
\end{align}

\end{appendix}

\end{document}